\documentclass[11pt, oneside]{article}   	
\usepackage{geometry}                		
\usepackage[indent=12pt,skip=1pt]{parskip}    	
	
\usepackage{graphicx}
\usepackage[dvipsnames]{xcolor}				
										
\usepackage{amssymb,enumerate,amsmath,amsthm,dsfont}

\usepackage{setspace,subcaption,multirow,comment}
\usepackage{natbib}
\usepackage{soul}
\usepackage{tikz,makecell}
\usepackage{pdfpages}

\newtheorem{proposition}{Proposition}
\newtheorem{theorem}{Theorem}
\newtheorem{assumption}{Assumption}

\newtheorem{cor}{Corollary}

\newtheorem*{assump-2a}{Assumption 2a}
\newtheorem*{assump}{Assumption 3a}
\newtheorem*{prop3}{Proposition 4}
\newtheorem{lem}{Lemma}
\newtheorem*{thm2}{Theorem 2}

\theoremstyle{definition}
\newtheorem{definition}{Definition}

\title{Semi-Parametric Sensitivity Analysis for Trials with \\ Irregular and Informative Assessment Times}

\author{Bonnie B. Smith$^{1,*}$, Yujing Gao$^2$, Shu Yang$^{2}$,  Ravi Varadhan$^{3}$, \\ Andrea J. Apter$^{4}$, and Daniel O. Scharfstein$^{5}$ }

\date{}

\begin{document}

\maketitle

\ 

\small 

\noindent \textsc{Summary:}  Many trials are designed to collect outcomes at or around pre-specified times after randomization.  If there is variability in the times when participants are actually assessed, this can pose a challenge to learning the effect of treatment, since not all participants have outcome assessments at the times of interest. Furthermore, observed outcome values may not be representative of all participants' outcomes at a given time.  Methods have been developed that account for some types of such irregular and informative assessment times;  however, since these methods rely on untestable assumptions, sensitivity analyses are needed. We develop a methodology that is benchmarked at the explainable assessment (EA) assumption, under which assessment and outcomes at each time are related only through data collected prior to that time.  Our method uses an exponential tilting assumption, governed by a sensitivity analysis parameter, that posits deviations from the EA assumption. Our inferential strategy is based on a new influence function-based, augmented inverse intensity-weighted estimator. Our approach allows for flexible semiparametric modeling of the observed data, which is separated from specification of the sensitivity parameter. We apply our method to a randomized trial of low-income individuals with uncontrolled asthma, and we illustrate implementation of our estimation procedure in detail.

\


\

\ 

\footnotesize

\noindent $^{1}$ Department of Biostatistics, Johns Hopkins Bloomberg School of Public Health, Baltimore, MD, USA \\
$^{2}$ Department of Statistics, North Carolina State University, Raleigh, NC, USA \\
$^{3}$ Department of Oncology, Johns Hopkins School of Medicine,  Baltimore, MD, USA \\
$^4$ Pulmonary Allergy Critical Care Division, Department of Medicine, Perelman  School of Medicine, \\ \hspace{0.2in} University of Pennsylvania,  Philadelphia, PA, USA \\
$^{5}$ Department of Population Health Sciences, University of Utah School of Medicine, Salt Lake City, \\ \hspace{0.2in} UT, USA \\
$^*$ \emph{email}:  bsmit179@jhmi.edu

\normalsize

\newpage

\section{Introduction}  \label{sec:intro}

Many randomized trials are designed to collect outcome information at or around certain pre-specified times after randomization.  In practice, however, there can be substantial variability in the times when participants' outcomes are actually assessed.  Such \emph{irregular assessment times} pose a challenge to learning the effect of treatment, similar to that posed by missing data.  While the goal is to learn population mean outcomes and treatment effects at certain target times, not all participants are assessed at those times; and the observed outcomes may not be representative. For example, participants may miss or postpone data collection appointments at times when their outcome is worse, such that outcomes in the study data tend to be better compared to the population distribution.  In other studies, participants may tend to have assessments at times when their outcome is worse---for example, if the study collects data at ``as-needed" appointments.  We say the assessment times are \emph{informative} if the distribution of observed outcomes at a given time differs from the population distribution of outcomes at that time.

A number of inferential methods have been developed for prospective studies with informative assessment times.  All approaches impose untestable assumptions about the joint distribution of the outcome and assessment time processes.  \citet{lin2001regression} posited semi-parametric regression models with time-varying covariates for the outcome and assessment time processes, and they used an assumption that the two processes are conditionally independent given these time-varying covariates to construct estimating equations for the outcome regression parameters.  Their approach was generalized by several authors to allow for dependence between the outcome and assessment time processes through latent variables (e.g. random effects and frailty terms) in addition to covariates in the outcome regression model; see for example \citet*{sun2007regression}, \citet*{liang2009jointmodeling}, \citet{sun2011semiparametric}, and \citet*{sun2011regression}.  \citet*{scharfstein2004dependent} instead developed an inverse weighting approach, also within an estimating equations framework, under which the outcome and assessment time processes can be associated through past observed outcomes and time-varying covariates which are not included in the outcome model.  Therefore, their approach allows inference for the marginal mean of the outcome process.  Inverse weighting approaches have also been developed by \citet*{buzkova2007dependent}, \citet*{buzkova2009repeated}, \citet*{pullenayegum2013dr}, and \citet{sun2016quantile}.  Other authors have used likelihood-based approaches coupled with assumptions that obviate the need for modeling the assessment time process: \citet{lipsitz2002dependent} used a parametric approach, while \citet*{chen2015regression} and \citet{shen2019regression} used composite likelihoods conditioned on order statistics to express the conditional density of observed outcomes in terms of the outcome density of interest.   As noted above, the key caveat for all of these approaches is that untestable assumptions are needed; therefore, sensitivity analysis would be a valuable addition to each method.  This is analogous to methods for trials with missing data, which require untestable assumptions such as missing at random.  There, sensitivity analysis has been recognized as an important component of the analysis; see, for example, the report, {\em The Prevention and Treatment of Missing Data in Clinical Trials} \citep{nationalacademy2010missingdata}.

Inverse intensity weighting approaches rely on the assumption that assessment and outcomes at each time $t$ are related only through study data observed before time $t$, such as baseline covariates, treatment assignment, times of earlier assessments, and outcomes and time-varying covariates observed at those earlier assessments. We refer to this assumption as \emph{explainable assessment}. While it is less restrictive than assuming that outcomes and assessment times are unrelated, or related only through baseline variables, the explainable assessment assumption may not hold in some studies.  For example, some participants could have a new downturn in their health that also prevents them from attending a data collection appointment.  Therefore, it is important to assess how inference changes under departures from this assumption.

Here we develop a sensitivity analysis methodology, anchored at the explainable assessment assumption, for estimating the population mean of the (possibly unobserved) outcome values at a fixed time after randomization.  Our method accounts for the possibility that participants with worse outcomes at a given time may be more (or less) likely than other participants to have assessments at that time, even after controlling for variables observed earlier in the study. Our estimation approach uses a new influence function-based \emph{augmented inverse intensity-weighted} estimator, which allows for flexible semi-parametric modeling while allowing for root-$n$ rates of convergence for our estimator.  Additionally, all modeling of the observed data is separate from the sensitivity parameter $\alpha$.

\begin{figure}

\begin{center}

\includegraphics[height=3in]{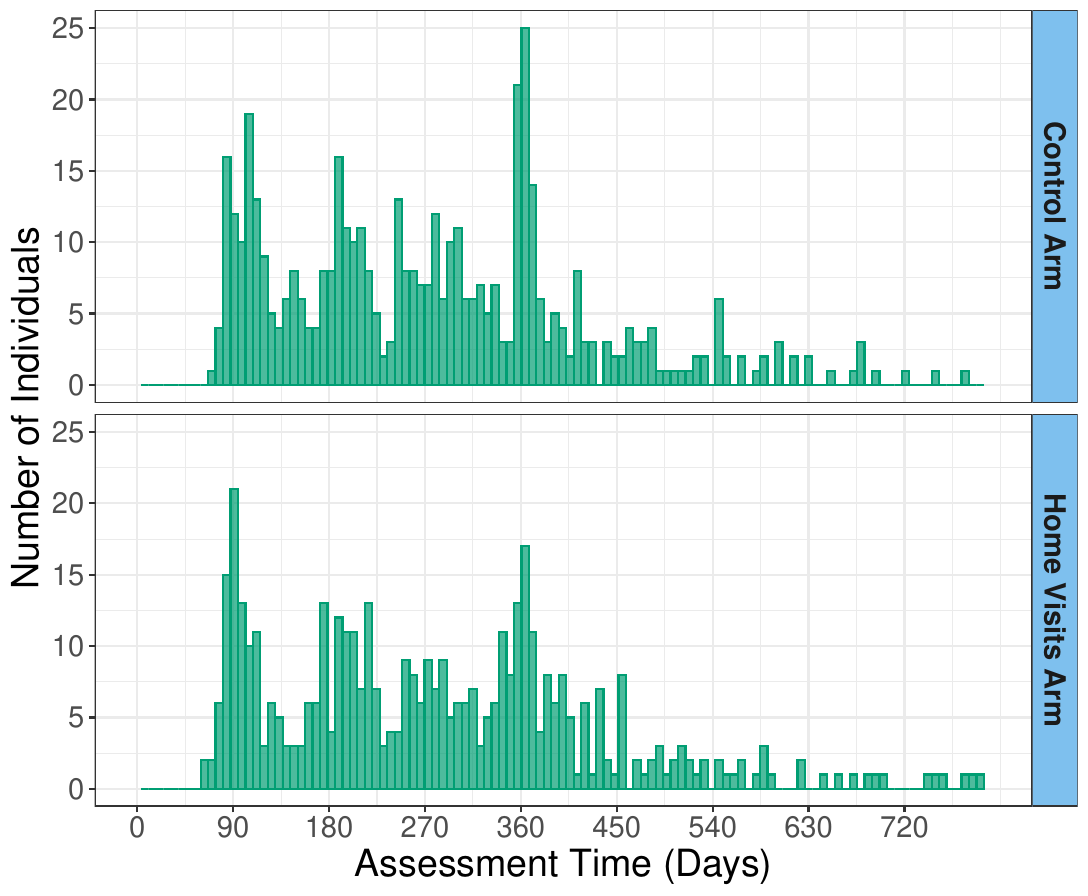}

\bigskip

\includegraphics[height=2.2in]{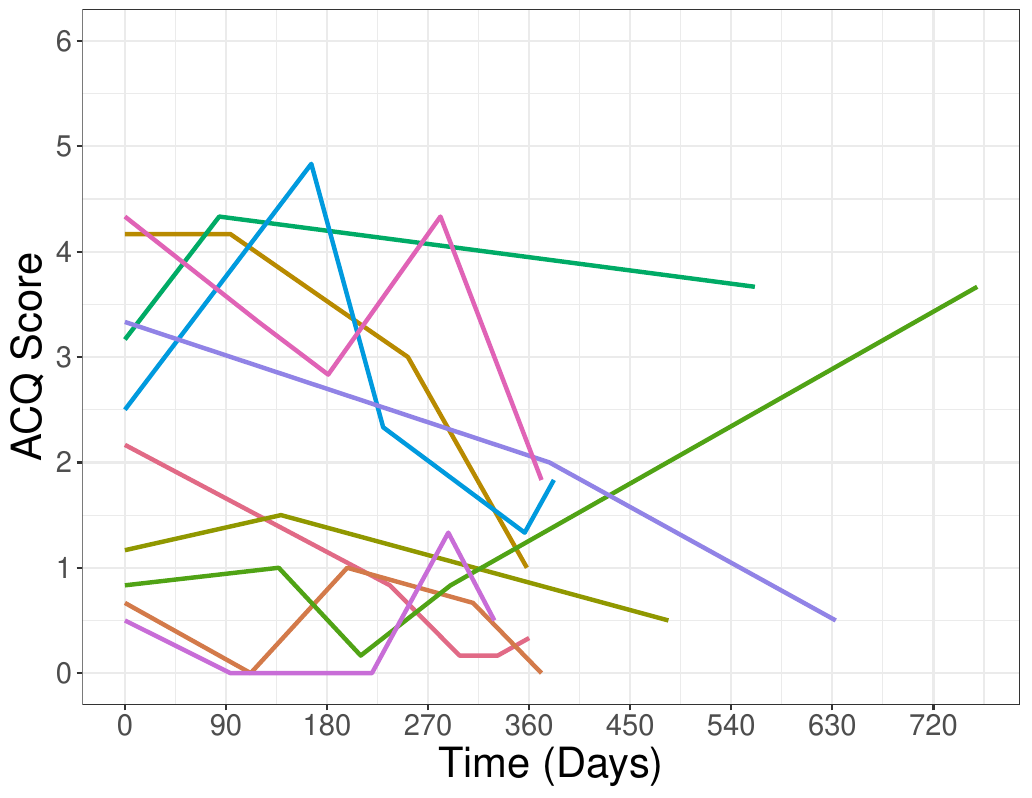}

\bigskip

\begin{singlespace}

\caption{Assessment times and outcomes in the ARC study.  Upper panel:  assessment times by treatment arm.  The protocol called for assessments at 3, 6, 9, and 12 months, but there was substantial variability in the actual times of assessment around these targeted times in each arm.   Lower panel:  outcome trajectories for a sample of participants, showing their score on the Asthma Control Questionnaire (ACQ) at each assessment time.  Outcomes fluctuated considerably over time for some participants.}

\label{fig:assessment_times}

\end{singlespace}

\end{center}
\end{figure}

We apply our methodology to the Asthma Research for the Community (ARC) study \citep{apter2019arc}, a pragmatic randomized trial of 301 low-income participants with uncontrolled asthma.   Participants in the active control group received usual care plus access to and training in a web-based portal designed to improve communication between participants and their healthcare providers.  Participants in the intervention group received home visits by community health workers to promote care coordination and help with the use of the patient portal, in addition to usual care and portal training.  The primary outcome was the score on the Asthma Control Questionnaire (ACQ) \citep{juniper1999questionnaire}, reflecting symptoms over the week prior to assessment.  The study protocol called for outcome data to be collected at 3, 6, 9, and 12 months after randomization;  however, research coordinators were often unable to schedule data collection appointments until substantially later than these targeted times. Figure \ref{fig:assessment_times} shows the actual times of assessments.  Additionally, in the intervention (control) arm, 4 (10) participants had zero post-baseline assessments, 9 (8) had only one, 24 (29) had only two, and 34 (27) had only three post-baseline assessments.

Data on specific reasons for delays  were not collected; however, investigators believe that difficulties in reaching participants were largely due to factors such as participants' competing work obligations and other demands on their time, where participants may have paid less close attention to requests for follow-up assessments during times when they were functioning well.  There were also some delays when participants waited to return contact from project staff because they were not feeling well enough or were seeking treatment or were hospitalized.  Figure \ref{fig:assessment_times} also shows outcome trajectories for a sample of participants, with substantial increases and decreases in ACQ Score over time for some individuals.  Given all these factors, it is possible that assessment at time $t$ may be associated with the outcome at time $t$, even after adjusting for variables such as previous outcome values.  The distribution of assessment times in Figure \ref{fig:assessment_times} is similar in both arms, as is the distribution of inter-assessment times (not shown); however, this does not indicate that treatment effect estimation would remain valid if we failed to account for informative assessment times in the analysis.  For example, the direction or strength of informativeness could be differential across treatment arms.

The rest of the paper is organized as follows:  in Section 2 we introduce notation and define explainable assessment.  In Section 3 we present our sensitivity analysis framework and model assumptions.  Section 4 details our estimation procedure. In Section 5 we discuss calibration of the sensitivity parameter.  A re-analysis of the ARC study is provided in Section 6.  A simulation study is presented in Section 7, and Section 8 concludes with a discussion.  A tutorial illustrating implementation of our estimator on simulated data is provided in Appendix A.1.

\section{Background} \label{sec:background}

\subsection{Setting and notation} \label{subsec:notation}

We consider a trial with a continuous outcome in which participants are randomized to either treatment or control.  Each participant's outcome is assessed at baseline and at some number of subsequent times, where the timing and possibly the number of post-baseline assessments varies by participant. The goal of the trial is to learn the population mean outcome under treatment versus control at one or more fixed follow-up times. For simplicity, we suppose that there is a time interval $[t_1,t_2]$ that includes all of these target follow-up times, and that assessments take place throughout this interval in each arm;  see Appendix A.2 for trials with gaps in time when few assessments occur. We assume that no participants drop out of the study (though they may have fewer assessments than the protocol specifies). 

Let $A$ be the treatment assignment for a random individual, with $A=1$ if the individual is assigned to treatment and $A=0$ if they are assigned to control.   Let $\tau$ be the end of follow-up.  For each $t \in [0,\tau]$, let $Y(t)$ be the (possibly unobserved) value of the participant's outcome at time $t$.  If fixed and/or time-varying auxiliary covariates are collected, let $\boldsymbol{X}(t)$ be the value of the participant's covariates at time $t$. Let $N(t)$ be the number of assessments that the individual has had up through time $t$, and let $\Delta N(t)=N(t)-N(t-)$ be the indicator that the individual has an assessment at time $t$.  Let $T_k$ be the time of the individual's $k$th post-baseline assessment.  We refer to $\{ Y(t): t \in [0,\tau] \}$ as the \emph{outcome process} and $\{ N(t): t \in [0,\tau] \}$ as the \emph{assessment process}.  For each $t \in [0,\tau]$, let $\overline{\boldsymbol{O}}(t)$ denote all of the participant's study data observed before time $t$, including baseline data, treatment assignment, times of assessments prior to $t$, and data collected at each assessment prior to $t$.  We call $\overline{\boldsymbol{O}}(t)$ the participant's \emph{observed past} before time $t$, with $\boldsymbol{O}=\overline{\boldsymbol{O}}(\tau)$ the participant's observed data over the entire study.  Finally, for each $t \in [0,\tau]$, let $Y^1(t)$ and $Y^0(t)$ be the outcome that the participant would have at time $t$ under assignment to treatment and control, respectively.  Let $\mu_1(t) = E \left\{Y^1(t)\right\}$ and $\mu_0(t)=E \left\{Y^0(t)\right\}$, the population mean outcome at time $t$ were all individuals assigned to treatment, or to control respectively. For the effect of treatment, we focus on the difference $\delta(t)=\mu_1(t)-\mu_0(t)$.

\subsection{Explainable assessment} \label{subsec:AAR}

The assumption of explainable assessment says (informally) that any relationship between assessment at time $t$ and the outcome $Y(t)$ is accounted for by the observed past $\overline{\boldsymbol{O}}(t)$.   This assumption has been referred to as sequential ignorability \citep{scharfstein2004dependent}, visiting at random \citep{pullenayegum2016review}, or assessment at random \citep{pullenayegum2022repeatedly}, and is analogous to the sequential exchangeability assumption that has been used in the longitudinal missing data literature, for example in \citet*{vansteelandt2007nonmonotone}.  To define explainable assessment formally, here we use the \emph{intensity function for the assessment process given the observed past}:
\begin{equation} \label{eqn:lambda}
 \lambda \left\{ t \mid \overline{\boldsymbol{O}}(t) \right\} = \lim_{\epsilon \to 0^+ }  \Big[ P \big\{ N(t+\epsilon)-N(t-)  = 1 \mid \overline{\boldsymbol{O}}(t) \big\} / \epsilon \Big],
 \end{equation}
where $N(t+\epsilon)-N(t-)$ is the indicator that the participant has an assessment during the time interval $[t,t+\epsilon]$.  Consider a time $t$ with $ \lambda \left\{ t \mid \overline{\boldsymbol{O}}(t) \right\}>0$.  We define $dF \left\{ y(t) \mid \Delta N(t)=1, \overline{\boldsymbol{O}}(t) \right\} = \lim_{\epsilon \to 0^+} dF \big\{ y(t) \mid N(t+\epsilon)-N(t-) =1, \overline{\boldsymbol{O}}(t) \big\}$ and \\ $dF \big\{ y(t) \mid  \Delta N(t)=0, \overline{\boldsymbol{O}}(t) \big\} = \lim_{\epsilon \to 0^+} dF \big\{ y(t) \mid N(t+\epsilon)-N(t-) =0, \overline{\boldsymbol{O}}(t) \big\}$, the distributions of $Y(t)$ among those who were, and who were not, assessed at time $t$, given $\overline{\boldsymbol{O}}(t)$.  

\medskip

\begin{definition}[Explainable assessment]  We say that assessment is \emph{explainable} (by the observed past) if $dF \left\{ y(t) \mid \Delta N(t)=1, \overline{\boldsymbol{O}}(t) \right\} = dF \big\{ y(t) \mid \Delta N(t)=0, \overline{\boldsymbol{O}}(t) \big\}$ for all $t$ with $\lambda \left\{ t \mid \overline{\boldsymbol{O}}(t) \right\}  > 0$.
\end{definition}

\medskip

\noindent That is, within strata of the observed past, under explainable assessment the distribution of $Y(t)$ is the same among those who were, and who were not, assessed at time $t$.

\citet{scharfstein2004dependent} developed the method of \emph{inverse intensity weighting} for studies with explainable assessment, extending weighting methods to the continuous-time setting by using weights based on the intensity function \eqref{eqn:lambda}. The weights create a pseudo-population in which assessment times and outcomes are no longer related if assessment is explainable.

\section{Sensitivity analysis framework and models}   \label{sec:model}

In some studies, dependence between assessment and outcomes at time $t$ may not be fully explained by variables from earlier assessments.  For example, in studies that collect outcomes at ``as-needed" appointments, a sudden downturn in health may lead participants to seek care.  Unfortunately, whether assessment is explainable cannot be determined from the study data, which contain no information about the distribution of unobserved outcomes, $dF \left\{ y(t) \mid \Delta N(t)=0, \overline{\boldsymbol{O}}(t) \right\}$. In particular, the current outcome $Y(t)$ could impact assessment at $t$ even if earlier outcomes do not impact assessment at $t$ strongly, particularly in studies where outcomes tend to fluctuate over time.   There may be alternate assumptions that are equally as plausible as explainable assessment, which could yield different inferences about the treatment effect.  Our sensitivity analysis provides an inferential strategy for the treatment effect $\delta(t)$ under a range of different plausible assumptions.

\subsection{Sensitivity analysis framework} \label{subsec:tilt}

Here we draw inference for $\mu_a(t)=E \left\{ Y^a(t)\right\}$ separately for each treatment assignment $a=0,1$.  We leverage the fact that, by randomization, $\mu_a(t)=E\{ Y(t) \mid A=a\}$, the mean outcome at time $t$ among participants assigned to treatment arm $a$, and we work separately by treatment arm.  That is, all assumptions, distributions, and estimators  are treatment arm-specific.  For ease of notation, we suppress dependence on the treatment arm until Section 4.3.  In addition to explainable assessment, we include assumptions under which outcomes among participants who are not assessed at a given time $t$ tend to be larger, or smaller, than outcomes among similar participants who are assessed at time $t$.  Specifically,  $dF \left\{ y(t) \mid \Delta N(t)=0, \overline{\boldsymbol{O}}(t) \right\}$ is assumed to be some ``tilted version" of the distribution $dF \left\{ y(t) \mid \Delta N(t)=1, \overline{\boldsymbol{O}}(t) \right\}$, with the magnitude and direction of the tilt determined by an arm-specific sensitivity parameter $\alpha$.  We assume that $E \left[ \exp \{ \alpha Y(t) \} \mid \Delta N(t) =1, \overline{\boldsymbol{O}}(t) \right]$ exists for all $\alpha$ in some neighborhood of $\alpha=0$.  Then, for each value of $\alpha$ in a range around $\alpha=0$ to be specified by the analyst (see Section \ref{sec:alpha_range}) and lying within this neighborhood:

\medskip

\begin{assumption}[Tilting assumption] \label{assumption:tilt} 
For each time $t$ with  $\lambda \left\{ t \mid \overline{\boldsymbol{O}}(t) \right\} >0$,  
\[ dF \left\{ y(t) \mid \Delta N(t)=0, \overline{\boldsymbol{O}}(t) \right\} = dF \left\{ y(t) \mid \Delta N(t)=1, \overline{\boldsymbol{O}}(t) \right\}   \times   \Big[ \exp\{ \alpha y(t) \} / c \{\overline{\boldsymbol{O}}(t) ; \alpha \} \Big] \]
where $c \left\{ \overline{\boldsymbol{O}}(t) ; \alpha \right\}  = E \left[ \exp \{ \alpha Y(t) \} \mid \Delta N(t) =1, \overline{\boldsymbol{O}}(t) \right]$ and we assume $c \left\{ \overline{\boldsymbol{O}}(t) ; \alpha \right\} <\infty$.

\end{assumption}

\medskip

\begin{figure}
\begin{center}

\includegraphics[width=5in]{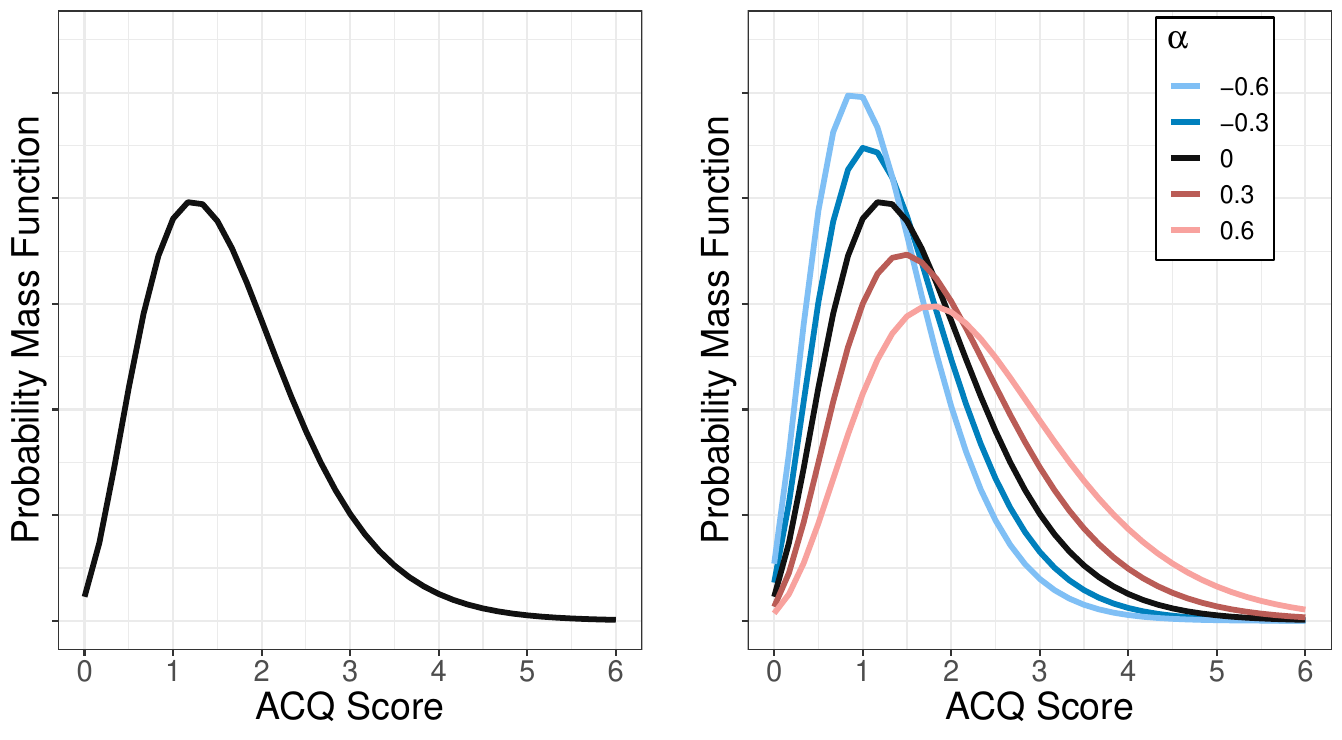}

\begin{singlespace}

\caption[Illustration of the tilt assumption]{Illustration of the tilt assumption (Assumption \ref{assumption:tilt}) in the context of the ARC trial.  Here, we consider the distribution of the ACQ Score $Y(t)$ at 6 months among participants with a certain observed past.  The distribution of  $Y(t)$ among participants who were assessed at $6$ months is shown in the left panel.  Posited distributions of $Y(t)$ among participants who had the same observed past but were not assessed at $6$ months are shown on the right.  Under the explainable assessment assumption ($\alpha=0$), the distribution for assessed versus non-assessed participants would be the same.  Under a positive (negative) value of $\alpha$, the distribution for non-assessed participants would be tilted with more weight on higher (lower) values of $Y(t)$.  Here we show smoothed depictions of the probability mass functions for this outcome.}

\label{fig:tilt_plots}

\end{singlespace}

\end{center}

\end{figure}

For a negative (positive) value of $\alpha$, the distribution of unobserved outcomes in the given arm is tilted to the left (right) relative to the distribution of observed outcomes in that arm, with smaller (larger) values of $Y(t)$ receiving greater weight.  A value of $\alpha=0$, or no tilt, is the explainable assessment assumption.  Figure \ref{fig:tilt_plots} illustrates tilting for values of $\alpha=-0.6, -0.3, 0, 0.3, 0.6$ in the context of the ARC trial. More generally, an alternate version of the tilting assumption can be used, with a different choice of function $q \{ t, Y(t); \alpha \}$ in place of $\alpha Y(t)$;  here we use $\alpha Y(t)$ for interpretability.  The construction that we use here, known as \emph{exponential tilting} \citep{barndorffnielsen1994inference}, has been used by  \citet{rotnitzky2001competing};  \citet*{birmingham2003patternmixture};  \citet{vansteelandt2007nonmonotone};  and  \citet{scharfstein2019globalsa}, among others, in sensitivity analyses for trials with missing or censored data.  It was also used by \citet*{franks2020observationalsa} for observational studies with possible unmeasured confounding, and for trials with irregular and informative assessment times by \citet{wang2020parametric}. 
\citet{wang2020parametric} developed a discrete-time framework with estimation carried out using g-computation with fully parametric models.  In contrast, our influence function-based approach allows more flexible modeling while maintaining $\sqrt{n}$ rates of convergence.

\medskip

\begin{proposition} \label{prop:mu_id}
For each time $t$ with $\lambda \left\{ t \mid \overline{\boldsymbol{O}}(t) \right\} >0$, the mean $\mu(t)$ in each arm is identified under Assumption \ref{assumption:tilt}, as
\[ \mu(t) = E \left(  \frac{E \big[ Y(t) \exp\{ \alpha Y(t) \} \mid \Delta N(t)=1,\overline{\boldsymbol{O}}(t) \big]}{E \big[  \exp\{ \alpha Y(t) \} \mid \Delta N(t)=1, \overline{\boldsymbol{O}}(t) \big] } \right). \]
\end{proposition}

\medskip

\noindent The proof is shown in Appendix B.1, and also shows the following:

\medskip

\begin{cor} \label{cor:cond_mean} For each $t$ with $\lambda \left\{ t \mid \overline{\boldsymbol{O}}(t) \right\}>0$, the conditional mean outcome given the observed past, $E \left\{ Y(t) \mid \overline{\boldsymbol{O}}(t) \right\}$, is identified from the observed data as:
\[ E \left\{ Y(t) \mid \overline{\boldsymbol{O}}(t) \right\} =\frac{E \big[ Y(t) \exp\{ \alpha Y(t) \}\  | \ \Delta N(t)=1,\overline{\boldsymbol{O}}(t) \big]}{E \big[  \exp\{ \alpha Y(t) \} \ | \ \Delta N(t)=1, \overline{\boldsymbol{O}}(t) \big] }. \]
\end{cor}

\subsection{Intensity weighting under the tilting assumption}

Our approach extends inverse intensity-weighting  to our sensitivity analysis framework.  Since assessment at time $t$ can depend on the current outcome $Y(t)$ under Assumption \ref{assumption:tilt}, we use weights based on the following intensity function:
\begin{equation}
 \rho \left\{ t \mid Y(t), \overline{\boldsymbol{O}}(t) \right\} =  \lim_{\epsilon \to 0^+ }  \Big[ P \big\{  N(t+\epsilon)-N(t-) = 1 \mid  Y(t),\overline{\boldsymbol{O}}(t) \big\} / \epsilon \Big] .
\end{equation}

\begin{assumption}[Positivity assumption] \label{assumption:pos}
There is some $c > 0$ such that, for all $t$ in $[t_1,t_2]$, $ \rho \left\{ t \mid Y(t), \overline{\boldsymbol{O}}(t) \right\} > c$ for all values of $Y(t)$ and $\overline{\boldsymbol{O}}(t)$.
\end{assumption}

\medskip

The intensity function $ \rho \left\{ t \mid Y(t), \overline{\boldsymbol{O}}(t) \right\}$ is related to the intensity function $\lambda \left\{ t \mid \overline{\boldsymbol{O}}(t) \right\}$ in equation \eqref{eqn:lambda} through the following:

\medskip

\begin{proposition} \label{prop:intens}
Under Assumptions \ref{assumption:tilt} and \ref{assumption:pos}, for each $t$ in $[t_1,t_2]$,
\[  \rho \left\{ t \mid Y(t), \overline{\boldsymbol{O}}(t) \right\} =  \lambda \left\{ t \mid \overline{\boldsymbol{O}}(t) \right\}  E \big[ \exp \{ \alpha Y(t) \} \mid \Delta N(t)=1, \overline{\boldsymbol{O}}(t) \big] / \exp\{ \alpha Y(t) \}  .\]
\end{proposition}

\medskip

\noindent The proof is given in Appendix B.1. We leverage this relationship in Section \ref{sec:inference} to keep observed data modeling separated from sensitivity parameters. Proposition \ref{prop:intens} also gives an interpretation of $\alpha$ as the log of the ratio of the intensities at time $t$ for participants who have the same observed past and whose outcomes $Y(t)$ differ by one unit:
\[ \log \left[ \frac{ \rho \left\{ t \mid Y(t)=y(t), \overline{\boldsymbol{O}}(t) \right\} }{ \rho \left\{ t \mid Y(t)=y(t)+1, \overline{\boldsymbol{O}}(t) \right\} }  \right]=  \alpha . \]

\subsection{Additional assumptions}

Proposition \ref{prop:mu_id} shows that, under Assumption \ref{assumption:tilt},  $\mu(t)$ at each time $t$ would theoretically be estimable from infinite data.  In order to estimate $\mu(t)$ from finite data, we also make the following smoothing assumption that allows us to borrow information across different times.  

\medskip

\begin{assumption}[Marginal mean assumption] \label{assumption:smooth} $\mu(t)  =  \boldsymbol{B}(t)^\prime \boldsymbol{\beta} \ \mbox{ for all } t \in [t_1,t_2]$, for some specified vector-valued basis function $\boldsymbol{B}(t)=(B_1(t),\ldots,B_p(t))^\prime$  with $\boldsymbol{V}= \int_{t=t_1}^{t_2} \boldsymbol{B}(t)\boldsymbol{B}(t)^\prime dt$ invertible, and $\boldsymbol{\beta} \in \mathbb{R}^p$ a parameter vector.
\end{assumption}

\medskip

\noindent Note that Assumption \ref{assumption:smooth} uses an identity link appropriate for a continuous outcome.  
 
\medskip

\begin{proposition} \label{prop:beta_id}
The parameter $\boldsymbol{\beta}$ is identified under Assumptions \ref{assumption:tilt}, \ref{assumption:pos}, and \ref{assumption:smooth}.
\end{proposition}

\begin{proof}
Under Assumptions \ref{assumption:tilt} and \ref{assumption:pos}, $\mu(t)$ is identified from the observed data for each $t \in [t_1,t_2]$, and under Assumption \ref{assumption:smooth}, $\boldsymbol{\beta} = \boldsymbol{V}^{-1} \int_{t=t_1}^{t_2} \boldsymbol{B}(t) \mu(t) dt$.
\end{proof}

Finally, we assume that assessment depends on future values of the outcome and covariates, the current value of covariates, and past unobserved values of the outcome and covariates only through past observed data and the current value of the outcome:

\medskip

\begin{assumption}[Non-future dependence assumption] \label{assumption:future} Let $\boldsymbol{L}=\{ Y(t), \boldsymbol{X}(t) : 0 \leq t \leq \tau \}$.  Then  $ \lim_{\epsilon \to 0^+ } \Big[ P \big\{ N(t+\epsilon)-N(t-) = 1 \mid \overline{\boldsymbol{O}}(t), \boldsymbol{L} \big\} / \epsilon \Big] = \lim_{\epsilon \to 0^+ } \Big[ P \big\{ N(t+\epsilon)-N(t-) = 1 \mid \overline{\boldsymbol{O}}(t), Y(t) \big\} /\epsilon \Big]$.
\end{assumption}

\medskip

Similar non-future dependence assumptions have been used in longitudinal settings with missing data \citep*{kenward2003patternmixture,wang2011note}.  Assumption \ref{assumption:future} aids in derivation of an influence function for $\boldsymbol{\beta}$.  However, investigators should consider whether it is tenable in their study.  An example where Assumption \ref{assumption:future} would likely not hold is a study where assessments occur at doctors' visits when participants are receiving care, which then impacts future outcomes.  In this case, after adjusting for the observed past and $Y(t)$, there could be dependence between future outcomes and assessment at time $t$, since both are related to receiving care at time $t$.  (Note that here receipt of care at time $t$ is not conditioned on, since our approach does not accommodate adjustment for variables that occur at the time of assessment $t$, except for $Y(t)$.)

\section{Estimation} \label{sec:inference}

\subsection{Observed data modeling} \label{subsec:modeling}

To implement our approach, researchers must fit two types of models.  First, in each arm the intensity function $\lambda \left\{ t \mid \overline{\boldsymbol{O}}(t) \right\}$ is modeled using an Andersen-Gill model \citep{andersengill1982}, or a stratified Andersen-Gill model  stratified by assessment number $\lambda \left\{ t \mid \overline{\boldsymbol{O}}(t) \right\} = \lambda_{0,k}(t) \exp \big\{ \boldsymbol{\gamma}^\prime \boldsymbol{Z}(t) \big\} D_k(t)$.  Here $\boldsymbol{Z}(t)$ is a specified (possibly vector-valued) function of the participant's observed past $\overline{\boldsymbol{O}}(t)$ containing key baseline covariates and time-varying factors that impact both assessment time and outcome, such as outcomes at previous assessments.  The function $\lambda_{0,k}(t)$ is an unspecified baseline intensity function for stratum $k$, $\boldsymbol{\gamma}$ is a parameter vector,  and $D_k(t)$ is an indicator that the participant is at risk for having the $k$th assessment at time $t$.   The baseline intensity functions $\lambda_{0,k}(t)$ are estimated by kernel smoothing the Breslow estimator of the cumulative baseline intensity functions \citep{breslow1972discussion}.

Second, the conditional distribution of observed outcomes in each arm given the observed past, $dF \left\{ y(t) \mid \Delta N(t)=1, \overline{\boldsymbol{O}}(t) \right\}$, is modeled using a single index model \citep{chiang2012new}. In the single index model, the conditional cumulative distribution function of $Y(t)$ given a vector of predictors, say $\boldsymbol{W}(t)$, is assumed to depend on $\boldsymbol{W}(t)$ only through a scalar $\boldsymbol{\theta}^\prime \boldsymbol{W}(t)$.  
Thus, this function is modeled as  $G \{\cdot, \boldsymbol{\theta}^\prime \boldsymbol{W}(t); \boldsymbol{\theta} \}$, where $G \{y,u;\boldsymbol{\theta} \}$ is a cumulative distribution function in $y$ for each $u$, and $\boldsymbol{\theta}$  is a vector of unknown parameters. The estimator of $G$ is a step function in $y$ that is kernel smoothed with respect to $\boldsymbol{\theta}^\prime \boldsymbol{W}(t)$ via a bandwidth parameter $h$; $\boldsymbol{\theta}$ and $h$ are jointly estimated by minimizing a pseudo sum of integrated squares \citep{chiang2012new}.  Specifically, in our context, for each $t$ and each value of $\overline{\boldsymbol{o}}(t)$, $\widehat{F} \left\{ y(t) \mid \Delta N(t)=1, \overline{\boldsymbol{O}}(t)=\overline{\boldsymbol{o}}(t) \right\}$ is a step function with jumps at all outcome values observed in the data.

\subsection{Estimation of the  mean outcome $\mu(t)$ under $\alpha$} \label{subsec:mean_est}

The following result provides a way of constructing estimators for $\boldsymbol{\beta}$ and $\mu(t)$ that incorporate the flexible models fit in Section \ref{subsec:modeling}, yet converge at fast parametric rates.

\medskip

\begin{theorem} \label{thm:IF}
Under Assumptions \ref{assumption:tilt}-\ref{assumption:future}, an influence function for $\boldsymbol{\beta}$ is given by:
\begin{align*} 
\boldsymbol{\varphi}(\boldsymbol{O}) = & \ \int_{t=t_1}^{t_2}  \boldsymbol{V}^{-1} \boldsymbol{B}(t) \frac{ \big[ Y(t)- E \big\{ Y(t) \mid \overline{\boldsymbol{O}}(t) \big\} \big] }{ \rho \left\{ t \mid Y(t), \overline{\boldsymbol{O}}(t) \right\}}  dN(t) \ + \\
& \ \ \ \ \ \int_{t=t_1}^{t_2}   \boldsymbol{V}^{-1} \boldsymbol{B}(t)  E\big\{ Y(t) \mid \overline{\boldsymbol{O}}(t)\big\}   dt  \ - \ \boldsymbol{\beta} \end{align*}
where $\boldsymbol{B}(t)$ and $\boldsymbol{V}$ are given in Assumption \ref{assumption:smooth}.
\end{theorem}

\medskip

\noindent The proof of Theorem 1 is given in Appendix B.3.

Suppose that we have data for $n$ independent individuals. We construct estimators $\widehat{\boldsymbol{\beta}}$ and $\widehat{\mu}(t)$ using the following steps, where we use a subscript $i$ to denote data for individual $i$.  
For each individual $i$:

\begin{enumerate}[1.]
\item For each assessment $k$ with $T_{ik}$ in the interval $[t_1,t_2]$, compute
\begin{align*}
\widehat{E} & \big[\exp \big\{ \alpha Y(T_{ik}) \big\} \mid \Delta N(T_{ik})=1, \overline{\boldsymbol{O}}_i(T_{ik})\big] = \\
& \ \ \ \ \ \int_{y \in \mathcal{Y}} \exp( \alpha y ) \ d\widehat{F} \big\{ Y(T_{ik})=y  \mid \Delta N(T_{ik})=1, \overline{\boldsymbol{O}}_i(T_{ik}) \big\},
\end{align*}
where $\mathcal{Y}$ is the set of all outcome values occurring in the data; the estimated conditional mean (see Corollary \ref{cor:cond_mean})
\[ \widehat{E} \left\{ Y(T_{ik}) \mid \overline{\boldsymbol{O}}_i(T_{ik}) \right\} = \frac{\widehat{E}[ Y(T_{ik}) \exp\{ \alpha Y(T_{ik})\} \mid \Delta N(T_{ik})=1, \overline{\boldsymbol{O}}_i(T_{ik})] }{\widehat{E}[ \exp\{ \alpha Y(T_{ik})\} \mid \Delta N(T_{ik})=1, \overline{\boldsymbol{O}}_i(T_{ik})] }; \]
 and the inverse intensity weight (see Proposition \ref{prop:intens})
 \begin{align*}
& \frac{1}{\widehat{\rho} \left\{ T_{ik} \mid Y_i(T_{ik}), \overline{\boldsymbol{O}}_i(T_{ik}) \right\} } = \\
& \ \ \ \ \ \ \ \frac{ \exp\{ \alpha Y_i(T_{ik}) \} }{ \widehat{\lambda}_{0,k}(T_{ik})\exp\{ \widehat{\boldsymbol{\gamma}} \boldsymbol{Z}_i(T_{ik})\} \ \widehat{E}[ \exp\{ \alpha Y(T_{ik}) \}  \mid \Delta N(T_{ik})=1,\overline{\boldsymbol{O}}_i(T_{ik})] } .
\end{align*}

\item  For each time $t$ in $[t_1,t_2]$, compute the predicted mean outcome at time $t$ given their observed past before time $t$: 
\[ \widehat{E} \left\{ Y(t) \mid \overline{\boldsymbol{O}}_i(t) \right\} = \frac{\widehat{E}[ Y(t) \exp\{ \alpha Y(t)\} \mid \Delta N(t)=1, \overline{\boldsymbol{O}}_i(t)] }{\widehat{E}[ \exp\{ \alpha Y(t)\} \mid \Delta N(t)=1, \overline{\boldsymbol{O}}_i(t)] }. \]

\item Compute $\widehat{\boldsymbol{\Psi}}(\boldsymbol{O}_i)=$
\[ 
 \sum_{k \in S_i}  \Bigg\{ \boldsymbol{V}^{-1} \boldsymbol{B}(T_{ik}) \frac{\big[ Y_i(T_{ik})- \widehat{E} \big\{ Y(T_{ik}) \mid \overline{\boldsymbol{O}}_i(T_{ik}) \big\}  \big] }{\widehat{\rho} \left\{ T_{ik} \mid Y_i(T_{ik}), \overline{\boldsymbol{O}}_i(T_{ik}) \right\} } \Bigg\}  +  \int_{t=t_1}^{t_2}   \boldsymbol{V}^{-1} \boldsymbol{B}(t)  \widehat{E} \big\{ Y(t) \mid \overline{\boldsymbol{O}}_i(t) \big\}   dt     
\]
where $S_i = \{ k: T_{ik} \in [t_1,t_2]\}$.
\end{enumerate}

Our \emph{augmented inverse intensity-weighted estimators} are: $\widehat{\boldsymbol{\beta}}= \frac{1}{n} \sum_{i=1}^{n}  \widehat{\boldsymbol{\Psi}}(\boldsymbol{O}_i)$ and $\widehat{\mu}(t)= \widehat{\boldsymbol{\beta}}^\prime \boldsymbol{B}(t)$.  

\subsection{Large-sample distribution of $\widehat{\boldsymbol{\beta}}$}

If the models for $\lambda \left\{ t \mid \overline{\boldsymbol{O}}(t) \right\}$ and $dF \left\{ y(t) \mid \Delta N(t)=1, \overline{\boldsymbol{O}}(t) \right\}$ are both correctly specified, and if Assumptions 1-4 and additional regularity conditions hold, then $\sqrt{n} \left( \widehat{\boldsymbol{\beta}}-\boldsymbol{\beta} \right) \Rightarrow N \big[\boldsymbol{0}, Var \{\boldsymbol{\varphi}(\boldsymbol{O}) \} \big]$.  See Appendix B.4;  there we derive the second-order remainder term in an expansion of $\sqrt{n} \left( \widehat{\boldsymbol{\beta}}-\boldsymbol{\beta} \right)$ following \citet{kennedy2016empiricalprocesses}, and we show conditions under which this remainder term is asymptotically negligible.

From this result, influence function-based variance estimators for $\widehat{\boldsymbol{\beta}}$ and $\widehat{\mu}(t)$ are given by $\widehat{Var} \Big( \widehat{\boldsymbol{\beta}} \Big) =$ $\displaystyle{ \frac{1}{n^2} \sum_{i=1}^{n}  \Big\{ \widehat{\boldsymbol{\Psi}}(\boldsymbol{O}_i) - \widehat{\boldsymbol{\beta}}  \Big\} \Big\{ \widehat{\boldsymbol{\Psi}}(\boldsymbol{O}_i) - \widehat{\boldsymbol{\beta}} \Big\}^{\prime}}$ and $\widehat{Var} \big\{ \widehat{\mu}(t) \big\} = \boldsymbol{B}(t)^{\prime} \widehat{Var} \big( \widehat{\boldsymbol{\beta}} \big) \boldsymbol{B}(t)$.  A Wald confidence interval for $\mu(t)$ can be constructed using this influence function-based variance estimator or using a jackknife variance estimator.  In simulations mimicking the ARC data, we found that nonparametric bootstrap was not a feasible of way of constructing confidence intervals; ties in bootstrapped datasets caused estimates of conditional cumulative distribution functions based on the single index model to be undefined.

\subsection{Inference for $\delta(t)$ }

Here we re-introduce subscripts for each treatment arm, and we also let $\alpha_1$ and $\alpha_0$ be sensitivity parameters for the treatment and control arms, respectively.  To conduct the sensitivity analysis for $\delta(t)$, the estimation procedure above is repeated in the treatment arm to estimate $\mu_1(t)$ under a range of $\alpha_1$ values, and separately in the control arm to estimate $\mu_0(t)$ under a range of $\alpha_0$ values.  These results are then combined to estimate the treatment effect $\delta(t)=\mu_1(t)-\mu_0(t)$ over a grid of sensitivity parameters $(\alpha_0,\alpha_1)$.

\section{Selection of a range of sensitivity parameter values} \label{sec:alpha_range}

The analyst must decide on a range of sensitivity parameter values to include in the sensitivity analysis.  Domain expertise should be used in making this decision, and how best to use such expertise is a key question for all sensitivity analyses.  \citet{cinelli2020sensitivity} have noted that ``perhaps [the] most fundamental obstacle to the use of sensitivity analysis is the difficulty in connecting the formal results to the researcher's substantive understanding about the object under study," and they write that the ``bounding procedure we should use depends on which \ldots quantities the investigator prefers and can most soundly reason about in their own research."  In keeping with this, we propose the following approach in which domain experts reason about the treatment arm-specific mean outcome:  We first query domain experts for extreme values $\mu_{min}$ and $\mu_{max}$ such that, in their judgment, a value of $\mu(t)$ outside of the bounds $(\mu_{min}, \mu_{max})$ at any time $t$ would be implausible.  We then treat any value $\alpha$ under which $\mu(t)$ falls outside of $(\mu_{min}, \mu_{max})$ for some $t$ as implausible, and retain all other values.  Other possible approaches could draw on bounding procedures that have been developed for sensitivity analyses for unmeasured confounding in observational studies.  Authors including \citet*{veitch2020sense, franks2020observationalsa,sjolander2022sacausal} have developed methods that use the strength of measured covariates' impact on the exposure and outcome to calibrate plausible values for the impact of unmeasured factors, after adjusting for measured covariates.  An approach for our setting that borrows from ideas of \citet{sjolander2022sacausal} could be to use some observable quantity that is related to the attenuation in $E\{ Y(t) \mid \Delta N(t)=1 \} -E \{Y(t) \mid \Delta N(t)=0 \}$ obtained by adjusting for the observed past $\overline{\boldsymbol{O}}(t)$.  This could in principle be used to calibrate plausible values for $E \{Y(t) \mid \Delta N(t)=1, \overline{\boldsymbol{O}}(t) \} -E \{ Y(t) \mid \Delta N(t)=0, \overline{\boldsymbol{O}}(t) \}$, the residual difference due to unmeasured factors $\boldsymbol{U}(t)$ after adjusting for $\overline{\boldsymbol{O}}(t)$.  As with other benchmarking approaches, this would use researchers' substantive beliefs that the strength of the impact of $\boldsymbol{U}(t)$ on $\Delta N(t)$ and $Y(t)$, after adjusting for $\overline{\boldsymbol{O}}(t)$, is no more than some factor, say $r$, times the marginal impact of $\overline{\boldsymbol{O}}(t)$ on $\Delta N(t)$ and $Y(t)$.  Careful consideration would be needed in selecting a plausible value of $r$, taking into account the time scale of the study, since $\boldsymbol{U}(t)$ could include the outcome just before time $t$, whereas the impact of $\overline{\boldsymbol{O}}(t)$ on $\Delta N(t)$ and $Y(t)$ could be dampened due to the time elapsed since the previous study visit.  Development of a method along these lines, and investigations that would guide when and how to implement it in practice, could be a direction for future research.

\section{Data analysis:  ARC trial} \label{sec:analysis}

Here, we analyze the ARC data using our sensitivity analysis methodology.  The Asthma Control Questionnaire (ACQ) is on a scale from $0$ (completely controlled asthma) to $6$ (extremely uncontrolled asthma) and takes values in $\{ 0, 1/6, 2/6, 3/6, \ldots , 6 \}$.  A positive value of $\alpha_a$ posits that unobserved values of the ACQ Score in treatment arm $a$ tend to be higher (that is, worse) than observed values of the outcome in that arm at each time $t$, after controlling for variables observed before time $t$.  This could be the case if participants tended to miss or postpone data collection appointments at times when their asthma was worse, so that some of the participants' higher ACQ Score values were not observed, while $\alpha_a$ could be negative if participants in arm $a$ tended to be more engaged with the study at times when their asthma was worse.  Since we do not know which, if either, of these is the case, we consider positive, negative, and zero values of $\alpha_a$. We consider  Assumption \ref{assumption:future} to be reasonable since the assessment process was not tied to clinical care that might affect future outcomes.

\begin{figure}
\centering
\begin{subfigure}[b]{0.43\textwidth}
    \centering
    \includegraphics[width=\textwidth]{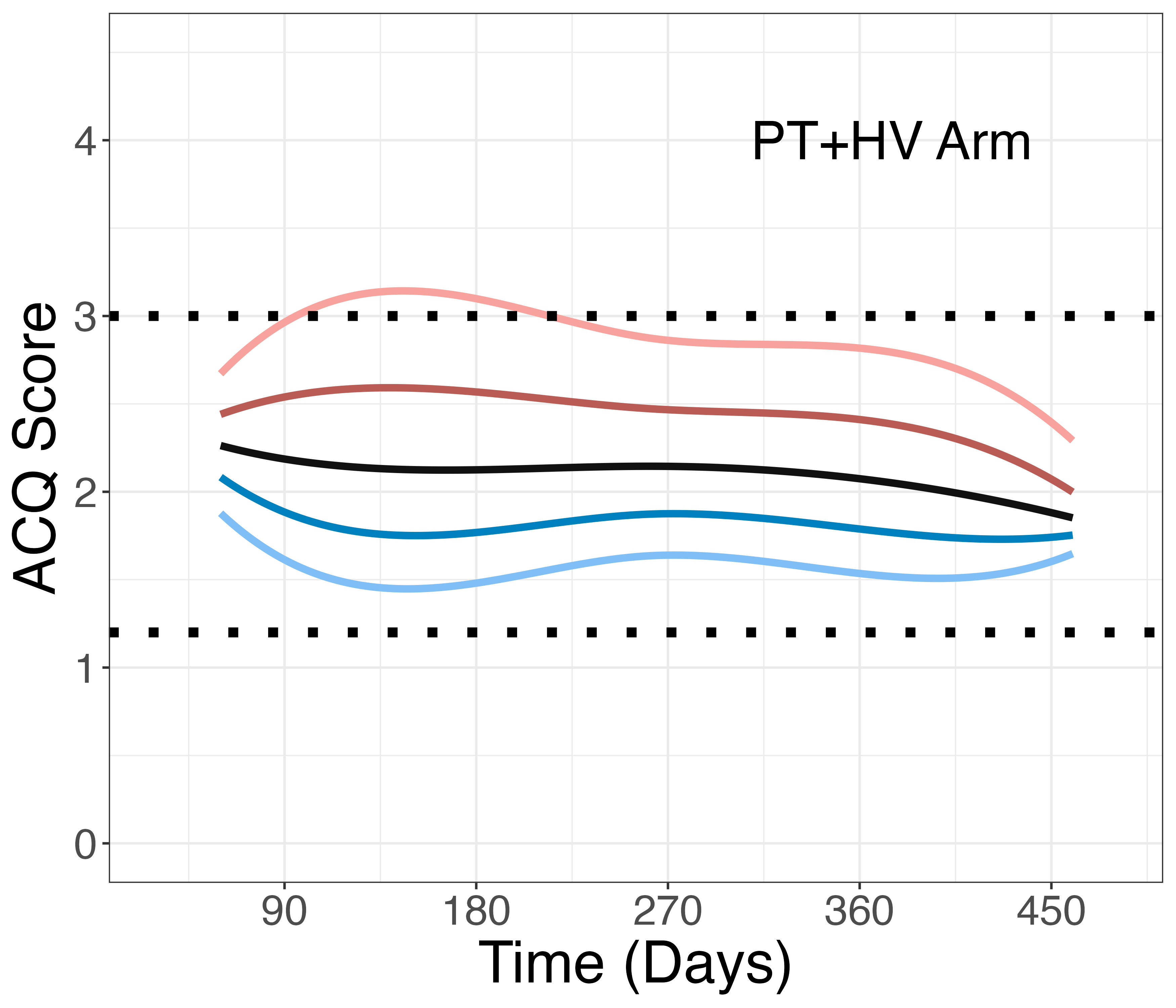}
\end{subfigure}
\hfill
\begin{subfigure}[b]{0.49\textwidth}  
    \centering 
    \includegraphics[width=\textwidth]{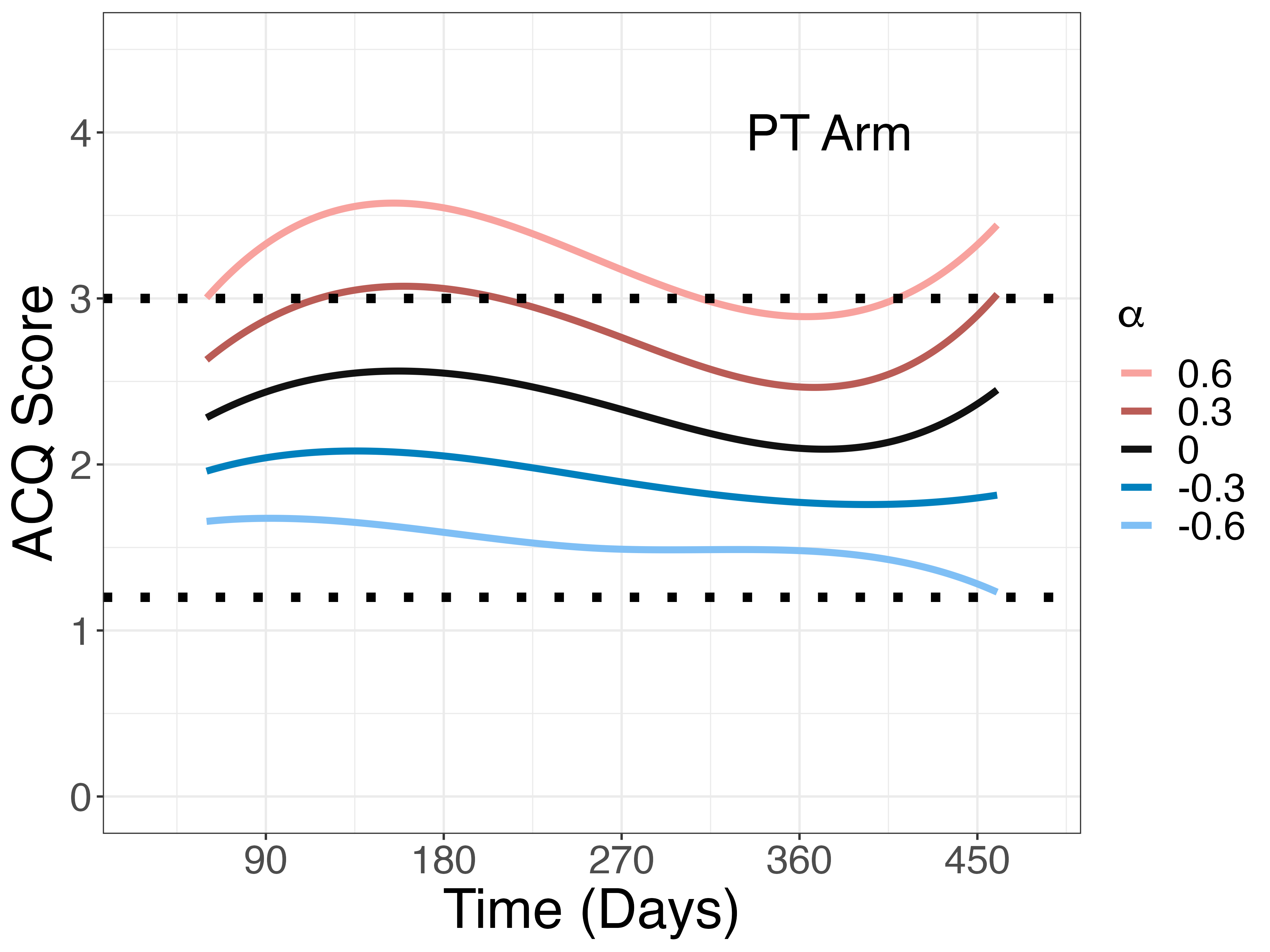}
\end{subfigure}

\medskip

\begin{subfigure}[b]{0.43\textwidth}   
    \centering 
    \includegraphics[width=\textwidth]{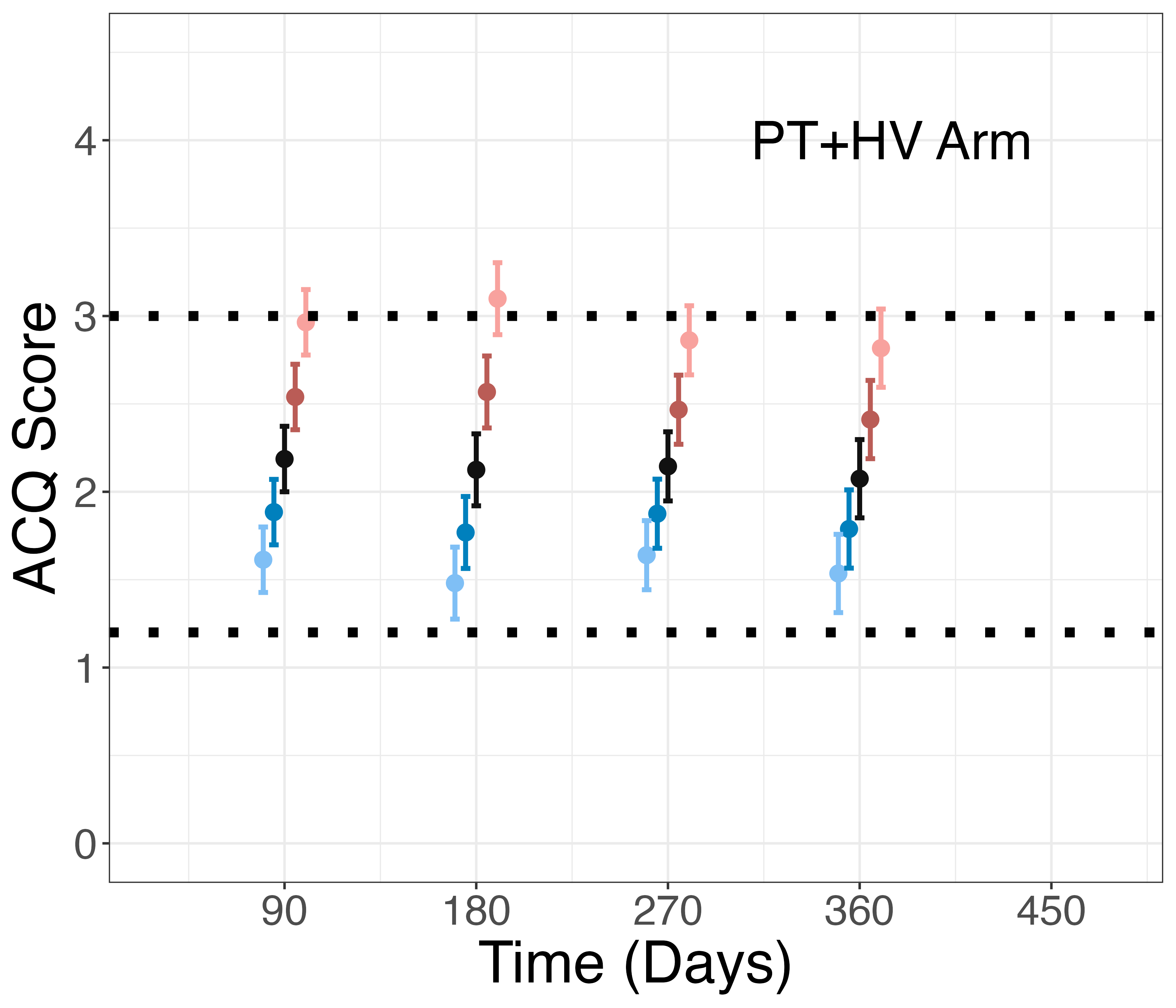}
   \end{subfigure}
\hfill
\begin{subfigure}[b]{0.49\textwidth}   
    \centering 
    \includegraphics[width=\textwidth]{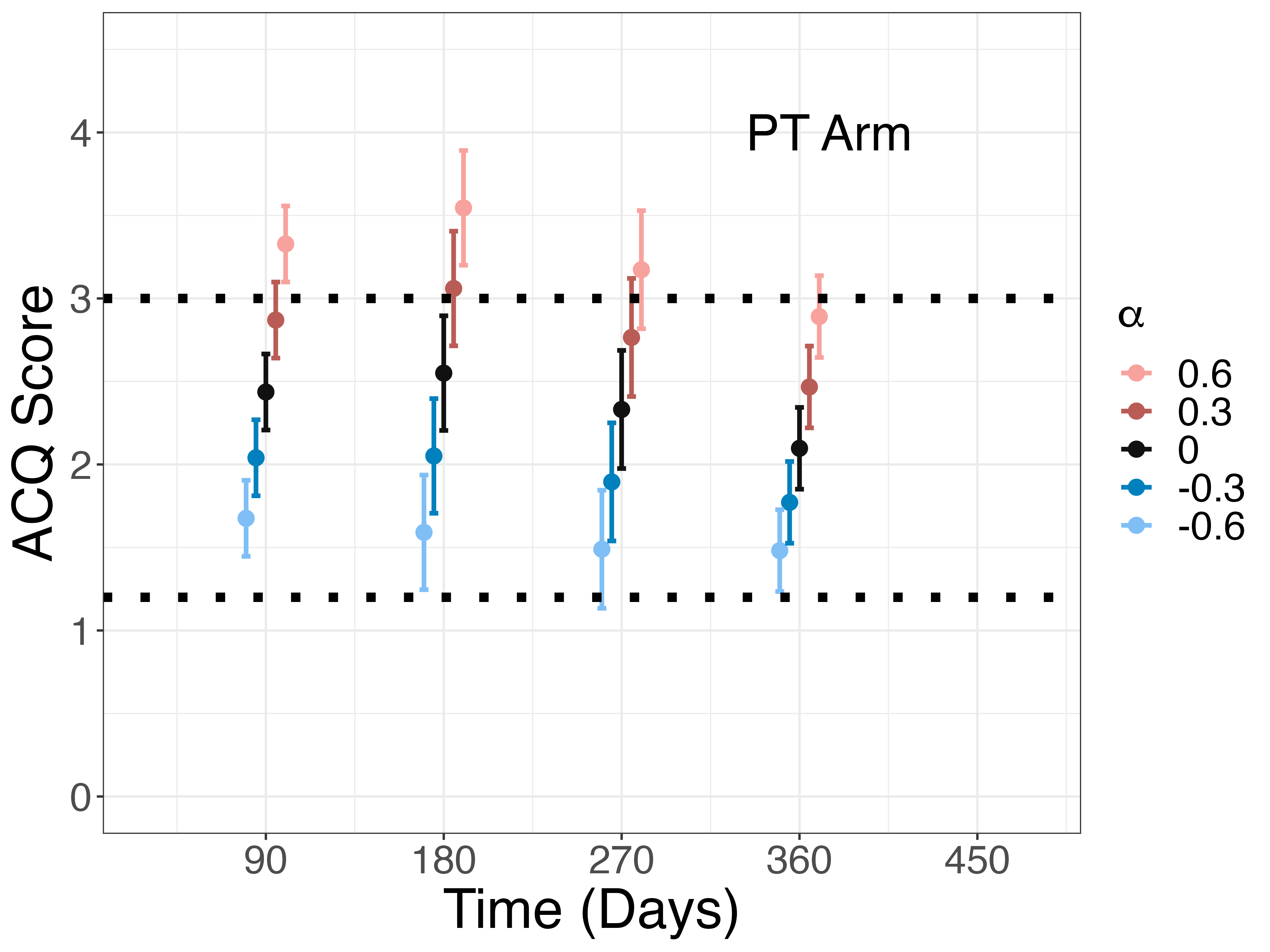}
\end{subfigure}

\caption{Population means of scores on the Asthma Control Questionnaire (ACQ) in the ARC trial under a range of sensitivity parameter values. Here we show inference for the population mean ACQ Score in the intervention (PT+HV) arm and the control (PT) arm.  Estimation is made under sensitivity parameter values of $\alpha = -0.6, -0.3, 0, 0.3, 0.6$, where a positive (negative) value of $\alpha$ posits that unobserved outcome values at time $t$ tend to be higher (lower) than observed values, after controlling for variables observed before time $t$. Upper panels: point estimates on the interval 60 to 460 days after randomization.  Lower panels: point estimates and 95\% Wald confidence intervals using the jackknife variance estimate at each target time of 90, 180, 270, and 360 days. For each arm, only those values of $\alpha$ under which the estimated curves lie completely between the dotted lines at $\mu_{\min} = 1.2$ and $\mu_{\max} = 3.0$ are considered plausible based on subject-matter expertise.} 
\label{fig:mean_curves}
\end{figure}

We estimate $\mu_1(t)$ and $\mu_0(t)$ over a time interval of 60 to 460 days, since this interval contains the target times of 90, 180, 270, and 360 days and  assessments occur throughout this period. For each $a=0,1$, we assume that $\mu_a(t)= \boldsymbol{\beta}_a^\prime \boldsymbol{B}(t)$ for $t \in [60, 460]$, with $\boldsymbol{B}(t)$ a cubic spline basis with one interior knot at $t=260$ days;  this choice of $\boldsymbol{B}(t)$ allows the marginal mean to be a fairly flexible smooth function of time.  We fit the models described in Section \ref{subsec:modeling}, modeling the intensity function $\lambda \left\{ t \mid \overline{\boldsymbol{O}}(t) \right\}$ separately for each treatment arm using a stratified Andersen-Gill model with the outcome at the previous assessment as the predictor. We also considered an intensity model that includes lag time since the previous visit as an additional predictor.  While lag time was a strong predictor in this model, the resulting inference for $\mu_a(t)$ was extremely similar under both models, and we therefore present the results of the simpler model.  The coefficient for the previous outcome is $-0.024$ (standard error $0.038$) in the intervention arm, and $0.042$ (standard error $0.036$) in the control arm.  We estimated the baseline intensity functions using kernel smoothing of the Breslow estimate of the cumulative baseline intensity, with an Epanechnikov kernel and a bandwidth of 30 days.  We modeled the conditional distribution of observed outcomes separately for each treatment arm using a single index model with the current time, lag time since the previous assessment, and a natural spline of the outcome at the previous assessment as predictors.  We then constructed the augmented inverse intensity weighted estimators given in Section \ref{subsec:mean_est}.

Figure \ref{fig:mean_curves} shows estimates of the curve $\mu_1(t), t \in [60, 460]$, under a range of $\alpha_1$ values and estimates of $\mu_0(t), t \in [60,460]$, under a range of $\alpha_0$ values, with higher $\mu_a(t)$ under higher values of $\alpha_a$.  Estimates and confidence intervals for $\mu_a(t)$ at the target times are also shown.  The minimal clinically important difference for the ACQ Score is $0.5$; therefore, one way of interpreting the magnitude of $\alpha_a$ in this study is that increasing $\alpha_a$ by $0.3$ corresponds to an increase in $\mu_a(t)$ that, at some times $t$, is approximately as much as the minimal clinically important difference for the outcome.  Next we consider the ranges of $\alpha_1$ and $\alpha_0$ values to include.  Our clinical collaborator (Author AJA) considered that a mean ACQ Score of $3.0$ or higher, or $1.2$ or lower, at any time would be extreme in either treatment arm.  These bounds are shown in Figure \ref{fig:mean_curves}.  A value of $\alpha_1 > 0.52$ led to a value of $\mu_1(t)$ that was greater than $3.0$, and a value of $\alpha_0 > 0.25$ led to a value of $\mu_0(t)$ that was greater than $3.0$;  therefore we use bounds of  $-0.6 \leq \alpha_1 \leq 0.52$ and $-0.6 \leq \alpha_0 \leq 0.25$ in our final sensitivity analysis.

 \begin{figure}
 
 \begin{center}
 \includegraphics[width=5in]{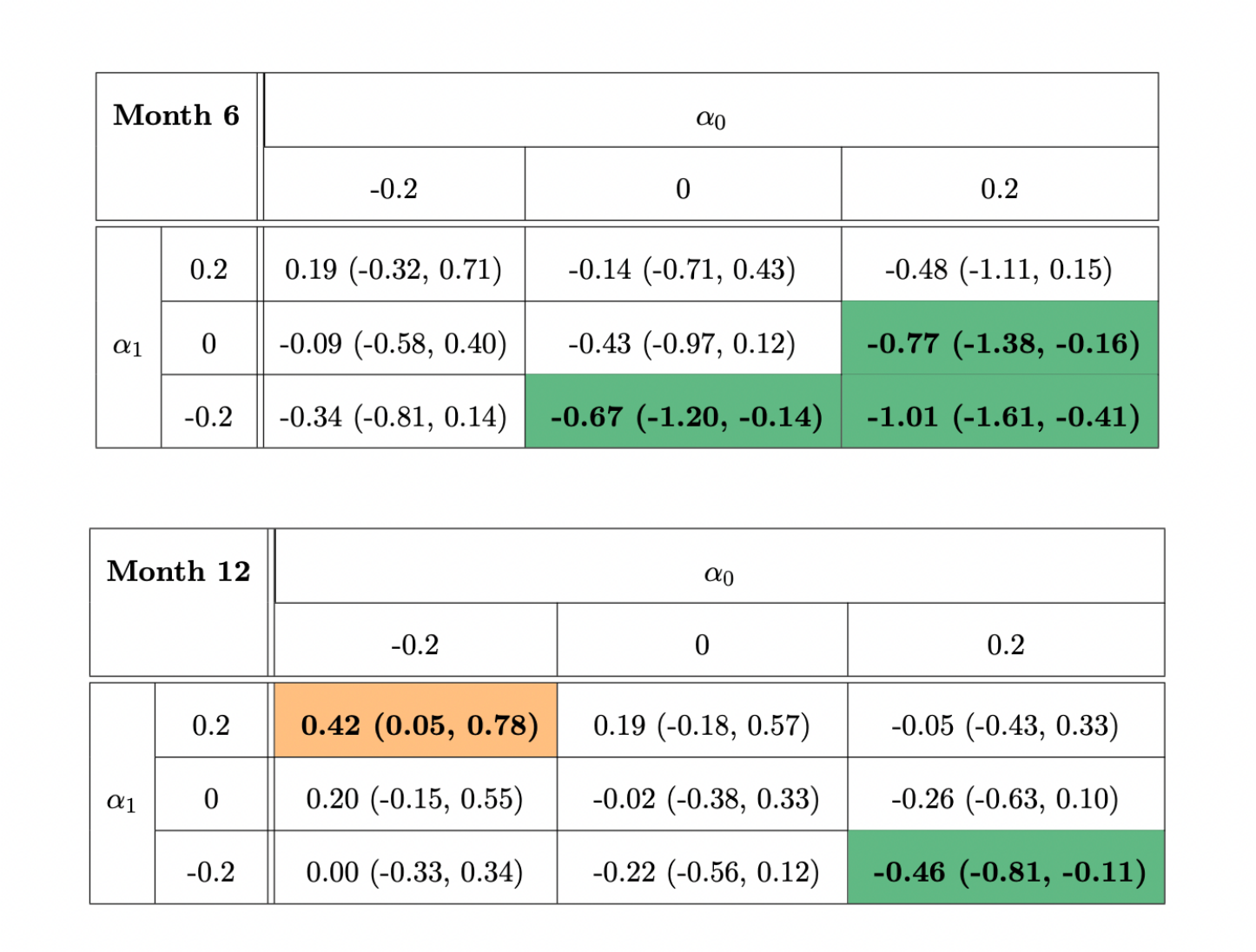}
 \end{center}
 
\caption{Treatment effects in the ARC data under different values of the sensitivity parameters $\alpha_1$ and $\alpha_0$.  Here, we show estimates (95\% confidence intervals) for the treatment effect $\delta(t) = \mu_1(t)-\mu_0(t)$ at 6 months (upper panel) and 12 months (lower panel).  Entries in green correspond to values of $\alpha_1$ and $\alpha_0$ under which there would be evidence that the home visits intervention reduces (that is, improves) the population mean score on the Asthma Control Questionnaire (ACQ) at that time, compared to portal training alone.  The entry in orange corresponds to values under which there would be evidence that the intervention raises the population mean ACQ Score.  Confidence intervals are Wald confidence intervals using the jackknife variance estimate.}
\label{fig:trt_effect_table}

\end{figure} 

\begin{figure}
\centering
\begin{subfigure}[b]{0.45\textwidth}
    \centering
    \includegraphics[width=\textwidth]{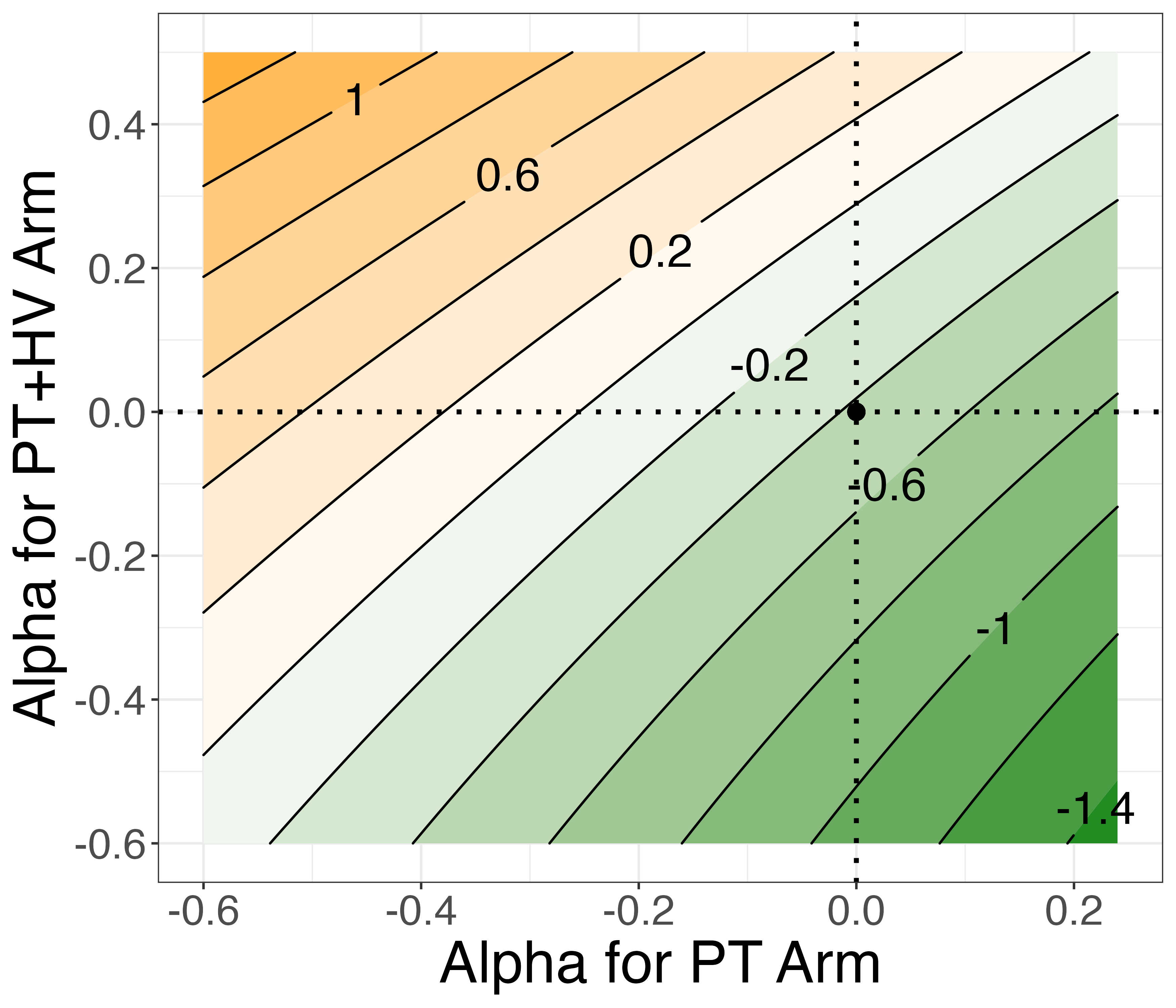}
    \caption[Estimates at 6 months]%
    {{\small Estimates at 6 months}}    
\end{subfigure}
\hfill
\begin{subfigure}[b]{0.51\textwidth}  
    \centering 
    \includegraphics[width=\textwidth]{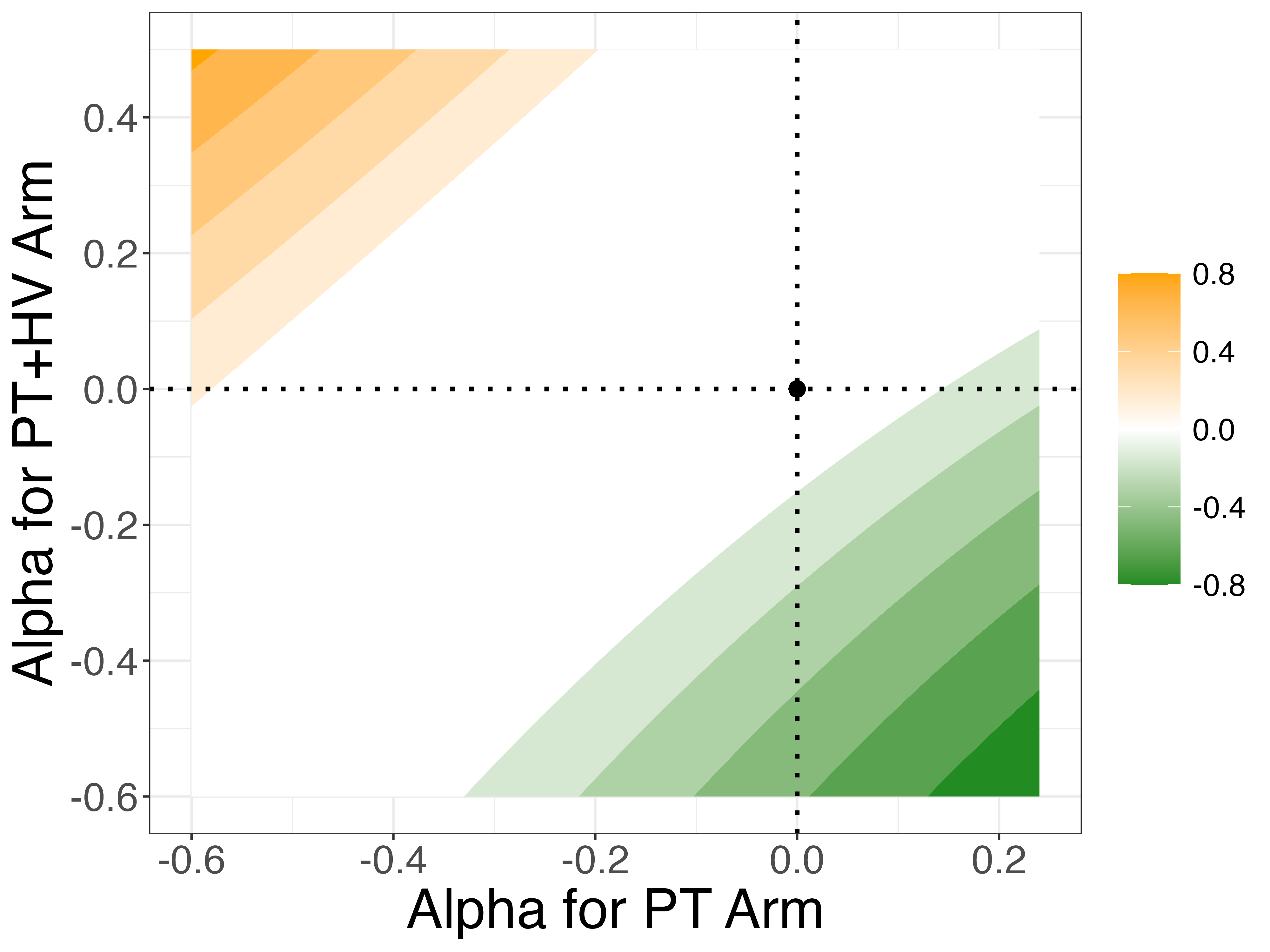}
    \caption[]%
    {{\small Uncertainty at 6 months}}    
 \end{subfigure}
 
 \bigskip
 
\begin{subfigure}[b]{0.45\textwidth}   
    \centering 
    \includegraphics[width=\textwidth]{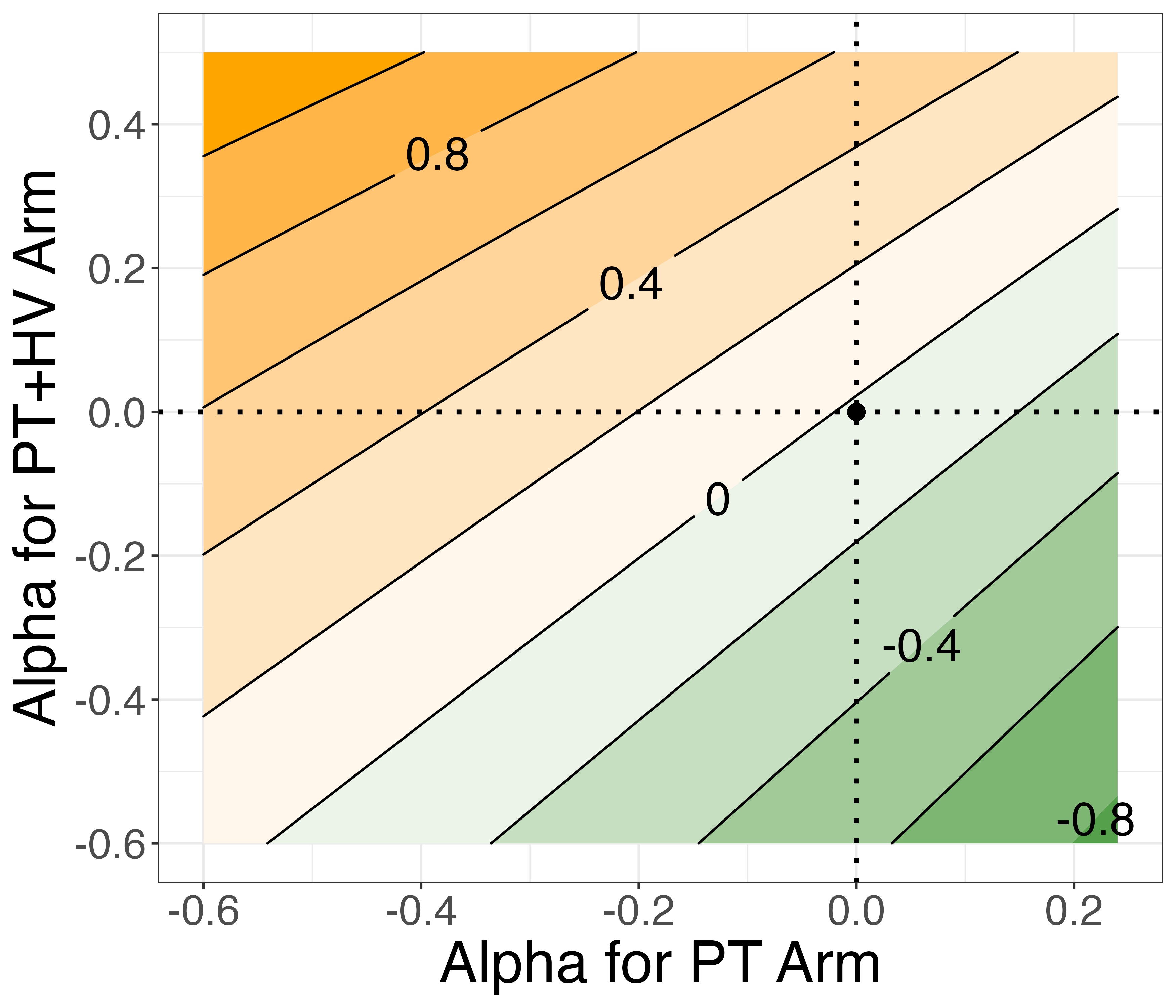}
    \caption[]%
    {{\small Estimates at 12 months}}    
  \end{subfigure}
\hfill
\begin{subfigure}[b]{0.51\textwidth}   
    \centering 
    \includegraphics[width=\textwidth]{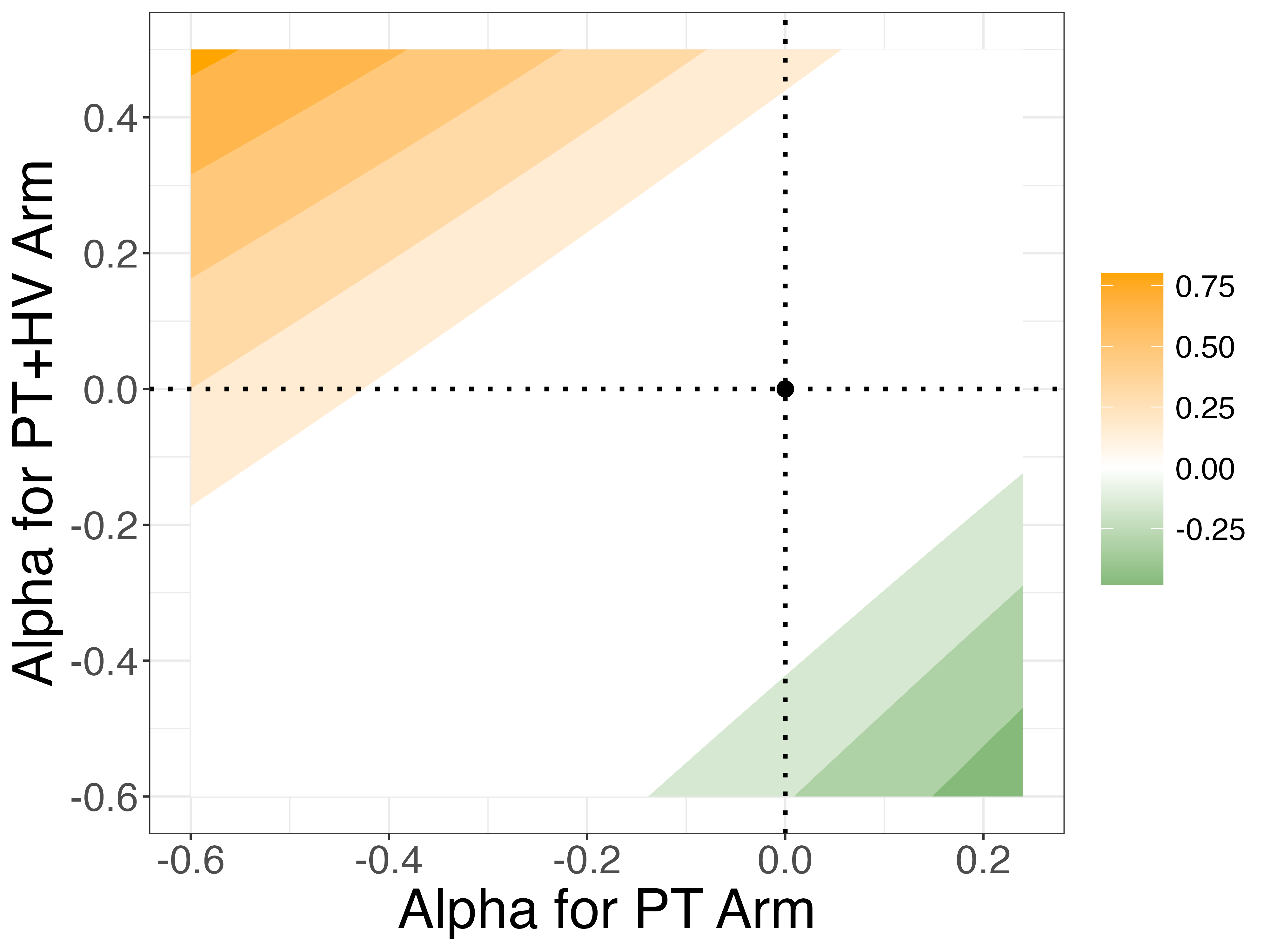}
    \caption[]%
    {{\small Uncertainty at 12 months}}    
  \end{subfigure}

  \vspace{0.2in}
\caption{Sensitivity analysis for the ARC trial.  Panels on the left show the point estimate of the treatment effect $\delta(t)$ at 6 months (upper left) and 12 months (lower left) under each pair of sensitivity parameter values.  Point estimates vary between $-1.48$ and $1.32$ at 6 months, and between $-0.85$ and $1.19$ at 12 months.  Panels on the right display information about 95\% confidence intervals for $\delta(t)$:  the regions in white correspond to sensitivity parameter values under which the confidence interval contains zero.  The regions in green (orange) in the lower right (upper left) correspond to values under which the confidence interval is entirely negative (positive).  The height in the panels on the right gives the value in the confidence interval that is closest to zero. Confidence intervals are Wald intervals using the jackknife variance estimate.}

\label{fig:contour_plots}
\end{figure}

Figure \ref{fig:trt_effect_table} shows estimates and confidence intervals for $\delta(t)$ at 6 and 12 months under selected values of $\alpha_0$ and $\alpha_1$.  If we assume $\alpha_0=\alpha_1$ (which includes explainable assessment in each treatment arm), there is not enough evidence to conclude a treatment effect at 6 months or at 12 months.  However, if we consider the possibility that informative assessments may operate differently in each arm, then we do have evidence of a treatment effect in some cases.  For example, if we assume $\alpha_0=0$, but assume that, in the intervention arm, unobserved values of the ACQ Score tend to be lower than observed values by an amount corresponding to $\alpha_1=-0.2$, then there would be evidence that the home visits intervention improves (that is, reduces) the population mean ACQ Score at 6 months compared to portal training alone, by an estimated six tenths of a point.  Estimation of $\delta(t)$ under a finer grid of $(\alpha_0,\alpha_1)$ values is presented via the contour plots in Figure \ref{fig:contour_plots}, showing point estimates and confidence interval information at 6 and 12 months.  Point estimates range between $-1.48$ and $1.32$ at 6 months, and between $-0.85$ and $1.19$ at 12 months.  If $\alpha_1$ and $\alpha_0$ are similar, or if $\alpha_1 > \alpha_0$ and the difference in values no more than $0.5$, then there is not enough evidence to conclude a treatment effect at 6 months.  On the other hand, if $\alpha_0 \geq \alpha_1 +0.2$, then in many cases there would be evidence that the intervention improves (reduces) the population mean ACQ Score relative to portal training alone at 6 months.  There are also values of $\alpha_0$ and $\alpha_1$ under which there would be evidence that the population mean ACQ Score is higher (that is, worse) under the intervention at 6 months;  however, informativeness would have to be strongly differential across treatment arms.  Informativeness would also have to be strongly differential across arms to have evidence of either a positive or negative treatment effect at 12 months.  In this study, such a large difference in the value of the sensitivity parameters between arms would likely not be plausible.

\section{Simulations} \label{sec:sims}

We generated realistic simulated data based on the ARC data, with a sample size of $N=200$ in each arm.  Details of our data-generating process are given in Appendix C.  Data were generated to follow our sensitivity analysis assumption (Assumption \ref{assumption:tilt}) with true $\alpha_1$ and $\alpha_0$ values of $-0.6$, $-0.3$, $0$, $0.3$, and $0.6$ and analyzed using our augmented inverse intensity-weighted estimators.  We first assessed the finite sample performance of our estimators by analyzing the simulated data using the true values of $\alpha_1$ and $\alpha_0$.  To demonstrate the benefit of our approach by showing the dangers of not accounting for informative assessment times in the analysis, we also analyze the same simulated data using the explainable assessment assumption that $\alpha_1=\alpha_0=0$ in each case.  This explores the performance of an approach which relies on the explainable assessment assumption in cases where that assumption does not hold.  Results for the treatment effects at 6 months and 12 months for each of these analyses are shown in Table \ref{table:sims_effect}.  Results for the mean outcome in each treatment arm at 3, 6, 9, and 12 months (including the true values of each mean) are given in Appendix C.  In the analysis estimating treatment effects using the true values of $\alpha_1$ and $\alpha_0$, bias is small, with an absolute value of less than $0.05$ in each case.  Confidence interval coverage is close to the nominal level of $0.95$, ranging between $0.930$ and $0.966$.  In the analysis assuming $\alpha_1=\alpha_0=0$, in many cases the bias is large and confidence interval coverage is poor, including coverage as low as $0.594$ even in cases with $\alpha_1=\alpha_0$.  This highlights the importance of considering a range of different assumptions through sensitivity analysis.

\begin{table}

\begin{center}

\footnotesize

 \renewcommand{\arraystretch}{1.5}

\begin{tabular}{ c c c c  c  c  c  c c c c c c }
 \Xhline{2pt}
 & & & \multicolumn{10}{c}{True $\alpha_0$ } \\
 \cline{4-13} 
   \multicolumn{3}{c}{~} & \multicolumn{2}{c}{-0.6} & \multicolumn{2}{c}{-0.3} & \multicolumn{2}{c}{0} & \multicolumn{2}{c}{0.3} & \multicolumn{2}{c}{0.6} \\
 \cline{4-13} 
 \multicolumn{3}{c}{~}  & $|$Bias$|$ &  Cov.  &  $|$Bias$|$  & Cov. &  $|$Bias$|$ & Cov.  & $|$Bias$|$  & Cov.   & $|$Bias$|$  & Cov.   \\
  \Xhline{2pt}
 \multirow{3}{0.4in}{\textbf{Month} \\ \ \ \  \textbf{6} } 
 &  
\multirow{2}{*}{ 0.6 } & S.A. & 0.021 & 0.936 & 0.016 & 0.940 & 0.010 &  0.944 &  0.005  & 0.958 &  0.000 &  0.954 \\
\cline{3-13}
& & Expl. & 1.459 & 0.000 & 1.162 & 0.002 &0.772  & 0.008 & 0.286 & 0.650 & 0.272 & 0.624 \\
\Xcline{2-13}{1.4pt}
 \multirow{8}{0.4in}{True \\ \ \ \ $\alpha_1$ } & \multirow{2}{*}{0.3} 
 & S.A. & 0.015 & 0.946 &  0.010 &  0.946 & 0.003 & 0.942 & 0.001 & 0.962 & 0.007 &  0.954 \\
 \cline{3-13}
&  & Expl. & 1.044 & 0.004 & 0.747 & 0.014 &0.357  & 0.462 & 0.129 & 0.888 & 0.687 & 0.052 \\
\Xcline{2-13}{1.4pt}
& \multirow{2}{*}{0} 
& S.A. & 0.010 & 0.952 & 0.005 &  0.934 & 0.001 &  0.942 & 0.006 &  0.956 & 0.012 &  0.952 \\
\cline{3-13}
& & Expl. & 0.686 & 0.022 & 0.389 & 0.398 & 0.001 & 0.942 & 0.487 & 0.208 & 1.045 & 0.006 \\
\Xcline{2-13}{1.4pt}
&\multirow{2}{*}{ -0.3} & S.A. & 0.005 &  0.954  & 0.000 & 0.944 & 0.006 &  0.954 &  0.011  &  0.954 &  0.016 &  0.938 \\
\cline{3-13}
& & Expl. & 0.390 & 0.396 & 0.093 &0.912 &0.297  & 0.576 & 0.783 & 0.028 & 1.341 &0.002  \\
\Xcline{2-13}{1.4pt}
& \multirow{2}{*}{ -0.6} 
& S.A. & 0.001 &  0.964 &  0.004 &  0.960 &  0.011 & 0.956 & 0.015 &  0.958 & 0.021 & 0.942 \\
\cline{3-13}
& & Expl. & 0.151&  0.876 &  0.146 &  0.854 & 0.536 & 0.160 & 1.022  & 0.008 & 1.580 & 0.000  \\
 \Xhline{2pt} 
 \multirow{3}{0.4in}{\textbf{Month} \\ \ \ \ \textbf{12} } 
 &  
\multirow{2}{*}{ 0.6 } & S.A. & 0.004 & 0.954 & 0.002 & 0.952 & 0.010 &  0.956  &  0.021  & 0.960  &0.036  &  0.956 \\
\cline{3-13}
& & Expl. & 1.356 & 0.000 & 1.080 & 0.000 & 0.717 & 0.008 & 0.257 & 0.652 & 0.283 & 0.594 \\
\Xcline{2-13}{1.4pt}
\multirow{8}{0.4in}{True \\ \ \ \ $\alpha_1$ } & \multirow{2}{*}{ 0.3 } 
& S.A. & 0.002 &  0.964 &  0.008 & 0.958 & 0.017 &  0.956 &  0.028 & 0.956 &0.042  & 0.944 \\
\cline{3-13}
& & Expl. & 0.958 & 0.000 & 0.682 & 0.018 & 0.319 & 0.496 & 0.141 & 0.870 & 0.681 & 0.016 \\
\Xcline{2-13}{1.4pt}
& \multirow{2}{*}{ 0 } 
& S.A. & 0.006 & 0.956 & 0.012 & 0.948 & 0.020 & 0.944 & 0.032 & 0.940 & 0.046 & 0.930 \\
\cline{3-13}
& & Expl. & 0.619 & 0.040 & 0.343 & 0.450 & 0.020 & 0.944 & 0.480 & 0.158 & 1.020 &0.000  \\
\Xcline{2-13}{1.4pt}
& \multirow{2}{*}{ -0.3 } 
& S.A. & 0.009 & 0.960 &  0.015 &  0.956 & 0.023 & 0.938 & 0.034 & 0.938 & 0.048 & 0.938 \\
\cline{3-13}
& & Expl. & 0.340 & 0.458 & 0.064 & 0.938 &0.299  & 0.568 & 0.759 & 0.008 & 1.299 &0.000  \\
\Xcline{2-13}{1.4pt}
& \multirow{2}{*}{ -0.6 } 
& S.A. & 0.011 & 0.966 & 0.017 & 0.960 & 0.025 &  0.952& 0.036 & 0.952 & 0.050 & 0.942 \\
\cline{3-13}
& & Expl. & 0.115 & 0.906 & 0.161 & 0.852 & 0.524  & 0.092  & 0.984 & 0.000 & 1.524 &0.000  \\
 \Xhline{2pt}
\end{tabular}

\end{center}

\bigskip

\normalsize

\begin{singlespace}

\caption{Simulation results.  Data were generated under our sensitivity analysis assumption using values of $\alpha_0,\alpha_1=-0.6,-0.3,0,0.3,0.6$.  The treatment effects at 6 and 12 months were then estimated using our augmented inverse intensity weighted estimator: (a) using the true values of $\alpha_0,\alpha_1$ (rows denoted ``S.A."), and (b) under the explainable assessment assumption that $\alpha_0=\alpha_1=0$ (rows denoted ``Expl.").  Shown are the absolute value of the empirical bias and the confidence interval coverage across 500 simulations. Confidence intervals are Wald confidence intervals using the jackknife variance estimate.  }

\label{table:sims_effect}

\end{singlespace}

\end{table}

\section{Discussion} \label{sec:discussion}

In many trials where the timing of outcome assessments varies by participant, assessment times may be related to underlying outcome values.  This dependence can give misleading conclusions about the effect of treatment if not correctly accounted for in the analysis.  Analysis methods for this setting make an untestable assumption about the informative assessment process; however, many assumptions can be consistent with the study data, and the treatment effect may differ across these assumptions.  In this sense, researchers face two sources of uncertainty:  the usual statistical uncertainty due to sample size, and the unknown degree to which assessments may be informative. Our sensitivity analysis methodology provides researchers with a tool that accounts for both of these factors.  By presenting inferences for treatment effects under a range of different assumptions, researchers will be able to provide a more accurate representation of the overall uncertainty in their study conclusions.  

Unfortunately, the question of whether assessment times are informative in a given study cannot necessarily be determined from the study data. In particular, assessment times may be informative even in studies without indications such as number of assessments varying by participant;  differential timing of assessments across treatment arms; or timing of assessment impacted by the outcome at previous assessments.  Substantive knowledge about the study should also be consulted in considering whether participants may be more (or less) likely to have an assessment when their outcome is worse.  If investigators anticipate having irregular follow-up times in their study, they can consider conducting participant interviews to learn whether the reasons for missed, delayed, or early appointments were related to participants' outcomes.  Participant responses could then be used to help assess whether sensitivity analysis is needed and inform the range of sensitivity parameter values to be included.

In this paper, we opted for an intensity modeling approach, as was used for example in \citet{scharfstein2004dependent}, \citet{sun2007regression}, \citet{buzkova2007dependent} and \citet{liang2009jointmodeling}.  It would also be possible to develop a discrete-time version of our approach using pooled logistic regression with smoothing of the time-specific intercepts.  Our estimation approach was developed for continuous outcomes and uses a mean model with the identity link (Assumption \ref{assumption:smooth}).  Future work will generalize our work to other link functions, such as the logit link appropriate for binary outcomes.  The ARC study had minimal dropout, and in our approach we have assumed that no participants are censored (though they may have fewer assessments than the study protocol specifies).  Future work will relax this assumption.  We also used an assumption of non-future dependence (Assumption \ref{assumption:future}) which may not be appropriate for some types of studies.  This assumption is not needed for identification, but is used in our estimation approach.  Finally, an important issue is selecting the range of sensitivity parameter values that will be included in the analysis.  Here, we have used a bounding approach that uses domain experts' knowledge in a direct way without the need for additional assumptions;  however, this approach may result in a wide range of values.  A key direction for future research is to develop and study other bounding procedures, including methods that would incorporate participant interviews described above.

\section*{Acknowledgements}

This project was funded by a Patient-Centered Outcomes Research Institute (PCORI) award ME-2021C3-24972. Shu Yang, Yujing Gao, Andrea Apter and Daniel Scharfstein  received funding from this award. The ARC study was funded through PCORI award AS-1307-05218.  Bonnie Smith's research was partially supported by NIH EB029977.  Additional funding support for Shu Yang, Andrea Apter and Daniel Scharfstein were grants NSF SES 2242776, NIH R18HL116285 and NIH R01DA046534, respectively.  Dr. Varadhan would like to acknowledge the funding support from the grant: NCI CCSG P30 CA006973.  We express our  gratitude to Ming-Yueh Huang for sharing his code for the single index model. We thank Russell Localio for sharing his insights about the ARC trial and Eleanor Pullenayegum for helpful discussions.

.\vspace*{-8pt}

\bibliographystyle{biom}
\bibliography{Irregular_bib}

\begin{thebibliography}{}

\bibitem[\protect\citeauthoryear{Andersen, Borgan, Gill, and Keiding}{Andersen
  et~al.}{1993}]{andersen1993bookABGK}
Andersen, P.~K., Borgan, O., Gill, R.~D., and Keiding, N. (1993).
\newblock {\em Statistical models based on counting processes}.
\newblock Springer-Verlag, New York.

\bibitem[\protect\citeauthoryear{Andersen and Gill}{Andersen and
  Gill}{1982}]{andersengill1982}
Andersen, P.~K. and Gill, R.~D. (1982).
\newblock Cox's regression model for counting processes: A large sample study.
\newblock {\em The Annals of Statistics} {\bf 10,} 1100 -- 1120.

\bibitem[\protect\citeauthoryear{Andersen, Ørnulf Borgan, Hjort, Arjas, Stene,
  and Aalen}{Andersen et~al.}{1985}]{andersenborgan1985countingprocesses}
Andersen, P.~K., Ørnulf Borgan, Hjort, N.~L., Arjas, E., Stene, J., and Aalen,
  O. (1985).
\newblock Counting process models for life history data: A review [with
  discussion and reply].
\newblock {\em Scandinavian Journal of Statistics} {\bf 12,} 97--158.

\bibitem[\protect\citeauthoryear{Apter, Localio, Morales, Han, Perez, Mullen,
  et~al\mbox{.}}{Apter et~al.}{2019}]{apter2019arc}
Apter, A.~J., Localio, A.~R., Morales, K.~H., Han, X., Perez, L., Mullen,
  A.~N., et~al. (2019).
\newblock Home visits for uncontrolled asthma among low-income adults with
  patient portal access.
\newblock {\em Journal of Allergy and Clinical Immunology} {\bf 144,}
  846--853.e11.

\bibitem[\protect\citeauthoryear{Barndorff-Nielsen and Cox}{Barndorff-Nielsen
  and Cox}{1994}]{barndorffnielsen1994inference}
Barndorff-Nielsen, O.~E. and Cox, D.~R. (1994).
\newblock {\em Inference and Asymptotics}.
\newblock Chapman and Hall.

\bibitem[\protect\citeauthoryear{Birmingham, Rotnitzky, and
  Fitzmaurice}{Birmingham et~al.}{2003}]{birmingham2003patternmixture}
Birmingham, J., Rotnitzky, A., and Fitzmaurice, G.~M. (2003).
\newblock Pattern–mixture and selection models for analysing longitudinal
  data with monotone missing patterns.
\newblock {\em Journal of the Royal Statistical Society: Series B (Statistical
  Methodology)} {\bf 65,} 275--297.

\bibitem[\protect\citeauthoryear{Breslow}{Breslow}{1972}]{breslow1972discussion}
Breslow, N.~E. (1972).
\newblock {Discussion on Professor Cox's paper}.
\newblock {\em Journal of the Royal Statistical Society: Series B
  (Methodological)} {\bf 34,} 216--217.

\bibitem[\protect\citeauthoryear{B\r{u}\v{z}kov\'{a} and
  Lumley}{B\r{u}\v{z}kov\'{a} and Lumley}{2007}]{buzkova2007dependent}
B\r{u}\v{z}kov\'{a}, P. and Lumley, T. (2007).
\newblock Longitudinal data analysis for generalized linear models with
  follow-up dependent on outcome-related variables.
\newblock {\em The Canadian Journal of Statistics / La Revue Canadienne de
  Statistique} {\bf 35,} 485--500.

\bibitem[\protect\citeauthoryear{Bůžková and Lumley}{Bůžková and
  Lumley}{2009}]{buzkova2009repeated}
Bůžková, P. and Lumley, T. (2009).
\newblock Semiparametric modeling of repeated measurements under
  outcome-dependent follow-up.
\newblock {\em Statistics in Medicine} {\bf 28,} 987--1003.

\bibitem[\protect\citeauthoryear{Chen, Ning, and Cai}{Chen
  et~al.}{2015}]{chen2015regression}
Chen, Y., Ning, J., and Cai, C. (2015).
\newblock Regression analysis of longitudinal data with irregular and
  informative observation times.
\newblock {\em Biostatistics} {\bf 16,} 727--739.

\bibitem[\protect\citeauthoryear{Chiang and Huang}{Chiang and
  Huang}{2012}]{chiang2012new}
Chiang, C.-T. and Huang, M.-Y. (2012).
\newblock New estimation and inference procedures for a single-index
  conditional distribution model.
\newblock {\em Journal of Multivariate Analysis} {\bf 111,} 271--285.

\bibitem[\protect\citeauthoryear{Cinelli and Hazlett}{Cinelli and
  Hazlett}{2020}]{cinelli2020sensitivity}
Cinelli, C. and Hazlett, C. (2020).
\newblock Making sense of sensitivity: Extending omitted variable bias.
\newblock {\em Journal of the Royal Statistical Society: Series B (Statistical
  Methodology)} {\bf 82,} 39--67.

\bibitem[\protect\citeauthoryear{Cook and Lawless}{Cook and
  Lawless}{2007}]{cooklawless2007book}
Cook, R.~J. and Lawless, J. (2007).
\newblock {\em The statistical analysis of recurrent events}.
\newblock Springer, New York.

\bibitem[\protect\citeauthoryear{Cox}{Cox}{1975}]{cox1975partiallikelihood}
Cox, D. (1975).
\newblock Partial likelihood.
\newblock {\em Biometrika} {\bf 62,} 269--276.

\bibitem[\protect\citeauthoryear{Cox}{Cox}{1972}]{cox1972regression}
Cox, D.~R. (1972).
\newblock Regression models and life-tables.
\newblock {\em Journal of the Royal Statistical Society: Series B
  (Methodological)} {\bf 34,} 187--202.

\bibitem[\protect\citeauthoryear{D\'{i}az~Mu{\~n}oz and van~der
  Laan}{D\'{i}az~Mu{\~n}oz and van~der Laan}{2011}]{diaz2011superlearner}
D\'{i}az~Mu{\~n}oz, I. and van~der Laan, M.~J. (2011).
\newblock Super learner based conditional density estimation with application
  to marginal structural models:.
\newblock {\em The International Journal of Biostatistics} {\bf 7,}.

\bibitem[\protect\citeauthoryear{Franks, D’Amour, and Feller}{Franks
  et~al.}{2020}]{franks2020observationalsa}
Franks, A.~M., D’Amour, A., and Feller, A. (2020).
\newblock Flexible sensitivity analysis for observational studies without
  observable implications.
\newblock {\em Journal of the American Statistical Association} {\bf 115,}
  1730--1746.

\bibitem[\protect\citeauthoryear{Juniper, O'byrne, Guyatt, Ferrie, and
  King}{Juniper et~al.}{1999}]{juniper1999questionnaire}
Juniper, E., O'byrne, P., Guyatt, G., Ferrie, P., and King, D. (1999).
\newblock Development and validation of a questionnaire to measure asthma
  control.
\newblock {\em European Respiratory Journal} {\bf 14,} 902--907.

\bibitem[\protect\citeauthoryear{Kennedy}{Kennedy}{2016}]{kennedy2016empiricalprocesses}
Kennedy, E.~H. (2016).
\newblock Semiparametric theory and empirical processes in causal inference.
\newblock In He, H., Wu, P., and Chen, D.-G.~D., editors, {\em Statistical
  Causal Inferences and Their Applications in Public Health Research}, ICSA
  Book Series in Statistics, pages 141--167. Springer International Publishing,
  Cham.

\bibitem[\protect\citeauthoryear{Kenward, Molenberghs, and Thijs}{Kenward
  et~al.}{2003}]{kenward2003patternmixture}
Kenward, M.~G., Molenberghs, G., and Thijs, H. (2003).
\newblock {Pattern‐mixture models with proper time dependence}.
\newblock {\em Biometrika} {\bf 90,} 53--71.

\bibitem[\protect\citeauthoryear{Liang, Lu, and Ying}{Liang
  et~al.}{2009}]{liang2009jointmodeling}
Liang, Y., Lu, W., and Ying, Z. (2009).
\newblock Joint modeling and analysis of longitudinal data with informative
  observation times.
\newblock {\em Biometrics} {\bf 65,} 377--384.

\bibitem[\protect\citeauthoryear{Lin and Ying}{Lin and
  Ying}{2001}]{lin2001regression}
Lin, D.~Y. and Ying, Z. (2001).
\newblock Semiparametric and nonparametric regression analysis of longitudinal
  data.
\newblock {\em Journal of the American Statistical Association} {\bf 96,}
  103--126.

\bibitem[\protect\citeauthoryear{Lin, Scharfstein, and Rosenheck}{Lin
  et~al.}{2004}]{scharfstein2004dependent}
Lin, H., Scharfstein, D.~O., and Rosenheck, R.~A. (2004).
\newblock Analysis of longitudinal data with irregular, outcome-dependent
  follow-up.
\newblock {\em Journal of the Royal Statistical Society: Series B (Statistical
  Methodology)} {\bf 66,} 791--813.

\bibitem[\protect\citeauthoryear{Lipsitz, Fitzmaurice, Ibrahim, Gelber, and
  Lipshultz}{Lipsitz et~al.}{2002}]{lipsitz2002dependent}
Lipsitz, S.~R., Fitzmaurice, G.~M., Ibrahim, J.~G., Gelber, R., and Lipshultz,
  S. (2002).
\newblock Parameter estimation in longitudinal studies with outcome-dependent
  follow-up.
\newblock {\em Biometrics} {\bf 58,} 621--630.

\bibitem[\protect\citeauthoryear{{National Research Council}}{{National
  Research Council}}{2010}]{nationalacademy2010missingdata}
{National Research Council} (2010).
\newblock {\em The Prevention and Treatment of Missing Data in Clinical
  Trials}.
\newblock The National Academies Press, Washington, DC.

\bibitem[\protect\citeauthoryear{Ogata}{Ogata}{1981}]{ogata1981simulation}
Ogata, Y. (1981).
\newblock {On Lewis' simulation method for point processes}.
\newblock {\em IEEE Transactions on Information Theory} {\bf 27,} 23--31.

\bibitem[\protect\citeauthoryear{Pullenayegum and Feldman}{Pullenayegum and
  Feldman}{2013}]{pullenayegum2013dr}
Pullenayegum, E.~M. and Feldman, B.~M. (2013).
\newblock Doubly robust estimation, optimally truncated inverse-intensity
  weighting and increment-based methods for the analysis of irregularly
  observed longitudinal data.
\newblock {\em Statistics in Medicine} {\bf 32,} 1054--1072.

\bibitem[\protect\citeauthoryear{Pullenayegum and Lim}{Pullenayegum and
  Lim}{2016}]{pullenayegum2016review}
Pullenayegum, E.~M. and Lim, L.~S. (2016).
\newblock Longitudinal data subject to irregular observation: A review of
  methods with a focus on visit processes, assumptions, and study design.
\newblock {\em Statistical Methods in Medical Research} {\bf 25,} 2992--3014.
\newblock PMID: 24855119.

\bibitem[\protect\citeauthoryear{Pullenayegum and Scharfstein}{Pullenayegum and
  Scharfstein}{2022}]{pullenayegum2022repeatedly}
Pullenayegum, E.~M. and Scharfstein, D.~O. (2022).
\newblock {Randomized trials with repeatedly measured outcomes: Handling
  irregular and potentially informative assessment times}.
\newblock {\em Epidemiologic Reviews} {\bf 44,} 121--137.

\bibitem[\protect\citeauthoryear{Ramlau-Hansen}{Ramlau-Hansen}{1983}]{ramlauhansen1983smoothing}
Ramlau-Hansen, H. (1983).
\newblock Smoothing counting process intensities by means of kernel functions.
\newblock {\em The Annals of Statistics} {\bf 11,} 453--466.

\bibitem[\protect\citeauthoryear{Rotnitzky, Scharfstein, Su, and
  Robins}{Rotnitzky et~al.}{2001}]{rotnitzky2001competing}
Rotnitzky, A., Scharfstein, D., Su, T.-L., and Robins, J. (2001).
\newblock Methods for conducting sensitivity analysis of trials with
  potentially nonignorable competing causes of censoring.
\newblock {\em Biometrics} {\bf 57,} 103--113.

\bibitem[\protect\citeauthoryear{Scharfstein and McDermott}{Scharfstein and
  McDermott}{2019}]{scharfstein2019globalsa}
Scharfstein, D.~O. and McDermott, A. (2019).
\newblock Global sensitivity analysis of clinical trials with missing
  patient-reported outcomes.
\newblock {\em Statistical Methods in Medical Research} {\bf 28,} 1439--1456.
\newblock PMID: 29557705.

\bibitem[\protect\citeauthoryear{Scharfstein, Rotnitzky, and
  Robins}{Scharfstein et~al.}{1999}]{scharfstein1999nonignorable}
Scharfstein, D.~O., Rotnitzky, A., and Robins, J.~M. (1999).
\newblock Adjusting for nonignorable drop-out using semiparametric nonresponse
  models.
\newblock {\em Journal of the American Statistical Association} {\bf 94,}
  1096--1120.

\bibitem[\protect\citeauthoryear{Shen, Liu, Chen, and Ning}{Shen
  et~al.}{2019}]{shen2019regression}
Shen, W., Liu, S., Chen, Y., and Ning, J. (2019).
\newblock Regression analysis of longitudinal data with outcome-dependent
  sampling and informative censoring.
\newblock {\em Scandinavian Journal of Statistics} {\bf 46,} 831--847.

\bibitem[\protect\citeauthoryear{Sjölander, Gabriel, and
  Ciocănea-Teodorescu}{Sjölander et~al.}{2022}]{sjolander2022sacausal}
Sjölander, A., Gabriel, E.~E., and Ciocănea-Teodorescu, I. (2022).
\newblock Sensitivity analysis for causal effects with generalized linear
  models.
\newblock {\em Journal of Causal Inference} {\bf 10,} 441--479.

\bibitem[\protect\citeauthoryear{Sun, Sun, and Liu}{Sun
  et~al.}{2007}]{sun2007regression}
Sun, J., Sun, L., and Liu, D. (2007).
\newblock Regression analysis of longitudinal data in the presence of
  informative observation and censoring times.
\newblock {\em Journal of the American Statistical Association} {\bf 102,}
  1397--1406.

\bibitem[\protect\citeauthoryear{Sun, Mu, Sun, and Tong}{Sun
  et~al.}{2011}]{sun2011semiparametric}
Sun, L., Mu, X., Sun, Z., and Tong, X. (2011).
\newblock Semiparametric analysis of longitudinal data with informative
  observation times.
\newblock {\em Acta Mathematicae Applicatae Sinica, English Series} {\bf 27,}
  29--42.

\bibitem[\protect\citeauthoryear{Sun, Song, and Zhou}{Sun
  et~al.}{2011}]{sun2011regression}
Sun, L., Song, X., and Zhou, J. (2011).
\newblock Regression analysis of longitudinal data with time-dependent
  covariates in the presence of informative observation and censoring times.
\newblock {\em Journal of Statistical Planning and Inference} {\bf 141,}
  2902--2919.

\bibitem[\protect\citeauthoryear{Sun, Peng, Manatunga, and Marcus}{Sun
  et~al.}{2016}]{sun2016quantile}
Sun, X., Peng, L., Manatunga, A., and Marcus, M. (2016).
\newblock Quantile regression analysis of censored longitudinal data with
  irregular outcome-dependent follow-up.
\newblock {\em Biometrics} {\bf 72,} 64--73.

\bibitem[\protect\citeauthoryear{Tsiatis}{Tsiatis}{2006}]{tsiatis2006book}
Tsiatis, A.~A. (2006).
\newblock {\em Semiparametric Theory and Missing Data}.
\newblock Springer, New York.

\bibitem[\protect\citeauthoryear{van~der Vaart}{van~der
  Vaart}{1998}]{vandervaart1998asymptotic}
van~der Vaart, A.~W. (1998).
\newblock {\em Asymptotic Statistics}.
\newblock Cambridge Series in Statistical and Probabilistic Mathematics.
  Cambridge University Press.

\bibitem[\protect\citeauthoryear{Vansteelandt, Rotnitzky, and
  Robins}{Vansteelandt et~al.}{2007}]{vansteelandt2007nonmonotone}
Vansteelandt, S., Rotnitzky, A., and Robins, J. (2007).
\newblock Estimation of regression models for the mean of repeated outcomes
  under nonignorable nonmonotone nonresponse.
\newblock {\em Biometrika} {\bf 94,} 841--860.

\bibitem[\protect\citeauthoryear{Veitch and Zaveri}{Veitch and
  Zaveri}{2020}]{veitch2020sense}
Veitch, V. and Zaveri, A. (2020).
\newblock Sense and sensitivity analysis: Simple post-hoc analysis of bias due
  to unobserved confounding.
\newblock In Larochelle, H., Ranzato, M., Hadsell, R., Balcan, M.~F., and Lin,
  H., editors, {\em Advances in Neural Information Processing Systems},
  volume~33, pages 10999--11009. Curran Associates, Inc.

\bibitem[\protect\citeauthoryear{Wang and Daniels}{Wang and
  Daniels}{2011}]{wang2011note}
Wang, C. and Daniels, M.~J. (2011).
\newblock {A note on MAR, identifying restrictions, model comparison, and
  sensitivity analysis in pattern mixture models with and without covariates
  for incomplete data}.
\newblock {\em Biometrics} {\bf 67,} 810--818.

\bibitem[\protect\citeauthoryear{Wang}{Wang}{2020}]{wang2020parametric}
Wang, Z. (2020).
\newblock Global sensitivity analysis for randomized trials with informative
  assessment times: A fully parametric approach.
\newblock Master's thesis, Johns Hopkins University.

\bibitem[\protect\citeauthoryear{Wells}{Wells}{1994}]{wells1994kernelestimation}
Wells, M.~T. (1994).
\newblock {Nonparametric kernel estimation in counting processes with
  explanatory variables}.
\newblock {\em Biometrika} {\bf 81,} 795--801.

\end{thebibliography}

\ 

\ 

\section*{Appendix A:  Implementation Details}

\subsection*{Appendix A.1:  Tutorial on implementing estimation of the mean outcome}

In the following tutorial we illustrate implementation of our estimation procedure for the mean outcome for a single treatment arm.  We use a simulated dataset, with similar structure to the data from the ARC trial, on $N=200$ intervention-arm participants.   Here we estimate the mean outcome in this arm under a small number of values of the sensitivity parameter for illustration purposes.

\newpage



# 1. Formatting the Data

The data should be in recurrent events format, with a row for each post-baseline assessment for each participant. Each row corresponds to a time period starting at the previous assessment time `Prev_time` and ending at the current assessment time `Time`. `Event` is the indicator of whether the participant has an assessment at time `Time`. In our simulated study, individuals exit the risk set upon their 4th post-baseline assessment if they have one, and otherwise they are censored at the end of the study period, which we set here to be at 829 days. `Visit` is the number of assessments (including baseline) that the participant had prior to time `Time`.

```
library(knitr)
library(dplyr)
library(ggplot2)

Trt_dat=readRDS("Tutorial_simulated_trt_data.rds")
```

Table 1: Data for the first 3 subjects in the simulated dataset

| Subject_ID | Visit | Prev_time | Time | Event | Asthma_control | Prev_outcome | Lag_time |
|---|---|---|---|---|---|---|---|
| 1 | 1 | 0 | 214 | 1 | 4.50 | 3.00 | 214 |
| 1 | 2 | 214 | 292 | 1 | 4.17 | 4.50 | 78 |
| 1 | 3 | 292 | 370 | 1 | 1.33 | 4.17 | 78 |
| 1 | 4 | 370 | 441 | 1 | 0.83 | 1.33 | 71 |
| 2 | 1 | 0 | 72 | 1 | 0.50 | 3.00 | 72 |
| 2 | 2 | 72 | 181 | 1 | 2.00 | 0.50 | 109 |
| 2 | 3 | 181 | 297 | 1 | 1.50 | 2.00 | 116 |
| 2 | 4 | 297 | 366 | 1 | 1.83 | 1.50 | 69 |
| 3 | 1 | 0 | 214 | 1 | 0.17 | 0.67 | 214 |
| 3 | 2 | 214 | 496 | 1 | 0.33 | 0.17 | 282 |
| 3 | 3 | 496 | 829 | 0 | NA | 0.33 | 333 |

# 2. Marginal Mean Model

We first plot the assessment times and specify an interval $[t_1, t_2]$ on which we will estimate the mean outcome. All target times should be within this interval, and the interval should be well-supported by data, since we will predict the mean outcome at each time in the interval. See Appendix 1 for a modification for studies where a single time interval cannot be used.

```
Trt_data_assessed=filter(Trt_dat,Event==1)

ggplot(Trt_data_assessed,aes(x=Time,colour=as.factor(Visit)))+
  geom_histogram(binwidth=7,alpha=0.7,colour="#009E73",fill="#009E73")+
  theme_bw()+
  expand_limits(x=c(0,850))+
  scale_x_continuous(breaks=seq(0,900,by=100))+
  labs(y="",title="Assessment Times, Simulated Data")
```

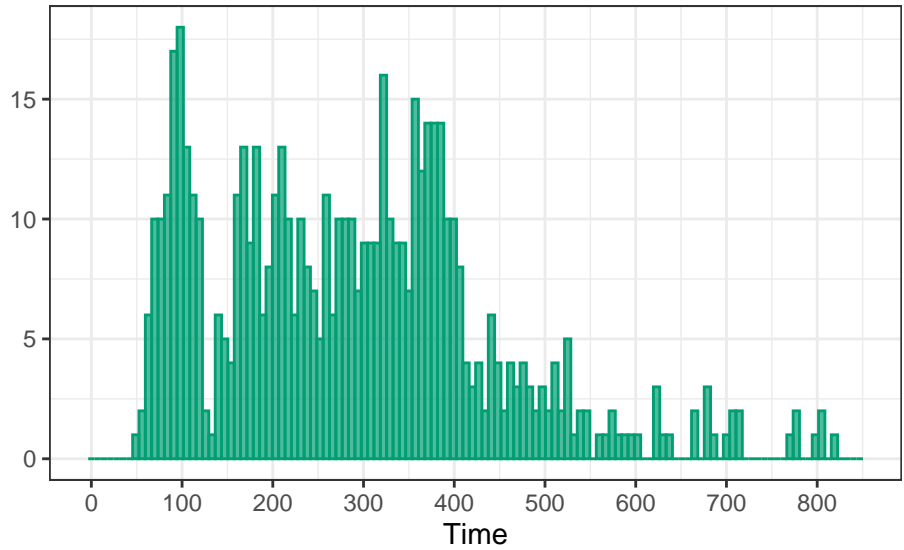

Assessment Times, Simulated Data

```
Trt_data_assessed=filter(Trt_dat,Event==1)
quantile(Trt_data_assessed$Time,c(0,0.05,0.33,0.67,0.95,1))
```

```
##     0%    5%   33%   67%   95%  100%
##   47.0  80.9 203.0 353.0 566.9 821.0
```

```
End=821
```

For the simulated data we will use a time interval of $[47, 560]$ days. Next we specify a functional form $B(t)$ for the marginal mean assumption:

$$\mu_j(t) = \beta' B(t), \ t \in [47, 560].$$

We construct a matrix `Basis_matrix` whose $(i, j)$ entry is the value of the $j$th component of $B(t)$ for the $i$th day of the time interval $[t_1, t_2]$.

Here we choose $B(t)$ to be a basis of B-splines, using cubic splines and two interior knots. This is a basis of length $p = 6$, so `Basis_matrix` has dimension $514 \times 6$.

```
library(splines2)

t1=47
t2=560

Basis_matrix=bSpline(x=t1:t2,knots=c(203,353),degree=3,intercept=TRUE)

#### length of beta
p=dim(Basis_matrix)[2]
```

Table 2: First 8 rows of the basis matrix.

| 1 | 2 | 3 | 4 | 5 | 6 |
|---:|---:|---:|---:|---:|---:|
| 1.000 | 0.000 | 0.000 | 0 | 0 | 0 |
| 0.981 | 0.019 | 0.000 | 0 | 0 | 0 |

|   | 1     | 2     | 3     | 4 | 5 | 6 |
|---|-------|-------|-------|---|---|---|
|   | 0.962 | 0.038 | 0.000 | 0 | 0 | 0 |
|   | 0.943 | 0.056 | 0.001 | 0 | 0 | 0 |
|   | 0.925 | 0.074 | 0.001 | 0 | 0 | 0 |
|   | 0.907 | 0.092 | 0.002 | 0 | 0 | 0 |
|   | 0.889 | 0.109 | 0.002 | 0 | 0 | 0 |
|   | 0.871 | 0.126 | 0.003 | 0 | 0 | 0 |

Below we plot the outcomes in the interval $[t_1, t_2]$ along with a nonparametric smooth (blue) and a curve fit using this B-spline basis (red).

```
Interval_data=filter(Trt_data_assessed,Time>=47,Time<=560)
fit=lm(Asthma_control~bSpline(Time,knots=c(203,353),degree=3,intercept=TRUE)-1,
       data=Interval_data)
B_predict=predict(fit)

ggplot(Interval_data,aes(x=Time,y=Asthma_control))+
  geom_point(colour="grey50",size=1.3)+
  geom_smooth(fill="lightblue")+
  geom_line(aes(x=Time,y=B_predict),colour="red")+
  expand_limits(x=c(0,600))+
  theme_bw()+
  labs(y="ACQ Score",title="Simulated Outcome Data Fit with a Nonparametric \nSmooth (Blue) and a B-Spl
  theme(title=element_text(size=9))
```

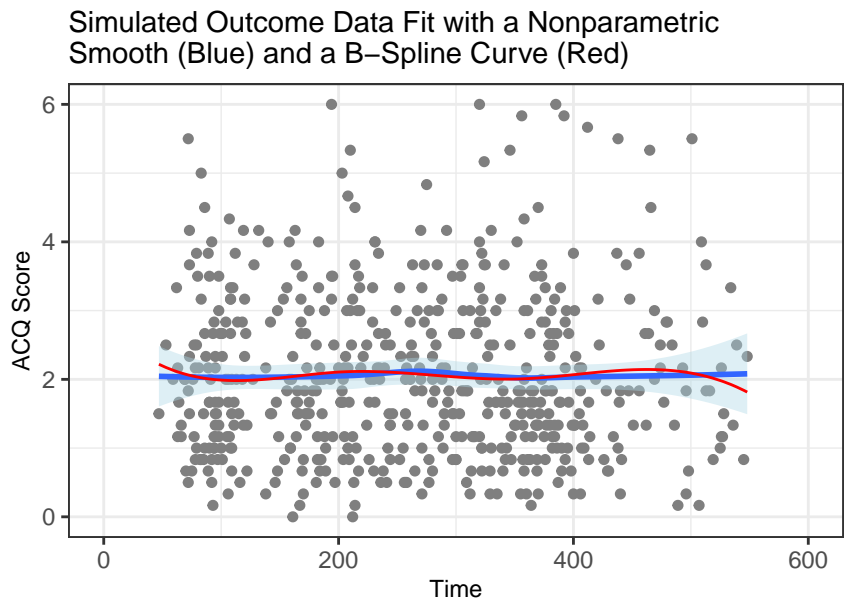

## 3. Observed Data Modeling

Next we fit a model for the intensity function $\lambda(t|\overline{O}(t))$ using a stratified Andersen-Gill model, where the stratification variable is `Visit`. We then estimate the baseline intensity function for each stratum using kernel smoothing with an Epanechnikov kernel, $K(\mu) = \frac{3}{4}(1-\mu)^2$, $|\mu| \leq 1$. We save the baseline intensity functions in a data frame `Base_intensities`.

In this model we use a single predictor, which is the outcome at the previous assessment, `Prev_outcome`. We use a bandwidth of 30 days for the kernel smoothing.

```r
library(survival)

AG_model=coxph(Surv(Prev_time,Time,Event)~Prev_outcome+strata(Visit),
               id=Subject_ID,data=Trt_dat)

gamma=AG_model$coefficients

AG_surv=survfit(AG_model,newdata=data.frame(Prev_outcome=0))
strata=AG_surv$strata
strata
```

```
## Visit=1 Visit=2 Visit=3 Visit=4
##     124     133     101      68
```

```r
v1=strata[1]
cumhaz_v1=data.frame(time=AG_surv$time[1:v1],
                     cumhaz=AG_surv$cumhaz[1:v1])

v2=strata[2]
cumhaz_v2=data.frame(time=AG_surv$time[(v1+1):(v1+v2)],
                     cumhaz=AG_surv$cumhaz[(v1+1):(v1+v2)])

v3=strata[3]
cumhaz_v3=data.frame(time=AG_surv$time[(v1+v2+1):(v1+v2+v3)],
                     cumhaz=AG_surv$cumhaz[(v1+v2+1):(v1+v2+v3)])

v4=strata[4]
cumhaz_v4=data.frame(time=AG_surv$time[(v1+v2+v3+1):(v1+v2+v3+v4)],
                     cumhaz=AG_surv$cumhaz[(v1+v2+v3+1):(v1+v2+v3+v4)])
```

```r
####  Function to obtain the baseline intensity function from a cumulative
####    baseline intensity function, using kernel smoothing with an Epanechnikov kernel
####    and baseline b
lambda0_fn=function(t,b,surv){
  
  times=surv$time
  cumhaz=surv$cumhaz
  
  increment=rep(0,length(times))
  increment[1]=cumhaz[1]
  for(j in 2:length(times)){ increment[j]=cumhaz[j]-cumhaz[j-1] }
  
  f_t=(1/b)*sum( 0.75*(1-( (t-times)/b )^2 )*( abs(t-times)<b )*increment )
  return(f_t)
  
}
```

```r
base_intens_v1=sapply(1:End,lambda0_fn,b=30,surv=cumhaz_v1)
```

```r
base_intens_v2=sapply(1:End,lambda0_fn,b=30,surv=cumhaz_v2)
base_intens_v3=sapply(1:End,lambda0_fn,b=30,surv=cumhaz_v3)
base_intens_v4=sapply(1:End,lambda0_fn,b=30,surv=cumhaz_v4)

Base_intensities=data.frame(Base_s1=base_intens_v1,Base_s2=base_intens_v2,
                            Base_s3=base_intens_v3,Base_s4=base_intens_v4)
row.names(Base_intensities)=1:End
```

Below we plot the baseline intensity function $\lambda_{0k}(t)$ for each stratum, with the assessment times for each visit number shown below for comparison.

```r
df_base=data.frame(Time=1:821,S1=base_intens_v1,S2=base_intens_v2,
                   S3=base_intens_v3,S4=base_intens_v4)

ggplot(df_base,aes(x=Time,y=S1))+
  geom_line(colour="red")+
  geom_line(aes(x=Time,y=S2),colour="orange")+
  geom_line(aes(x=Time,y=S3),colour="forestgreen")+
  geom_line(aes(x=Time,y=S4),colour="purple")+
  theme_bw()+
  labs(y="Baseline Intensity",title="Baseline Intensity Function for Each Visit")
```

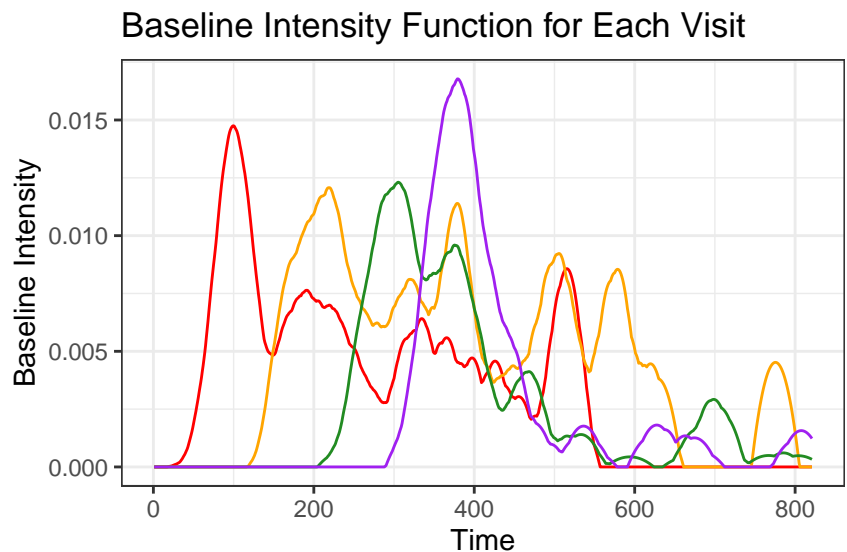

```r
ggplot(Trt_data_assessed,aes(x=Time,colour=as.factor(Visit),fill=as.factor(Visit)))+
  geom_histogram(binwidth=7,alpha=0.5)+
  scale_color_manual(values=c("red","orange","forestgreen","purple"))+
  scale_fill_manual(values=c("red","orange","forestgreen","purple"))+
  theme_bw()+
  expand_limits(x=c(0,821))+
  labs(y="Count",title="Assessment Times by Visit Number")+
  theme(legend.position = "none")
```

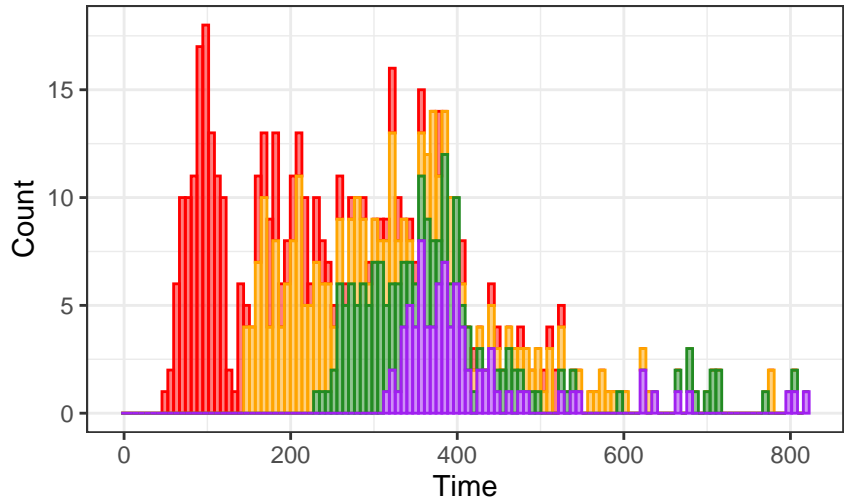

Assessment Times by Visit Number

The second model we fit is for the conditional outcome distribution $dF(y(t)|A(t) = 1, \bar{O}(t))$. For this we recommend using a single index model (Chiang and Huang, 2012), which provides a flexible way of modeling this distribution. Let $W = W(t, \bar{O}(t))$ be a vector of predictors included in the outcome model. The single index model assumes that the conditional cumulative distribution function (cdf) of observed outcomes given the observed past depends on $W$ only through a scalar $\theta'W$, where $\theta$ is a vector of unknown parameters, say as a function $G(\cdot, \theta'W)$. Let $Y_{ik}$ and $W_{ik}$ be the outcome and vector of predictors corresponding to participant $i$'s $k$th assessment. The predicted value of the conditional cdf given $W = w$ at each value $y$ is given by the Nadarya-Watson estimator:

$$\hat{G}(y, \theta'w) = \frac{\sum_{i,k} 1(Y_{ik} \leq y) K_h(\theta'W_{ik} - \theta'w)}{\sum_{i,k} K_h(\theta'W_{ik} - \theta'w)}$$

where the sums are over all assessments for all participants, $K(\cdot)$ is a kernel function, $h$ is a bandwidth, and $K_h(u) = K(u/h)/h$. The bandwidth $h$ and the coefficient vector $\theta$ are jointly estimated by minimizing a pseudo sum of integrated squares.

The predictors that we use here are current time `Time`, the lag time from the previous assessment `Lag_time`, and a natural cubic spline of the outcome at the previous assessment, which we denote as `ns1`, `ns2`, and `ns3`. Since the scales of these predictors vary widely, before we fit the single index model, we standardize `Time` and `Lag_time`. We use the Gaussian kernel for the single index model.

To fit the single index model, begin by forming the stacked design matrix denoted by `Xi` and the stacked outcome vector denoted by `Yi`. These are input into the function `cumuSIR(X,Y)` below to obtain the initial value of coefficients using the sliced inverse regression for the single index model. Then, the function `SIDR(X,Y,initial)` is used to obtain the estimated coefficients $\hat{\theta}$ and bandwidth $\hat{h}$ for the single index model. These will be used in the next stage to obtain predicted means.

```
library(splines)
library(dfoptim)

Trt_dat_outcome <- cbind(Trt_data_assessed,
                         ns1 = ns(Trt_data_assessed$Prev_outcome, df = 3)[, 1],
                         ns2 = ns(Trt_data_assessed$Prev_outcome, df = 3)[, 2],
                         ns3 = ns(Trt_data_assessed$Prev_outcome, df = 3)[, 3],
                         Time_scaled = scale(Trt_data_assessed$Time),
```

```r
                             Lag_time_scaled = scale(Trt_data_assessed$Lag_time))


##### Design matrix
Xi <- dplyr::select(Trt_dat_outcome, ns1,ns2,ns3,Time_scaled,Lag_time_scaled)
Xi <- as.matrix(Xi)

####  Outcome vector
Yi <- as.matrix(Trt_dat_outcome$Asthma_control, ncol = 1)


cumuSIR <- function(X, Y, eps = 1e-7){
    X <- as.matrix(X)
    Y <- as.matrix(Y)

    number_n <- dim(X)[1]
    number_p <- dim(X)[2]

    Y.CP <- matrix(
      Y[rep(1:number_n, times = number_n), ]<=
        Y[rep(1:number_n, each = number_n), ],
      nrow = number_n, ncol = number_n
    )

    # centralizing covariates
    X.cs <- t(t(X)-colMeans(X))

    # calculating m(y)=\E[X_i 1(Y_i\leq y)]
    m.y <- t(X.cs) %*% Y.CP/number_n
    # calculating K=\E[m(Y_i)m(Y_i)^T]
    Km <- m.y %*% t(m.y)/number_n

    Bhat <- eigen(solve(var(X) + eps*diag(number_p), Km))$vectors

    return(list(basis = Bhat))
}


SIDR <- function(X, Y,
                  Y.CP = NULL,
                  initial = NULL,
                  kernel = "dnorm",
                  method = "optim",
                  optim_method = "BFGS",
                  abs.tol = 1e-4,
                  bandwidth = NULL,
                  wi.boot = NULL)
{
  X <- as.matrix(X)
  Y <- as.matrix(Y)

  number_n <- dim(X)[1]
  number_p <- dim(X)[2]
```

```r
  if (is.null(initial))
  {
    initial <- c(1, rep(0, number_p-1))
  }else
  {
    initial <- as.vector(initial)
    initial <- initial/initial[1]
  }

  if (is.null(bandwidth))
  {
    if (kernel=="K2_Biweight")
    {
      if (is.null(wi.boot))
      {
        Eij3 <- function(parameter){
          K <- function(x, h) 15/16*(1-(x/h)^2)^2 * (abs(x) <= h) * (x!=0)

          b <- c(1, parameter[1:(number_p-1)])
          h <- exp(parameter[number_p])

          x <- c(X%*%b)
          y <- Y

          n <- length(y)
          yo <- order(y)
          ys <- y[yo]
          uy <- rle(ys)[[1]]
          cols <- cumsum(uy)
          ei <- rep(0, n)
          for (i in 1:n){
            Kih <- K(x-x[i],h=h)
            denom <- sum(Kih)
            ei[i] <- sum(uy*(1*(y[i] <= ys)[cols] -
                        (denom != 0)* cumsum(Kih[yo])[cols] / (denom + (denom == 0))))^2
          }
          return(sum(ei)/n^2)
        }

      }else
      {
        stop("There's no weighted version of the K2_Biweight kernel.")
      }
    }else if (kernel == "dnorm")
    {
      if (is.null(wi.boot))
      {
        Eij3 <- function(parameter){
          K <- function(x, h) dnorm(x/h, 0, 1)* (x!=0)

          b <- c(1, parameter[1:(number_p-1)])
          h <- exp(parameter[number_p])
```

8

```r
      x <- c(X%*%b)
      y <- Y
      
      n <- length(y)
      yo <- order(y)
      ys <- y[yo]
      uy <- rle(ys)[[1]]
      cols <- cumsum(uy)
      ei <- rep(0, n)
      for (i in 1:n){
        Kih <- K(x-x[i],h=h)
        denom <- sum(Kih)
        ei[i] <- sum(uy*(1*(y[i] <= ys)[cols] -
                    (denom != 0)* cumsum(Kih[yo])[cols] / (denom + (denom == 0)))^2)
      }
      return(sum(ei)/n^2)
    }
    
  }else
  {
    
    stop("There's no weighted version of the dnorm kernel.")
  }
}else if (kernel=="K4_Biweight")
{
  if (is.null(wi.boot))
  {
    Eij3 <- function(parameter){
      K <- function(x, h) 105/64*(1-3*((x/h)^2))*(1-(x/h)^2)^2 * (abs(x) <= h) * (x!=0)
      
      b <- c(1, parameter[1:(number_p-1)])
      h <- exp(parameter[number_p])
      
      x <- c(X%*%b)
      y <- Y
      
      n <- length(y)
      yo <- order(y)
      ys <- y[yo]
      uy <- rle(ys)[[1]]
      cols <- cumsum(uy)
      ei <- rep(0, n)
      for (i in 1:n){
        Kih <- K(x-x[i],h=h)
        denom <- sum(Kih)
        ei[i] <- sum(uy*(1*(y[i] <= ys)[cols] -
                    (denom != 0)* cumsum(Kih[yo])[cols] / (denom + (denom == 0)))^2)
      }
      return(sum(ei)/n^2)
    }
    
  }else
  {
```

```r
      stop("There's no weighted version of the K4_Biweight kernel.")
    }
  }

  if(method == "nlminb")
  {
    esti <- nlminb(start = c(initial[-1], 0),
                   objective = Eij3,
                   control = list(abs.tol = abs.tol))
  }else if (method == "optim")
  {
    # the new optimize function using optim, you can change the lower and upper
    esti <- optim(par = c(initial[-1], 0),
                  fn = Eij3,
                  method = optim_method,
                  control = list(abstol = abs.tol))
  }else if (method == "nmk")
  {
    esti <- nmk(par = c(initial[-1], 0),
                fn = Eij3,
                control = list(tol = abs.tol))
  }

  results <- list(coef = c(1, esti$par[1:(number_p-1)]),
                  bandwidth = exp(esti$par[number_p]),
                  details = esti)
}else
{
  if (kernel=="K2_Biweight")
  {
    if (is.null(wi.boot))
    {
      Eij3 <- function(parameter){
        K <- function(x, h) 15/16*(1-(x/h)^2)^2 * (abs(x) <= h) * (x!=0)

        b <- c(1, parameter[1:(number_p-1)])
        h <- bandwidth

        x <- c(X%*%b)
        y <- Y

        n <- length(y)
        yo <- order(y)
        ys <- y[yo]
        uy <- rle(ys)[[1]]
        cols <- cumsum(uy)
        ei <- rep(0, n)
        for (i in 1:n){
          Kih <- K(x-x[i],h=h)
          denom <- sum(Kih)
          ei[i] <- sum(uy*(1*(y[i] <= ys)[cols] -
                    (denom != 0)* cumsum(Kih[yo])[cols] / (denom + (denom == 0))))^2
        }
```

```r
      return(sum(ei)/n^2)
    }

  }else
  {
    stop("There's no weighted version of the K2_Biweight kernel.")
  }
}else if (kernel=="dnorm")
{
  if (is.null(wi.boot))
  {
    Eij3 <- function(parameter){
      K <- function(x, h) dnorm(x/h,0,1) * (x!=0)

      b <- c(1, parameter[1:(number_p-1)])
      h <- bandwidth

      x <- c(X%*%b)
      y <- Y

      n <- length(y)
      yo <- order(y)
      ys <- y[yo]
      uy <- rle(ys)[[1]]
      cols <- cumsum(uy)
      ei <- rep(0, n)
      for (i in 1:n){
        Kih <- K(x-x[i],h=h)
        denom <- sum(Kih)
        ei[i] <- sum(uy*(1*(y[i] <= ys)[cols] -
                  (denom != 0)* cumsum(Kih[yo])[cols] / (denom + (denom == 0))))^2)
      }
      return(sum(ei)/n^2)
    }

  }else
  {

    stop("There's no weighted version of the dnorm kernel.")
  }
}else if (kernel=="K4_Biweight")
{
  if (is.null(wi.boot))
  {
    Eij3 <- function(parameter){
      K <- function(x, h) 105/64*(1-3*((x/h)^2))*(1-(x/h)^2)^2 * (abs(x) <= h) * (x!=0)

      b <- c(1, parameter[1:(number_p-1)])
      h <- bandwidth

      x <- c(X%*%b)
      y <- Y
```

```r
        n <- length(y)
        yo <- order(y)
        ys <- y[yo]
        uy <- rle(ys)[[1]]
        cols <- cumsum(uy)
        ei <- rep(0, n)
        for (i in 1:n){
          Kih <- K(x-x[i],h=h)
          denom <- sum(Kih)
          ei[i] <- sum(uy*(1*(y[i] <= ys)[cols] -
                    (denom != 0)* cumsum(Kih[yo])[cols] / (denom + (denom == 0)))^2)
        }
        return(sum(ei)/n^2)
      }

    }else
    {
      stop("There's no weighted version of the K4_Biweight kernel.")
    }
  }

  if(method == "nlminb")
  {
    esti <- nlminb(start = initial[-1],
                   objective = Eij3,
                   control = list(abs.tol = abs.tol))
  }else if (method == "optim")
  {
    # the new optimize function using optim, you can change the lower and upper
    esti <- optim(par = initial[-1],
                  fn = Eij3,
                  method = optim_method,
                  control = list(abstol = abs.tol))
  }else if (method == "nmk")
  {
    esti <- nmk(par = initial[-1],
                fn = Eij3,
                control = list(tol = abs.tol))
  }
  results <- list(coef = c(1, esti$par[1:(number_p-1)]),
                  bandwidth = bandwidth,
                  details = esti)
  }

  return(results)
}


##### Initial values
SDR1 <- cumuSIR(X = Xi, Y = Yi)

##### Single index model estimation;  returns coefficient vector theta and bandwidth b
SID2 <- SIDR(X = Xi, Y = Yi, initial = SDR1$basis[, 1], kernel = "dnorm", method = "nmk")
```

```
SID2
## $coef
## [1]  1.00000000  4.78544373  5.10021095 -0.06230537  0.12073346
##
## $bandwidth
## [1] 0.2138144
##
## $details
## $details$par
## [1]  4.78544373  5.10021095 -0.06230537  0.12073346 -1.54264709
##
## $details$value
## [1] 0.1461918
##
## $details$feval
## [1] 92
##
## $details$restarts
## [1] 0
##
## $details$convergence
## [1] 0
##
## $details$message
## [1] "Successful convergence"
```

Using these fitted values $\hat{\theta}$ and $\hat{h}$, we will use the following function to estimate the conditional distribution function of $Y(t)$ given predictors:

```
#####   This function is to estimate the distribution function based on the
#####        single index model.
#####   Here y_seq is the support of the outcome variable, or of the empirical
#####        distribution of the outcome variable
NW <- function(Xb, Y, xb, y_seq, h){
  
  K <- function(x, h) dnorm(x/h, 0, 1) # Gaussian
  # K <- function(x, h) 15/16*(1-(x/h)^2)^2 * (abs(x) <= h) # K2_biweight
  # K <- function(x, h) 105/64*(1-3*((x/h)^2))*(1-(x/h)^2)^2 * (abs(x) <= h) # K4_biweight
  
  Kxb <- sapply(xb, function(x, Xb) K(Xb-x, h), Xb=Xb)
  
  Ylty <- sapply(y_seq, function(x, Y) 1*(Y <= x), Y=Y)
  
  denom <- colSums(Kxb)
  
  fyxb <- (denom!=0)*crossprod(Kxb, Ylty)/(denom + (denom==0))
  
  return(fyxb)
  
}
```

Note that the observed data modeling has not involved the sensitivity parameter $\alpha_1$. In the next section, we use the fitted observed data models together with posited values of $\alpha_1$ to estimate $\mu_1(t)$.

## 4. Constructing the Estimators $\hat{\beta}$ and $\hat{\mu}_1(t)$

Here we estimate $\beta$ and $\mu_1(t)$ under sensitivity parameter values $\alpha_1 = -0.4$, 0, and 0.4. The estimator $\hat{\beta}$ has two components: an inverse weighted piece and an augmentation term.

The inverse weighted piece uses a weight $1/\rho(t|Y(t), \bar{O}(t))$, where:

$$\rho(t|Y(t), \overline{O}(t)) = \lambda(t|\overline{O}(t)) \exp\{-\alpha_1 Y(t)\} E[\exp\{\alpha_1 Y(t)\}|A(t) = 1, \overline{O}(t)].$$

To compute the inverse weighted piece, for each participant, for each of their assessment times $T_k \in [t_1, t_2]$, we compute the quantity:

$$V^{-1} B(T_k) \frac{Y_i(T_k) - \hat{E}[Y(T_k)|\overline{O}_i(T_k)]}{\hat{\rho}(T_k|Y_i(T_k), \overline{O}_i(T_k))}.$$

The function below estimates the conditional expected values under a given $\alpha_1$, by first computing the single index model distribution function. The first output in the following function is the estimator for

$$E[Y(t)|\overline{O}(t)] = \frac{E[Y(t) \exp\{\alpha_1 Y(t)\}|A(t) = 1, \overline{O}(t)]}{E[\exp\{\alpha_1 Y(t)\}|A(t) = 1, \overline{O}(t)]},$$

and the second output is the estimator for $E[\exp\{\alpha_1 Y(t)\}|A(t) = 1, \overline{O}(t)]$.

```
####  The input y_seq is the support of the outcome variable
Cond_mean_fn_single <- function(alpha, X, Y, x, y_seq, beta, bandwidth){

   # conditional distribution
   Fhat <- NW(Xb = X %*% beta, Y = Y,
              xb = x %*% beta, y_seq = y_seq,
              h = bandwidth)

   # density function
   Fhat1 <- c(0, Fhat[1:(length(y_seq) - 1)])
   pmf <- Fhat - Fhat1

   E_exp_alphaY <- sum( exp(alpha*y_seq)*pmf )
   E_Yexp_alphaY <- sum( y_seq*exp(alpha*y_seq)*pmf )
   E_Y_past <- E_Yexp_alphaY/E_exp_alphaY

   return(list(E_Y_past,E_exp_alphaY))

}
```

The augmentation term of the estimator involves the integral $\int_{t=t_1}^{t_2} V^{-1} B(t) \hat{E}[Y(t)|\overline{O}(t)] dt$, which we approximate for each participant as:

$$\sum_{t=t_1}^{t_2} V^{-1} B(t) \hat{E}[Y(t)|\overline{O}_i(t)].$$

Thus for each participant $i$, we must obtain the predicted value $g_i(t) = \hat{E}[Y(t)|\overline{O}_i(t)]$ for each $t$ in $[t_1, t_2]$.

In our outcome model, the predictors are based on values at the previous assessment time; therefore $g_i(t)$ is a piecewise function with jumps at each of participant $i$'s assessment times. We illustrate this for an example participant below.

```r
Time_mean <- mean(Trt_dat_outcome$Time)
Time_sd <- sd(Trt_dat_outcome$Time)
Lag_time_mean <- mean(Trt_dat_outcome$Lag_time)
Lag_time_sd <- sd(Trt_dat_outcome$Lag_time)

Trt_dat1 <- cbind(Trt_dat,
                    ns1 = ns(Trt_dat$Prev_outcome, df = 3)[, 1],
                    ns2 = ns(Trt_dat$Prev_outcome, df = 3)[, 2],
                    ns3 = ns(Trt_dat$Prev_outcome, df = 3)[, 3],
                    time_scale = (Trt_dat$Time - Time_mean) / Time_sd,
                    Lag_time_scale = (Trt_dat$Lag_time - Lag_time_mean) / Lag_time_sd)

i=74

df_i1=filter(Trt_dat1, Subject_ID==i)

    # for each time in [t1,t2], generate participant i's covariate values at time t
    # this may require computing the most recent assessment before time t

    length_time <- t2-t1+1

    df_est <- data.frame(Time = seq(t1,t2,by=1),
                         Lag_time = rep(0, length_time),
                         ns1 = rep(0, length_time),
                         ns2 = rep(0, length_time),
                         ns3 = rep(0, length_time),
                         Time_scaled = rep(0, length_time),
                         Lag_time_scaled = rep(0, length_time))

    {
      for(k in 1:length_time){
        t <- t1+k-1

        min_time <- min(df_i1$Time)
        max_time <- max(df_i1$Time)

        if(t <= min_time){

          df_est[k, 2:7] <- c(t,
                              df_i1$ns1[1],
                              df_i1$ns2[1],
                              df_i1$ns3[1],
                              (t - Time_mean) /Time_sd,
                              (t - Lag_time_mean) / Lag_time_sd)

        }else if(t >= max_time){

          temp <- df_i1$Asthma_control[nrow(df_i1)]
```

```r
              df_est[k, 2:7] <- c(t - max_time,
                                  ns(temp + 0.01 , df = 3)[1],
                                  ns(temp + 0.01, df = 3)[2],
                                  ns(temp + 0.01, df = 3)[3],
                                  (t - Time_mean) / Time_sd,
                                  (t - max_time - Lag_time_mean) / Lag_time_sd)

        }else{

          # t within min_time and max_time
          t_index1 <- max(which(df_i1$Time < t))
          t_index2 <- min(which(df_i1$Time > t))

          df_est[k, 2:7] <- c(t - df_i1$Time[t_index1],
                              df_i1$ns1[t_index2],
                              df_i1$ns2[t_index2],
                              df_i1$ns3[t_index2],
                              (t - Time_mean) / Time_sd,
                              (t - df_i1$time[t_index1] - Lag_time_mean) / Lag_time_sd)
        }

      }
    }

#### next, get the predicted mean outcome at each time t for participant i

Alpha_seq <- c(-0.4,0,0.4)

Time_means_single  <- matrix(0, nrow = length_time, ncol = length(Alpha_seq))

for(j in 1:length(Alpha_seq)){
  alpha <- Alpha_seq[j]
  y_seq <- seq(0, 6, by = 1/6)

  for(l in 1:length_time){

    temp <- Cond_mean_fn_single(alpha,
                                X = Xi,
                                Y = Yi,
                                x = c(df_est$ns1[l],
                                      df_est$ns2[l],
                                      df_est$ns3[l],
                                      df_est$Time_scaled[l],
                                      df_est$Lag_time_scale[l]),
                                y_seq=y_seq,
                                beta=SID2$coef,
                                bandwidth=SID2$bandwidth)

    Time_means_single[l, j] <- temp[[1]]
  }
}
```

```r
df <- data.frame(Time = rep(t1:t2, 3),
                 Prediction = c(Time_means_single[, 1],
                                Time_means_single[, 2],
                                Time_means_single[, 3]),
                 Alpha = c(rep("-0.4", length(t1:t2)),
                           rep("0", length(t1:t2)),
                           rep("0.4", length(t1:t2)))
)

ggplot(df, aes(x=Time, y=Prediction, color = Alpha)) +
  geom_line(linewidth=1)+
  theme_bw()+
  labs(y="ACQ Score",
       title=expression(Predicted~Mean~Outcome~Function~Under~alpha[1]),
       subtitle="Given an Example Participant's Observed Past",
       colour=expression(alpha[1]))+
  expand_limits(y=c(0,6),x=c(0,600))+
  theme(title=element_text(size=9))+
  geom_vline(xintercept = c(85, 283, 707), linetype = "dashed") +
  annotate("text", x = 45, y = -0.5, label = "Visit 1", size = 3)+
  annotate("text", x = 240, y = -0.5, label = "Visit 2", size = 3)+
  annotate("text", x = 665, y = -0.5, label = "Visit 3", size = 3)
```

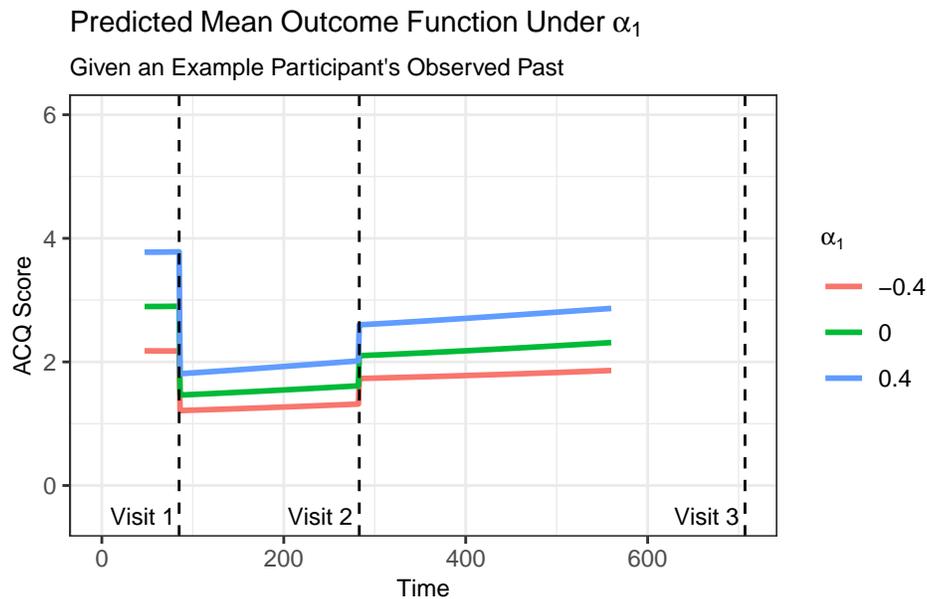

Table 3: Simulated data for the example participant

| Subject_ID | Visit | Prev_time | Time | Event | Asthma_control | Prev_outcome | Lag_time |
|---:|---:|---:|---:|---:|---:|---:|---:|
| 74 | 1 | 0 | 85 | 1 | 0.83 | 4.17 | 85 |
| 74 | 2 | 85 | 283 | 1 | 2.50 | 0.83 | 198 |
| 74 | 3 | 283 | 707 | 1 | 0.67 | 2.50 | 424 |

Below, for each $\alpha_1 = -0.4, 0, 0.4$, we:

1. Compute the inverse weighted term for each participant/assessment
2. Compute the augmentation term for each participant
3. Obtain $\hat{\beta}$ by summing the inverse weighted terms and the augmentation term for each participant, then averaging over participants.

4. Obtain the variance estimator $\widehat{Var}(\hat{\beta})$

```r
##### sensitivity parameter values
Alpha_seq=c(-0.4,0,0.4)


length_time <- t2-t1+1
V_inverse=solve(t(Basis_matrix)%*%Basis_matrix)
Weights_term2=V_inverse%*%t(Basis_matrix)

u=unique(Trt_dat1$Subject_ID)
N=length(u)


####  Support of the ACQ Score outcome
y_seq <- seq(0, 6, by = 1/6)

Visits_df=filter(Trt_dat_outcome, Time >= t1,Time <= t2)
K=dim(Visits_df)[1]


Beta_hat=matrix(ncol=length(Alpha_seq),nrow=p)
Var_betahat=list(length=length(Alpha_seq))

for(a in 1:length(Alpha_seq)){
  
  Visits_df_a=Visits_df
  alpha=Alpha_seq[a]
  
  
  ############   1. Compute the inverse weighted term for each participant/assessment:
  Term1_mat=matrix(nrow=p,ncol=K)
  
  for(k in 1:K){
    
    df_k=Visits_df_a[k,]
    
    #####  Conditional expected value at time t_k
    temp <- Cond_mean_fn_single(alpha,
                                X = Xi,
                                Y = Yi,
                                x = c(df_k$ns1,
                                      df_k$ns2,
                                      df_k$ns3,
                                      df_k$Time_scaled,
                                      df_k$Lag_time_scaled),
```

18

```r
                                        y_seq=y_seq,
                                        beta = SID2$coef,
                                        bandwidth=SID2$bandwidth)

   E_Y_past=temp[[1]]
   E_exp_alphaY=temp[[2]]


   #### B(t) at time t_k
   B_tk=matrix(Basis_matrix[df_k$Time-t1+1,],ncol=1)

   #### baseline intensity at time t_k from the appropriate stratum
   base_lambda=Base_intensities[df_k$Time,df_k$Visit]

   Term1_mat[,k]=(V_inverse%*%B_tk)*(df_k$Asthma_control-E_Y_past)/
     (base_lambda*exp(gamma*df_k$Prev_outcome)*exp(-alpha*df_k$Asthma_control)*E_exp_alphaY)

 }


################   2. Compute the augmentation term for each participant
Time_mean <- mean(Trt_dat_outcome$Time)
Time_sd <- sd(Trt_dat_outcome$Time)
Lag_time_mean <- mean(Trt_dat_outcome$Lag_time)
Lag_time_sd <- sd(Trt_dat_outcome$Lag_time)

Trt_dat1 <- cbind(Trt_dat,
                         ns1 = ns(Trt_dat$Prev_outcome, df = 3)[, 1],
                         ns2 = ns(Trt_dat$Prev_outcome, df = 3)[, 2],
                         ns3 = ns(Trt_dat$Prev_outcome, df = 3)[, 3],
                         time_scale = (Trt_dat$Time - Time_mean) / Time_sd,
                         Lag_time_scale = (Trt_dat$Lag_time - Lag_time_mean) / Lag_time_sd)

Term2_mat=matrix(nrow=p, ncol=N)

for(i in 1:N){

   df_i1=filter(Trt_dat1, Subject_ID==u[i])

   ###  For each time in [t1,t2], generate participant i's covariate values at time t;
   ###    this may require computing the most recent assessment before time t
   df_est <- data.frame(Time = seq(t1,t2,by=1),
                        Lag_time = rep(0, length_time),
                        ns1 = rep(0, length_time),
                        ns2 = rep(0, length_time),
                        ns3 = rep(0, length_time),
                        Time_scaled = rep(0, length_time),
                        Lag_time_scaled = rep(0, length_time))

  {
    for(k in 1:length_time){

        t <- t1+k-1
```

```r
      min_time <- min(df_i1$Time)
      max_time <- max(df_i1$Time)
      
      if(t <= min_time){
        
        df_est[k, 2:7] <- c(t,
                            df_i1$ns1[1],
                            df_i1$ns2[1],
                            df_i1$ns3[1],
                            (t - Time_mean) /Time_sd,
                            (t - Lag_time_mean) / Lag_time_sd)
        
      }else if(t >= max_time){
        
        temp <- df_i1$Asthma_control[nrow(df_i1)]
        
        df_est[k, 2:7] <- c(t - max_time,
                            ns(temp + 0.01 , df = 3)[1],
                            ns(temp + 0.01, df = 3)[2],
                            ns(temp + 0.01, df = 3)[3],
                            (t - Time_mean) / Time_sd,
                            (t - max_time - Lag_time_mean) / Lag_time_sd)
        
      }else{
        
        # t within min_time and max_time
        t_index1 <- max(which(df_i1$Time < t))
        t_index2 <- min(which(df_i1$Time > t))
        
        df_est[k, 2:7] <- c(t - df_i1$Time[t_index1],
                            df_i1$ns1[t_index2],
                            df_i1$ns2[t_index2],
                            df_i1$ns3[t_index2],
                            (t - Time_mean) / Time_sd,
                            (t - df_i1$time[t_index1] - Lag_time_mean) / Lag_time_sd)
      }
      
    }
}

#### Predicted mean outcome at each time t for participant i
Time_means_single  <- matrix(0, nrow = length_time, ncol = 1)

for(l in 1:length_time){
  
  temp <- Cond_mean_fn_single(alpha,
                              X = Xi,
                              Y = Yi,
                              x = c(df_est$ns1[l],
                                    df_est$ns2[l],
                                    df_est$ns3[l],
                                    df_est$Time_scaled[l],
                                    df_est$Lag_time_scale[l]),
```

```r
                                  y_seq=y_seq,
                                  beta=SID2$coef,
                                  bandwidth=SID2$bandwidth)

      Time_means_single[l] <- temp[[1]]

    }

    Term2_mat[,i] <- Weights_term2%*%Time_means_single

  }


  ################   3. Sum the inverse-weighted terms and augmentation terms
  ################         and average over participants
  Psi_mat=matrix(nrow=p,ncol=N)

  for(i in 1:N){

    w=which(Visits_df$Subject_ID == u[i])

    if(length(w) ==0){ temp = 0 }
    if(length(w) ==1){ temp=Term1_mat[,w] }
    if(length(w) > 1){ temp=rowSums(Term1_mat[,w]) }

    Psi_mat[,i]=temp + Term2_mat[,i]

  }


  Beta_hat[,a] <- rowMeans(Psi_mat)


  ###############   4.  Variance estimate of beta-hat
  beta_hat=matrix(Beta_hat[,a],ncol=1)
  Var_array <- array(dim=c(p, p, N))

  for(i in 1:N){

      temp <- Psi_mat[ , i] - beta_hat
      Var_array[ , , i] <-  temp%*%t(temp)

  }

  Var_betahat[[a]] <- (1/N^2)*rowSums(Var_array, dims=2)


}
```

Finally, we compute $\hat{\mu}_1(t) = \hat{\beta}' B(t)$ and $\widehat{Var}(\hat{\mu}_1) = B(t)'\widehat{Var}(\hat{\beta})B(t)$.

```
 mean_fn=function(t,a){

    Bt=matrix(Basis_matrix[t-t1+1,],ncol=1)
    betahat=matrix(Beta_hat[,a],ncol=1)

    t(betahat)%*%Bt

 }

var_est_fn=function(t,a){

    Bt=matrix(Basis_matrix[t-t1+1,],ncol=1)
    var_betahat=Var_betahat[[a]]

    t(Bt)%*%var_betahat%*%Bt

}

##### Variance estimates of mu-hat at the target times under alpha=0.4
sapply(c(90,180,270,360),var_est_fn,a=3)
```

```
## [1] 0.01158709 0.01577595 0.01589921 0.02271194
```

The estimated mean outcome curves $\hat{\mu}_1(t), t \in [47, 560]$ under each $\alpha_1 = -0.4, 0, 0.4$ are shown below, along with a nonparametric smooth fit to the observed outcomes.

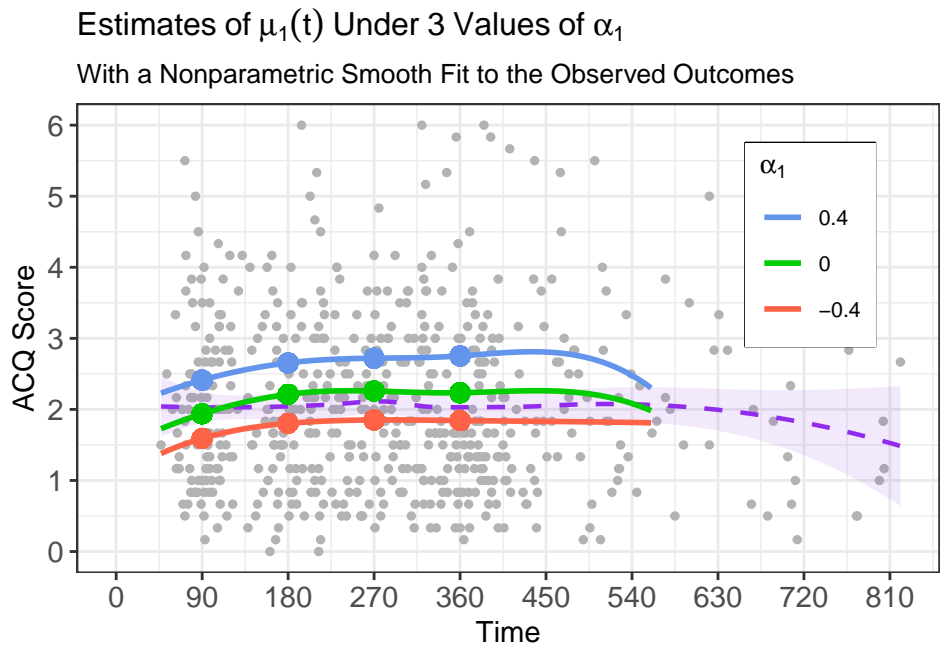



\subsection*{Appendix A.2: Studies with gaps in assessment times}

In the main manuscript we assumed that there was a single time interval $[t_1,t_2]$, containing all target assessment times, with assessments taking place throughout $[t_1,t_2]$.  However, in some studies, assessments may take place over periods of time around each target time, with gaps between these periods.  In this case, our approach should be adapted as described below, so that inference is drawn only for the intervals during which assessments are concentrated. All assumptions and estimation are separate by treatment arm, and we suppress dependence on the treatment arm below.

Suppose that around each target assessment time $t_m^*$, there is an interval $[t_{1m},t_{2m}]$  with $\lambda \left\{ t \mid \overline{\boldsymbol{O}}(t) \right\} > 0$ on  $[t_{1m},t_{2m}]$.  We make a positivity assumption for these intervals, as well as a separate marginal mean assumption for each interval.

\medskip

\begin{assump-2a}[Modified positivity assumption]
There is some $c > 0$ such that, for each $m$, for all $t$ in $[t_{1m},t_{2m}]$, $\rho \left\{ t \mid \overline{\boldsymbol{O}}(t),Y(t) \right\} > c$ for all values of $Y(t)$ and $\overline{\boldsymbol{O}}(t)$.
\end{assump-2a}

\medskip

\begin{assump}[Interval-specific marginal mean assumptions]
For each interval $m$, $\mu(t)=\boldsymbol{B}_m(t)^\prime \boldsymbol{\beta}^{(m)}$ for all $t \in [t_{1m},t_{2m}]$, for some specified interval-specific vector-valued function $\boldsymbol{B}_m(t)=(B_{m,1}(t),\ldots,B_{m,p_m}(t))^\prime$ such that $\boldsymbol{V}_m=\int_{t=t_{1m}}^{t_{2m}} \boldsymbol{B}_m(t)\boldsymbol{B}_m(t)^\prime dt$ is invertible, and some interval-specific parameter vector $\boldsymbol{\beta}^{(m)} \in \mathbb{R}^{p_m}$.
\end{assump}

\medskip

Under Assumptions 1, 2a, and 3a, each $\boldsymbol{\beta}^{(m)}$ is identified from the observed data.  Next, models for $\lambda \left\{ t \mid \overline{\boldsymbol{O}}(t) \right\}$ and $dF \left\{ y(t) \mid \Delta N(t)=1,\overline{\boldsymbol{O}}(t) \right\}$ are fit.  Each of these uses all data for the given treatment arm; that is, we do not fit separate models for each time period. Each parameter vector $\boldsymbol{\beta}^{(m)}$ is then estimated separately.  For each individual $i$, let $S_{im} = \{ k: T_{ik} \in [t_{1m},t_{2m}] \}$.  Depending on the participant, $S_{im}$ could include zero, one, or more than one assessment(s).  Let $\widehat{\boldsymbol{\Psi}}_{im}= $
\[ 
 \sum_{k \in S_{im}}  \Bigg( \boldsymbol{V}_m^{-1} \boldsymbol{B}_m(T_{ik}) \frac{\big[ Y_i(T_{ik})- \widehat{E} \big\{ Y(T_{ik}) \mid \overline{\boldsymbol{O}}_i(T_{ik}) \big\} \big] }{\widehat{\rho} \{ T_{ik} \mid Y_i(T_{ik}), \overline{\boldsymbol{O}}_i(T_{ik}) \} } \Bigg)  + \int_{t=t_{1m}}^{t_{2m}}   \boldsymbol{V}_m^{-1} \boldsymbol{B}_m(t)  \widehat{E} \big\{ Y(t) \mid \overline{\boldsymbol{O}}_i(t) \big\}   dt .\]
Then the augmented inverse intensity-weighted estimators are $\widehat{\boldsymbol{\beta}}^{(m)}= \frac{1}{n} \sum_{i=1}^n  \widehat{\boldsymbol{\Psi}}_{im}$ and $\widehat{\mu}(t)= \left(\widehat{\boldsymbol{\beta}}^{(m)} \right)^\prime\boldsymbol{B}_m(t)$ for $t \in [t_{1m},t_{2m}]$.

\section*{Appendix B:  Theory}

\subsection*{Appendix B.1: Proofs of propositions}

Proof of Proposition 1:

\begin{proof}
Let $t$ be a time with $\lambda \{t \mid \overline{\boldsymbol{O}}(t) \} >0$.  For each $\epsilon > 0$,
\begin{align}
\mu(t)=  & \ \int_{\overline{\boldsymbol{o}}(t)} \int_{y(t)} y(t) dF \big\{ y(t) \mid \overline{\boldsymbol{O}}(t)= \overline{\boldsymbol{o}}(t)  \big\} dF \big\{ \overline{\boldsymbol{o}}(t) \big\} \notag \\
= & \ \int_{\overline{\boldsymbol{o}}(t)} \int_{y(t)} y(t) \Big[ dF \Big\{ y(t) \mid N(t+\epsilon)-N(t-)=0, \overline{\boldsymbol{O}}(t)=\overline{\boldsymbol{o}}(t) \Big\} \times \notag \\
& \hspace{0.7in} P \Big\{  N(t+\epsilon)-N(t-)=0  \mid \overline{\boldsymbol{O}}(t)=\overline{\boldsymbol{o}}(t) \Big\} \  + \notag \\
& \hspace{0.6in} dF \Big\{ y(t) \mid  N(t+\epsilon)-N(t-)=1, \overline{\boldsymbol{O}}(t)=\overline{\boldsymbol{o}}(t) \Big\} \times \notag \\
& \hspace{1in} P \Big\{  N(t+\epsilon)-N(t-)=1  \mid \overline{\boldsymbol{O}}(t)=\overline{\boldsymbol{o}}(t) \Big\} \Big] dF\big\{ \overline{\boldsymbol{o}}(t)  \big\}. \notag 
\end{align}

Taking the limit as $\epsilon \to 0^+$, we have:
\begin{align}
 \mu(t)= & \  \int_{\overline{\boldsymbol{o}}(t)} \int_{y(t)} y(t) \Big[ dF \Big\{ y(t) \mid \Delta N(t)=0, \overline{\boldsymbol{O}}(t)=\overline{\boldsymbol{o}}(t) \Big\} \times 1 \ \ + \notag \\
& \hspace{0.8in} dF \Big\{ y(t)\  \big| \Delta N(t)=1, \overline{\boldsymbol{O}}(t)=\overline{\boldsymbol{o}}(t) \Big\}\times 0 \Big] dF \big\{ \overline{\boldsymbol{o}}(t)  \big\}   \notag  \\
= & \int_{\overline{\boldsymbol{o}}(t)}  \int_{y(t)}  \frac{ y(t) \exp \big\{ \alpha y(t) \big\}  dF \left\{ y(t) \mid \Delta N(t)=1, \overline{\boldsymbol{O}}(t) =\overline{\boldsymbol{o}}(t) \right\} }{ E \left[ \exp \big\{\alpha Y(t) \big\} \mid \Delta N(t) =1, \overline{\boldsymbol{O}}(t)=\overline{\boldsymbol{o}}(t) \right] }  dF\big\{ \overline{\boldsymbol{o}}(t) \big\} \notag \\
= & \  E \left(  \frac{E \big[ Y(t) \exp\{ \alpha Y(t) \} \mid \Delta N(t)=1,\overline{\boldsymbol{O}}(t) \big]}{E \big[  \exp\{ \alpha Y(t) \} \mid \Delta N(t)=1, \overline{\boldsymbol{O}}(t) \big] } \right) . \notag
\end{align}
\end{proof}

\ 

\noindent Proof of Proposition 2:

\begin{proof}   Fix a time $t \in [t_1,t_2]$ and a value $y(t)$.  For each $\epsilon > 0$ let: 

\begin{align*}
A_\epsilon := & \ dF\{ y(t) \mid  N(t+\epsilon)-N(t-)=0, \overline{\boldsymbol{O}}(t) \} \\
 B_\epsilon := & \ dF \{ y(t) \mid N(t+\epsilon)-N(t-) =1, \overline{\boldsymbol{O}}(t) \} \frac{ \exp \{ \alpha Y(t) \} }{ E \big[ \exp\{ \alpha Y(t) \} \mid  N(t+\epsilon)-N(t-)=1, \overline{\boldsymbol{O}}(t) \big] }.
\end{align*}

\noindent For each $\epsilon >0$, by Bayes' rule we have:
\begin{align}
 P\big\{ & N(t+\epsilon)-N(t-) =   1 \mid Y(t)=y(t), \overline{\boldsymbol{O}}(t) \big\} \notag \\
= & \   P \big\{ N(t+\epsilon)-N(t-) =1 \mid\overline{\boldsymbol{O}}(t) \big\} dF \big\{ y(t) \mid N(t+\epsilon)-N(t-)=1, \overline{\boldsymbol{O}}(t) \big\}  \bigg/  \notag \\
& \ \ \ \ \ \Big[ P \big\{ N(t+\epsilon)-N(t-)=1 \mid \overline{\boldsymbol{O}}(t) \big\} dF \big\{ y(t) \mid  N(t+\epsilon)-N(t-)=1, \overline{\boldsymbol{O}}(t) \big\} +  \notag \\
& \hspace{1in} \ P \big\{  N(t+\epsilon)-N(t-)=0 \mid \overline{\boldsymbol{O}}(t) \big\} A_\epsilon \Big] \notag \\
 = & \ \frac{   P \big\{ N(t+\epsilon)-N(t-)=1 \mid \overline{\boldsymbol{O}}(t) \big\}   }{  P \big\{ N(t+\epsilon)-N(t-)=1 \mid \overline{\boldsymbol{O}}(t) \big\} +  P \big\{ N(t+\epsilon)-N(t-)=0 \mid \bar{\boldsymbol{O}}(t) \big\} D_\epsilon  \ } , \label{bayes}
 \end{align}

\noindent for $D_\epsilon :=  A_\epsilon / dF \{ y(t) \mid  N(t+\epsilon)-N(t-)=1, \overline{\boldsymbol{O}}(t) \big\}$.  

We divide each side of \eqref{bayes} by $\epsilon$ and take the limit as $\epsilon \to 0^+$:  On the left hand side, $\displaystyle{\lim_{\epsilon \to 0^+}  \Big[ P \big\{ N(t+\epsilon)-N(t-) =1 \mid Y(t)=y(t), \overline{\boldsymbol{O}}(t) \big\} / \epsilon \Big] =\rho \left\{ t \mid \overline{\boldsymbol{O}}(t),Y(t) \right\} }$, while in the numerator of the right hand side, $\lim_{\epsilon \to 0^+} \Big[   P \big\{ N(t+\epsilon)-N(t-)=1 \mid \overline{\boldsymbol{O}}(t) \big\} /\epsilon \Big] = \lambda \left\{ t \mid \overline{\boldsymbol{O}}(t) \right\}$.  In the denominator, write:
\begin{align*}
D_\epsilon =  & \  \frac{ B_\epsilon}{ dF \{ y(t) \mid N(t+\epsilon)-N(t-)=1 , \overline{\boldsymbol{O}}(t) \big\} } + \frac{ A_\epsilon- B_\epsilon }{ dF \{ y(t) \mid  N(t+\epsilon)-N(t-)=1, \overline{\boldsymbol{O}}(t) \big\} }\\
= & \ \frac{ \exp\{ \alpha Y(t) \} }{ E \big[ \exp \{ \alpha Y(t) \} \mid N(t+\epsilon)-N(t-) =1, \overline{\boldsymbol{O}}(t) \big]  }  + \frac{ A_\epsilon- B_\epsilon}{ dF \{ y(t) \mid N(t+\epsilon)-N(t-)=1, \overline{\boldsymbol{O}}(t) \big\} } 
\end{align*}

\noindent and note that:
\begin{align*}
\lim_{\epsilon \to 0^+} (A_\epsilon  - B_\epsilon) = & \  dF \left\{ y(t) \mid \Delta N(t)=0,\overline{\boldsymbol{O}}(t) \right\} - \\
& \ \ \ dF \left\{ y(t) \mid \Delta N(t)=1,\overline{\boldsymbol{O}}(t) \right\} \frac{ \exp \{ \alpha Y(t) \} }{ E \big[ \exp \{ \alpha Y(t) \} \mid  \Delta N(t)=1, \overline{\boldsymbol{O}}(t) \big] } \\
= & \ 0
\end{align*}
\noindent
by Assumption 1.  Therefore, $\lim_{\epsilon \to 0^+} D_{\epsilon} = \exp\{ \alpha Y(t) \} / E \big[ \exp \{ \alpha Y(t) \} \mid \Delta N(t) =1, \overline{\boldsymbol{O}}(t) \big]$, and the result follows.

\end{proof}

\subsection*{Appendix B.2: Intensity functions}

Let $\{ \boldsymbol{\mathcal{F}}_t \}_{t=0}^\tau$ be the filtration defined by $\boldsymbol{\mathcal{F}}_t= \big( \{ \boldsymbol{X}(u),Y(u) : \Delta N(u)=1, 0 \leq u \leq t \}, \{ N(u): 0 \leq u \leq t \} \big)$.  
Here $\boldsymbol{\mathcal{F}}_{t}$ is the observed past up through time $t$, and $\boldsymbol{\mathcal{F}}_{t-}= \overline{\boldsymbol{O}}(t)$ is the observed past prior to, but not including, time $t$.  The intensity function $\lambda \left\{ t \mid \overline{\boldsymbol{O}}(t) \right\}$ (equation (1) in the main manuscript) is the intensity for the counting process $\{ N(t): t \in [0,\tau] \}$ with respect to the filtration $\{ \boldsymbol{\mathcal{F}}_t \}_{t=0}^\tau$.  Additionally, let $\boldsymbol{L} = \{ \boldsymbol{X}(t),Y(t): 0 \leq t \leq \tau \}$ and let $\{ \boldsymbol{\mathcal{G}}_t \}_{t=0}^\tau$ be the filtration defined by $\boldsymbol{\mathcal{G}}_t= \big(\boldsymbol{L}, \{ N(u) : 0 \leq u \leq t \} \big) = ( \boldsymbol{L}, \boldsymbol{\mathcal{F}}_t )$.  Under Assumption 4, $\rho \{ t \mid Y(t), \overline{\boldsymbol{O}}(t) \}$ (equation (2) in the main manuscript) is the intensity function for $\{ N(t) \}_{t=0}^\tau$ with respect to the filtration $\{ \boldsymbol{\mathcal{G}}_t \}_{t=0}^\tau$.  We use this to show the following result, which we use in the proof of Theorem 1.

\ 

\begin{lem} \label{martingale}
Define $M(t) = N(t) - \int_{u=0}^t \rho \{u \mid Y(u), \overline{\boldsymbol{O}}(u) \}du$, and let $H(t)$ be a predictable process.  Under Assumption 4, $E \left[ \left\{ \int_{u=t_1}^{t_2} H(u) dM(u) \right\}  \mid \boldsymbol{L} \right] = 0$.
\end{lem}

\begin{proof}
Under Assumption 4, $\{ M(t) \}_{t=0}^\tau$ is a martingale adapted to the filtration $\{ \boldsymbol{\mathcal{G}}_t \}_{t=0}^\tau$.  Then $U(t) = \int_{u=0}^t H(u) dM(u)$ is also a martingale adapted to $\{ \boldsymbol{\mathcal{G}}_t \}_{t=0}^\tau$, so $E \big\{ U(t) \mid \boldsymbol{\mathcal{G}}_s \big\}= U(s)$ for all $s < t$.  In particular, $U(t)$ has mean zero.  Also,  $E \left[ \left\{ \int_{u=0}^t H(u) dM(u) \right\} \mid \boldsymbol{L} \right] = E \big\{ U(t) \mid \boldsymbol{\mathcal{G}}_{0} \big\} = U(0)= 0$ for all $t$ since $\boldsymbol{\mathcal{G}}_{0}=\boldsymbol{L}$.  Therefore:
\begin{align*}
E \left[ \left\{ \int_{u=t_1}^{t_2} H(u) dM(u) \right\} \mid \boldsymbol{L} \right]= & \ E \left[ \ \left\{ \int_{u=0}^{t_2} H(u) dM(u) \right\} \mid \boldsymbol{L} \right] -E \left[ \left\{ \int_{u=0}^{t_1} H(u) dM(u) \right\} \mid \boldsymbol{L} \right] \notag \\
 = & \ 0 - 0 =0.
\end{align*} 
\end{proof}

A sufficient condition under which $H(t)$ is a predictable process is that $H(t)$ is left continuous in $t$ and that, for each $t$, $H(t)$ is $\boldsymbol{\mathcal{G}}_t$-measurable.  In particular, with an assumption that $Y(t)$ is left continuous, $H(t)$ can be any function of $\overline{\boldsymbol{O}}(t)$ and $Y(t)$.

\subsection*{Appendix B.3: Influence function derivation} \label{append:IF}

In this appendix we prove Theorem 1 using the semi-parametric theory presented in \citet{scharfstein1999nonignorable} and \citet{tsiatis2006book}.  We first briefly review elements of this theory applied to our context.

We will refer here to $\boldsymbol{L} = \{ \boldsymbol{X}(t),Y(t): 0 \leq t \leq \tau \}$ as the full data.  We write $\boldsymbol{R}= \{ N(t) : t_1 \leq t \leq t_2 \}$, and we refer to $\boldsymbol{Z}=(\boldsymbol{L},\boldsymbol{R})$ as the total data.  Let $\mathcal{H}^L$, $\mathcal{H}^O$, and $\mathcal{H}^Z$ be the Hilbert spaces of mean-zero, finite-variance, $p$-dimensional (where $p$ is the length of $\boldsymbol{\beta}$) functions of the full, observed, and total data respectively, with the covariance inner product.  Full-data influence functions are normalized elements of the orthogonal complement of the full-data nuisance tangent space $\Lambda(F_L)$.  Observed data influence functions are normalized elements of $\Lambda^{O,\perp}= \Lambda_1^{O,\perp} \cap \Lambda_2^{O,\perp}$, where $\Lambda_1^O$ is the image of $\Lambda(F_L)$, and $\Lambda_2^O$ is the image of the coarsening nuisance tangent space $\Lambda(F_{R|L})$, under the map $A:  \mathcal{H}^Z \to \mathcal{H}^O$ given by $A( \cdot ) = E \big\{ \cdot \mid \boldsymbol{O} \big\}$.   Therefore, the orthogonal complement $\Lambda_1^{O,\perp}$ is equal to the null space of the adjoint of the map $A_1: \Lambda(F_L) \to \mathcal{H}^O$ given by $A_1( \cdot ) = E \{ \cdot \mid \boldsymbol{O} \}$.  Similarly, $\Lambda_2^{O,\perp}$ is the null space of the adjoint of the map $A_2: \Lambda(F_{R|L}) \to \mathcal{H}^O$ given by $A_2( \cdot) = E\{ \cdot \mid \boldsymbol{O} \}$.  The adjoint of $A_1$ is the map $A_1^*: \mathcal{H}^O \to \Lambda(F_L)$ given by $A_1^*(\cdot)=\Pi \big[  E \{ \cdot \mid \boldsymbol{L} \} \mid \Lambda(F_L) \big]$.  Therefore, $\Lambda_1^{O,\perp}= \big\{ h(\boldsymbol{O}) \in \mathcal{H}^O : \mathcal{K}(g(O)) \in \Lambda(F_L)^{\perp} \big\}$, where $\mathcal{K}: \mathcal{H}^O \to \mathcal{H}^L$ is the map defined by $\mathcal{K}(g(\boldsymbol{O}))= E \big\{ g(\boldsymbol{O}) \mid\boldsymbol{L} \big\}$.   Again using adjoints, $\Lambda_2^{O,\perp} = \big\{ g(\boldsymbol{O}) \in \mathcal{H}^O :  g(\boldsymbol{O}) \in \Lambda(F_{R|L})^\perp \big\} = \mathcal{H}^O \cap \Lambda(F_{R|L})^\perp$, and therefore, $\Lambda^{ O,\perp} = \Lambda_1^{O,\perp} \cap \Lambda(F_{R|L})^\perp$.

Based on this theory, we use the following procedure to derive an observed-data influence function:  We first derive a full-data influence function $\boldsymbol{\varphi}(\boldsymbol{L})$.   We obtain elements of $\Lambda_1^{O,\perp}$ by inverse-weighting $\boldsymbol{\varphi}(\boldsymbol{L})$ and adding elements of the augmentation space (defined below).  We then identify one of these elements that is also orthogonal to $\Lambda(F_{R|L})$.  By the above theory, this element is in the space $\Lambda^{O,\perp}$.  We further show that this element inherits the property of being correctly normalized from $\boldsymbol{\varphi}(\boldsymbol{L})$, and is therefore an observed data influence function.

\noindent Proof of Theorem 1:

\begin{proof} \

\underline{Part 1: A full-data influence function.}

We obtain a full-data influence function for $\boldsymbol{\beta}$ following the approach of \cite{diaz2011superlearner}.  
We specify a function $\boldsymbol{B}(t)$, with $\boldsymbol{V}=\int_{t=t_1}^{t_2} \boldsymbol{B}(t)\boldsymbol{B}(t)^{\prime}dt$ invertible, that we will use when we make Assumption 3.  However, we initially do not assume that the equality in Assumption 3 holds, and instead work in the nonparametric model on $\boldsymbol{L}$.  We define a parameter $\boldsymbol{\gamma}_0$ in this model as:
\begin{equation} \label{cp}
\boldsymbol{\gamma}_0 = \boldsymbol{V}^{-1} \int_{t=t_1}^{t_2} \boldsymbol{B}(t) \mu(t) dt. 
\end{equation}
Here $\boldsymbol{\gamma}_0$ is the unique minimizer of the squared-error loss $ \mathcal{L}(\boldsymbol{\gamma})=  \int_{t=t_1}^{t_2} \Big\{ \mu(t) - \boldsymbol{B}(t)^\prime \boldsymbol{\gamma} \Big\}^2 dt$ from approximating $\mu(t)$ by $\boldsymbol{B}(t)^{\prime} \boldsymbol{\gamma}$ for some $\boldsymbol{\gamma}$, and is equal to $\boldsymbol{\beta}$ when Assumption 3 does hold.  Below we compute the influence function for $\boldsymbol{\gamma}_0$ in the nonparametric model, which we denote by $\boldsymbol{\varphi}^{NP}(\boldsymbol{L} ; \boldsymbol{\gamma}_0)$.   Then $\boldsymbol{\varphi}^{NP}(\boldsymbol{L}; \boldsymbol{\gamma}_0)$ will also be a valid full-data influence function for $\boldsymbol{\beta}$ in our model.

Let $\mathcal{P}$ be any parametric submodel (parametrized by $\epsilon$ such that $\epsilon=0$ corresponds to the true distribution of $\boldsymbol{L}$) and let $a(\boldsymbol{L})$ denote the score vector for $\mathcal{P}$. The influence function for $\boldsymbol{\gamma}_0$ is the mean-zero, $p$-dimensional function of $\boldsymbol{L}$ satisfying:  $\frac{ \partial \boldsymbol{\gamma}_0}{ \partial \boldsymbol{\epsilon} } \Big|_{\epsilon=0} = E \big\{ \boldsymbol{\varphi}^{NP}(\boldsymbol{L};\boldsymbol{\gamma}_0) \ a(\boldsymbol{L}) \big\}$.  For a fixed $a(\boldsymbol{L})$, we consider the parametric submodel $\mathcal{P}_a = \big\{ P_a(\epsilon) : \epsilon \in D \subset \mathbb{R} \big\}$, where $P_a(\epsilon)= dF_0 (\boldsymbol{L}) \big\{ 1 + \epsilon \ a(\boldsymbol{L}) \big\}$, and where the score of $P_a(\epsilon)$ evaluated at $\epsilon=0$ is $a(\boldsymbol{L})$.  In $\mathcal{P}_a$, equation \eqref{cp} becomes:
\[ \boldsymbol{\gamma}_0(\epsilon) = \boldsymbol{V}^{-1} \int_{t=t_1}^{t_2} \boldsymbol{B}(t) \left[ \int_\ell y(t) dF_0 \{1+\epsilon a(\boldsymbol{\ell}) \} \right] dt . \]
Taking the derivative with respect to $\epsilon$ and evaluating at $\epsilon=0$ gives:
\begin{align*} 
\frac{d \boldsymbol{\gamma}_0}{d \epsilon} \Big|_{\epsilon=0} = & \ \boldsymbol{V}^{-1} \int_{t=t_1}^{t_2} \boldsymbol{B}(t) \left\{ \int_\ell y(t) dF_0(\boldsymbol{\ell}) a(\boldsymbol{\ell}) \right\} dt \\
= & \ \boldsymbol{V}^{-1} \int_{t=t_1}^{t_2} \boldsymbol{B}(t)E \big\{ Y(t) a(\boldsymbol{L}) \big\} dt \\
= & \ E \left[ \left\{ \boldsymbol{V}^{-1} \int_{t=t_1}^{t_2} \boldsymbol{B}(t)Y(t)dt \right\} a(\boldsymbol{L})  \right] \\
= & \ E \left\{ \left( \boldsymbol{V}^{-1} \int_{t=t_1}^{t_2} \boldsymbol{B}(t) \Big[ Y(t)- E \{Y(t) \} \Big] dt \right) a(\boldsymbol{L})  \right\}.
\end{align*}

\noindent Therefore $\boldsymbol{\varphi}^{NP}(\boldsymbol{L} ; \boldsymbol{\gamma}_0)=\boldsymbol{V}^{-1} \int_{t=t_1}^{t_2} \boldsymbol{B}(t) \Big[Y(t)- E \{Y(t) \} \Big] dt$.  If we now make Assumption 3, then this simplifies as:
\begin{equation} \label{eqn:full_if}
\boldsymbol{\varphi}(\boldsymbol{L} ) =  \boldsymbol{V}^{-1} \int_{t=t_1}^{t_2} \boldsymbol{B}(t) \Big\{ Y(t)- \boldsymbol{B}(t)^\prime \boldsymbol{\beta} \Big\} dt.
\end{equation}

\medskip


\underline{Part 2: an inverse-weighted element and the augmentation space.}  

Next we find an element $\boldsymbol{g}^*(\boldsymbol{O}) \in \mathcal{H}^O$ which maps to $\boldsymbol{\varphi}(\boldsymbol{L})$ in equation \eqref{eqn:full_if} under the map $\mathcal{K}( \cdot )$.  One such element is the following inverse-weighted element:
\begin{equation}
\boldsymbol{g}^*(\boldsymbol{O}) = \boldsymbol{V}^{-1}\int_{t=t_1}^{t_2} \frac{ \boldsymbol{B}(t) \big\{ Y(t) -   \boldsymbol{B}(t)^\prime \boldsymbol{\beta}  \big\} }{ \rho \left\{ t \mid \overline{\boldsymbol{O}}(t),Y(t) \right\} }   dN(t)  , \label{inv-wtd-g}
\end{equation}

\noindent as we verify: By Lemma \ref{martingale},

\begin{align*}
E & \big\{ \boldsymbol{g}^*(\boldsymbol{O}) \mid \boldsymbol{L} \big\}  \\
= & \   E \left\{  \left(  \boldsymbol{V}^{-1} \int_{t=t_1}^{t_2}   \frac{ \boldsymbol{B}(t)\{ Y(t) - \boldsymbol{B}(t)^\prime \boldsymbol{\beta} \}  }{ \rho \left\{ t \mid \overline{\boldsymbol{O}}(t),Y(t) \right\} } \Big[ dN(t) - \rho \left\{ t \mid \overline{\boldsymbol{O}}(t),Y(t) \right\} dt \Big] \right) \mid \boldsymbol{L}  \right\} \\
& \hspace{1in} + E \left\{ \ \left( \boldsymbol{V}^{-1} \int_{t=t_1}^{t_2}   \boldsymbol{B}(t) \big\{ Y(t) - \boldsymbol{B}(t)^\prime \boldsymbol{\beta} \big\} dt \right) \mid \boldsymbol{L} \right\} \\
= &  \ 0+  \boldsymbol{\varphi}(\boldsymbol{L})
\end{align*}

\noindent as claimed.

Next consider the \emph{augmentation space}, $Aug = \big\{ \boldsymbol{g}(\boldsymbol{O}) \in \mathcal{H}^O : E [ \boldsymbol{g}(\boldsymbol{O}) \mid \boldsymbol{L} ] =\boldsymbol{0} \big\}$.  By Lemma \ref{martingale}, we have the following containment: $Aug \supseteq$
\begin{equation} \label{eqn:aug}   \left\{ \ \left( \int_{t=t_1}^{t_2}  \frac{ \boldsymbol{h} \{t, \overline{\boldsymbol{O}}(t) \} }{ \rho \left\{ t \mid \overline{\boldsymbol{O}}(t),Y(t) \right\}} \Big[ dN(t) - \rho \left\{ t \mid \overline{\boldsymbol{O}}(t),Y(t) \right\} dt \Big] \right)  \ : \ \mbox{ any } \underbrace{ \boldsymbol{h} \{ t,  \overline{\boldsymbol{O}}(t) \}}_{ p \times 1} \ \right\}. 
\end{equation}
Thus, for the inverse-weighted element $\boldsymbol{g}^*(\boldsymbol{O})$ in equation \eqref{inv-wtd-g}, $\Lambda_1^{O,\perp} \supseteq$
\[ \left\{ \boldsymbol{g}^*(\boldsymbol{O}) + \left( \int_{t=t_1}^{t_2}  \frac{ \boldsymbol{h} \{t, \bar{\boldsymbol{O}}(t) \} }{\rho \left\{ t \mid \overline{\boldsymbol{O}}(t),Y(t) \right\}} \Big[ dN(t) - \rho \left\{ t \mid \overline{\boldsymbol{O}}(t),Y(t) \right\} dt \Big] \right)  \ : \ \mbox{ any } \underbrace{ \boldsymbol{h} \{ t,  \overline{\boldsymbol{O}}(t) \} }_{ p \times 1} \ \right\}. \]

\medskip


\underline{Part 3:  the coarsening tangent space.}

The coarsening tangent space $\Lambda(F_{R|L})$ is the set of all elements of $\mathcal{H}^Z$ that are linear combinations of scores of $\boldsymbol{R} | \boldsymbol{L}$ in a parametric submodel of our model.  Following \citet{cooklawless2007book}, we can compute the likelihood for $\boldsymbol{R}|\boldsymbol{L}$ by first partitioning the interval $[t_1,t_2]$ into a finite number of subintervals $[a_k, a_k+\epsilon)$,  $k=0,\ldots, K$, each of length $\epsilon$, and then taking the limit as $K \to \infty$.  For fixed $\epsilon >0$, let $\Delta(a_k)$ denote the number of visits during the interval $[a_k,a_k+\epsilon)$;  for small $\epsilon$, this number of visits is assumed to be 0 or 1.  For fixed $K$, the conditional (on $\boldsymbol{L}$) likelihood of $\boldsymbol{R}=(N(a_0),\ldots,N(a_K))$ is the same as the conditional (on $\boldsymbol{L}$) likelihood of $(\Delta(a_0),\ldots,\Delta(a_K))$,
which can be written as:
\[ P \{ \Delta(a_0) \mid\boldsymbol{L} \} \prod_{k=1}^K P \big\{ \Delta(a_k) \mid \Delta(a_0), \ldots, \Delta(a_{k-1}), \boldsymbol{L} \big\}.  \]
For each $k=1,\ldots, K$:
\begin{align*}
P \big\{ \Delta(a_k) \mid \Delta(a_0), \ldots, \Delta(a_{k-1}), \boldsymbol{L} \big\} = & \  \rho \big\{ a_k \mid \boldsymbol{L}, \overline{\boldsymbol{O}}(a_k) \big\} \epsilon + o(\epsilon) \\
= & \ \rho \big\{ a_k \mid Y(a_k), \overline{\boldsymbol{O}}(a_k) \big\} \epsilon + o(\epsilon)
\end{align*}
where the last equality holds by Assumption 4.  Similarly, 
\[ P \{ \Delta(a_0) \mid \boldsymbol{L} \} =  \rho \{ a_0 \mid \boldsymbol{L}, \overline{\boldsymbol{O}}(a_0)\} \epsilon + o(\epsilon) = \rho \{a_0 \mid Y(a_0), \overline{\boldsymbol{O}}(a_0) \} \epsilon + o(\epsilon). \]  
Therefore, for fixed $K$, the conditional (on $\boldsymbol{L}$) likelihood of $(\Delta(a_0),\ldots,\Delta(a_K))$ is:
\[ \prod_{k=0}^K   \Big[  \rho \{a_k \mid Y(a_k), \overline{\boldsymbol{O}}(a_k) \} \epsilon + o(\epsilon) \Big]^{\Delta(a_k) } \Big[ 1- \rho \{a_k \mid Y(a_k), \overline{\boldsymbol{O}}(a_k) \}\epsilon + o(\epsilon)\Big]^{1-\Delta(a_k) } . \]

\noindent After dividing by the constant $\prod_{k=0}^K \epsilon^{\Delta(a_k)}$, the conditional likelihood is proportional to:
\begin{equation}  \prod_{k=0}^K   \Big[  \rho \{a_k \mid Y(a_k), \overline{\boldsymbol{O}}(a_k) \} + o(\epsilon)/\epsilon \Big]^{\Delta(a_k) } \Big[] 1- \rho \{ a_k \mid Y(a_k), \overline{\boldsymbol{O}}(a_k) \} \epsilon + o(\epsilon)\Big]^{1-\Delta(a_k) } . \label{like-h} 
\end{equation}

\medskip

The continuous-time conditional (on $\boldsymbol{L}$) likelihood is the limit of \eqref{like-h} as $K \to \infty$.  We compute this by viewing \eqref{like-h} as a product of three factors and noting that:
\[ \lim_{K \to \infty} \prod_{k=0}^K   \Big[  \rho \{ a_k \mid Y(a_k), \overline{\boldsymbol{O}}(a_k) \}  + o(\epsilon)/\epsilon \Big]^{\Delta(a_k) } = \prod_{t=t_1}^{t_2}  \left[ \rho \{ t \mid Y(t), \overline{\boldsymbol{O}}(t) \}  \right]^{\Delta N(t) } . \] 
By a result on product integrals \citep{cooklawless2007book},
\[ \lim_{K \to \infty} \prod_{k=0}^K  \Big[ 1- \rho \{a_k \mid Y(a_k), \overline{\boldsymbol{O}}(a_k) \} \epsilon + o(\epsilon)\Big] =  \exp \left[ - \int_{t=0}^{\tau} \rho \{ t \mid Y(t), \overline{\boldsymbol{O}}(t) \} dt \right] . \]
Finally,  since $\Delta(a_k)=1$ at only a fixed number of values $a_k$ as $K$ varies, and since \\ $\lim_{\epsilon \to 0^+} [ 1- \rho \{ a_k \mid Y(a_k), \overline{\boldsymbol{O}}(a_k) \} \epsilon + o(\epsilon) ] =1$ for each $a_k$, we have:
\[\lim_{K \to \infty}\prod_{k=0}^K  \Big[ 1- \rho \{ t_k \mid Y(t_k), \overline{\boldsymbol{O}}(a_k) \} \epsilon + o(\epsilon) \Big]^{\Delta(a_k)}  =1 .\]
Therefore, the continuous-time conditional (on $\boldsymbol{L}$) likelihood $\mathcal{L}$ is:
\[ \mathcal{L} = \left[ \prod_{t=t_1}^{t_2} \rho \{t \mid  Y(t), \bar{\boldsymbol{O}}(t) \}^{\Delta N(t)} \right] \exp \left[  - \int_{t=t_1}^{t_2} \rho \{ t \mid Y(t), \bar{\boldsymbol{O}}(t) \} dt \right] .\]

To compute $\Lambda(F_{R|L})$, we now consider the conditional likelihood in a parametric submodel of our model for $dF(\boldsymbol{R} \mid \boldsymbol{L})$.  Any such parametric submodel is induced by a model for the intensity function $\rho \{ t \mid Y(t), \overline{\boldsymbol{O}}(t) \}$, say $\Big\{ \rho \{ t \mid Y(t), \overline{\boldsymbol{O}}(t) ; \boldsymbol{\gamma} \} : \boldsymbol{\gamma} \in \Gamma \Big\}$ with $\boldsymbol{\gamma}$ a finite-dimensional parameter and $\boldsymbol{\gamma}=\boldsymbol{0}$ corresponding to the truth.   The conditional likelihood in this parametric submodel is:
\[ \mathcal{L}(\boldsymbol{\gamma}) =  \left[ \prod_{t=t_1}^{t_2} \rho \{ t \mid Y(t), \overline{\boldsymbol{O}}(t) ; \boldsymbol{\gamma} \}^{\Delta N(t)} \right] \exp \left[  - \int_{t=t_1}^{t_2} \rho \{ t \mid Y(t), \overline{\boldsymbol{O}}(t) ; \boldsymbol{\gamma} \} dt \right] . \]

\noindent The log-conditional likelihood is:
\[ \ell( \boldsymbol{\gamma}) = \int_{t=t_1}^{t_2}   \log \Big[ \rho \{ t \mid Y(t), \overline{\boldsymbol{O}}(t) ; \boldsymbol{\gamma} \} \Big]dN(t)  \ - \ \int_{t=t_1}^{t_2} \rho \{ t \mid Y(t), \overline{\boldsymbol{O}}(t) ; \boldsymbol{\gamma} \} dt , \]

\noindent and the score is:
\begin{align*}
 \boldsymbol{S}_{\boldsymbol{\gamma}}(\boldsymbol{O}) =  & \ \int_{t=t_1}^{t_2}     \left[ \frac{   \frac{ \partial  }{ \partial \gamma} \rho \{ t \mid Y(t), \overline{\boldsymbol{O}}(t) ; \boldsymbol{\gamma} \} }{ \rho \{ t \mid Y(t), \overline{\boldsymbol{O}}(t) ; \boldsymbol{\gamma} \} } \right]dN(t)  \ - \ \int_{t=t_1}^{t_2}   \frac{ \partial }{ \partial \boldsymbol{\gamma}} \rho \{ t \mid Y(t), \overline{\boldsymbol{O}}(t) ; \boldsymbol{\gamma} \} dt  \\
= & \  \int_{t=t_1}^{t_2} \left[  \frac{   \frac{ \partial }{ \partial \gamma}  \rho \{ t \mid Y(t), \overline{\boldsymbol{O}}(t) ; \boldsymbol{\gamma} \}  }{ \rho \{ t \mid Y(t), \overline{\boldsymbol{O}}(t) ; \boldsymbol{\gamma} \} } \right] \Big[ dN(t) -\rho \{ t \mid Y(t), \overline{\boldsymbol{O}}(t) ; \boldsymbol{\gamma} \}dt \Big] .
\end{align*}

In order to satisfy the relation in Proposition 2, we must have $\rho \{ t \mid Y(t), \overline{\boldsymbol{O}}(t) ; \boldsymbol{\gamma} \} = r \{ t, \overline{\boldsymbol{O}}(t) ; \boldsymbol{\gamma} \} \exp\{-\alpha Y(t) \}$ for some function  $r \{ t, \overline{\boldsymbol{O}}(t) ; \boldsymbol{\gamma} \}$.  Since
\[ \frac{   \frac{ \partial }{ \partial \gamma}  \rho \{ t \mid Y(t), \overline{\boldsymbol{O}}(t) ; \boldsymbol{\gamma} \}  }{ \rho \{ t \mid Y(t), \overline{\boldsymbol{O}}(t) ; \boldsymbol{\gamma} \} } =  \frac{   \frac{ \partial }{ \partial \gamma}  r \{ t, \overline{\boldsymbol{O}}(t) ; \boldsymbol{\gamma} \}  }{  r \{ t, \overline{\boldsymbol{O}}(t) ; \boldsymbol{\gamma} \}  }  \] 
is a function of $t$ and $\overline{\boldsymbol{O}}(t)$ only, evaluating the score at the truth $\boldsymbol{\gamma}=\boldsymbol{0}$ shows any score of a parametric submodel of $\Lambda(F_{R|L})$ must be of the form $\int_{t=t_1}^{t_2} \boldsymbol{h} \{ t, \bar{\boldsymbol{O}}(t) \} \Big[dN(t) - \rho \left\{ t \mid \overline{\boldsymbol{O}}(t),Y(t) \right\} dt \Big]$ for some function $\boldsymbol{h} \{ t,\overline{\boldsymbol{O}}(t) \}$.  That is, we have shown the following containment: $\Lambda(F_{R|L}) \subseteq$
\begin{equation} \label{eqn:coarsening_space}
 \bigg\{  \bigg( \int_{t=t_1}^{t_2} \boldsymbol{h} \{ t, \bar{\boldsymbol{O}}(t) \}  \Big[dN(t) - \rho \left\{ t \mid \overline{\boldsymbol{O}}(t),Y(t) \right\}dt \Big]  \bigg)  \in \mathcal{H}^Z \ : \mbox{ any }  \underbrace{ \boldsymbol{h} \{ t, \bar{\boldsymbol{O}}(t)\} }_{ p \times 1} \bigg \}  .
\end{equation}

\medskip


\underline{Part 4: an observed-data influence function.} 

Consider the inverse-weighted element in equation \eqref{inv-wtd-g},
\[ \boldsymbol{g}^*(\boldsymbol{O}) = \boldsymbol{V}^{-1}\int_{t=t_1}^{t_2} \frac{ \boldsymbol{B}(t) \big\{ Y(t) -   \boldsymbol{B}(t)^\prime \boldsymbol{\beta}  \big\} }{ \rho \left\{ t \mid \overline{\boldsymbol{O}}(t),Y(t) \right\} }   dN(t) \]
and consider the function:
\[ \boldsymbol{h}^*(\boldsymbol{O}) =  \boldsymbol{V}^{-1} \int_{t=t_1}^{t_2} \frac{ \boldsymbol{B}(t) \big[ E \big\{ Y(t) \mid \overline{\boldsymbol{O}}(t) \big\} - \boldsymbol{B}(t)^\prime \boldsymbol{\beta}  \big] }{\rho \left\{ t \mid \overline{\boldsymbol{O}}(t),Y(t) \right\}} \Big[ dN(t) - \rho \left\{ t \mid \overline{\boldsymbol{O}}(t),Y(t) \right\} dt \Big]. \]

\noindent The function $\boldsymbol{\varphi}(\boldsymbol{O})$ of Theorem 1 is equal to $\boldsymbol{g}^*(\boldsymbol{O})-\boldsymbol{h}^*(\boldsymbol{O})$.  To prove Theorem 1, we must show that $\boldsymbol{g}^*(\boldsymbol{O})-\boldsymbol{h}^*(\boldsymbol{O})$ is an element of $\Lambda^{O,\perp} = \Lambda_1^{O,\perp} \cap \Lambda(F_{R|L})^\perp$ and that it is properly normalized.  Here $\boldsymbol{h}^*(\boldsymbol{O}) \in Aug$ by \eqref{eqn:aug}, and therefore $\boldsymbol{g}^*(\boldsymbol{O}) - \boldsymbol{h}^*(\boldsymbol{O}) \in \Lambda_1^{O,\perp}$.  The function $\boldsymbol{g}^*(\boldsymbol{O}) - \boldsymbol{h}^*(\boldsymbol{O})$ is orthogonal to $\Lambda(F_{R|L})$ (in the covariance inner product) if $E \Big[ \{ \boldsymbol{g}^*(\boldsymbol{O})-\boldsymbol{h}^*(\boldsymbol{O})  \}^\prime \boldsymbol{w}(\boldsymbol{Z}) \Big] = 0$ for every $\boldsymbol{w}(\boldsymbol{Z}) \in \Lambda(F_{R|L})$.

Let $\boldsymbol{w}(\boldsymbol{Z})= \big( W_1(\boldsymbol{Z}), \ldots, w_p(\boldsymbol{Z}) \big)$ be any element of $\Lambda(F_{R|L})$, and for each $j$, write $w_j(\boldsymbol{Z}) = \int_{t=t_1}^{t_2} h_j\{ t, \overline{\boldsymbol{O}}(t)\}  \Big[dN(t) - \rho \left\{ t \mid \overline{\boldsymbol{O}}(t),Y(t) \right\} dt \Big] $ (where here we are using the containment \eqref{eqn:coarsening_space}).  Denote the $j$th components of $\boldsymbol{g}^*(\boldsymbol{O})$ and $\boldsymbol{h}^*(\boldsymbol{O})$ as 
\[  g_j^*(\boldsymbol{O}) = \int_{t=t_1}^{t_2} \frac{ a_j(t) \big\{ Y(t) -   \boldsymbol{B}(t)^\prime \boldsymbol{\beta}  \big\} }{ \rho \left\{ t \mid \overline{\boldsymbol{O}}(t),Y(t) \right\} }   dN(t) \]
\[ h_j^*(\boldsymbol{O})= \int_{t=t_1}^{t_2} \frac{ a_j(t) \big[ E \big\{ Y(t) \mid \overline{\boldsymbol{O}}(t) \big\} - \boldsymbol{B}(t)^\prime \boldsymbol{\beta}  \big] }{\rho \left\{ t \mid \overline{\boldsymbol{O}}(t),Y(t) \right\}} \Big[ dN(t) - \rho \left\{ t \mid \overline{\boldsymbol{O}}(t),Y(t) \right\} dt \Big] \]
where $a_j(t)$ is the $j$th component of $\boldsymbol{V}^{-1}\boldsymbol{B}(t)$.

We show orthogonality using the martingale covariance property \citep{andersen1993bookABGK}, which says that, for predictable processes $H_1(t)$ and $H_2(t)$, 
 \[ E \left\{ \left( \int_{t=t_1}^{t_2} H_1(t) \Big[ dN(t) - \rho \left\{ t \mid \overline{\boldsymbol{O}}(t),Y(t) \right\} dt \Big] \right)  \left( \int_{t=t_1}^{t_2} H_2(t) \Big[ dN(t) - \rho \left\{ t \mid \overline{\boldsymbol{O}}(t),Y(t) \right\} dt \Big] \right) \right\}  \]
 \begin{equation} =  E \left[  \int_{t=t_1}^{t_2} H_1(t) H_2(t) \rho \{t \mid Y(t),\overline{\boldsymbol{O}}(t) \} dt  \right].  \label{mart-covar}
\end{equation}
By \eqref{mart-covar},
$E \big\{   h_j^*(\boldsymbol{O}) w_j(\boldsymbol{Z}) \big\} 
=  E \left( \int_{t=t_1}^{t_2} a_j(t) h_j \{ t, \overline{\boldsymbol{O}}(t) \}   \big[ E \big\{ Y(t) \mid  \overline{\boldsymbol{O}}(t) \big\} - \boldsymbol{B}(t)^\prime \boldsymbol{\beta} \big] dt \right)$.
On the other hand, by adding and subtracting we can write:
\begin{align}
E  \big\{ & g_j^*(\boldsymbol{O})w_j(\boldsymbol{Z})\big\}  \notag \\
= & \ E \left\{\left(  \int_{t=t_1}^{t_2} \frac{ a_j(t)  \big\{ Y(t)  - \boldsymbol{B}(t)^\prime \boldsymbol{\beta} \big\} }{ \rho \left\{ t \mid \overline{\boldsymbol{O}}(t),Y(t) \right\} } \Big[ dN(t) - \rho \left\{ t \mid \overline{\boldsymbol{O}}(t),Y(t) \right\}dt \Big] \right) w_j(\boldsymbol{Z}) \right\} \label{mean-gw} \\
& \hspace{0.5in} +E \left( \left[ \int_{t=t_1}^{t_2} a_j(t) \big\{ Y(t)  - \boldsymbol{B}(t)^\prime \boldsymbol{\beta}  \big\} dt \right] w_j(\boldsymbol{Z}) \right) . \notag 
\end{align}

\noindent  Using iterated expectation conditioning on $\boldsymbol{L}$, the second term in \eqref{mean-gw} equals:
 \[ \hspace{-2in} E   \bigg[   \left( \int_{t=t_1}^{t_2}  a_j(t) \big\{ Y(t) - \boldsymbol{B}(t)^\prime \boldsymbol{\beta} \big\} dt \right) \times \]
 \[ \hspace{0.5in} \underbrace{ E \left\{  \left( \int_{t=t_1}^{t_2} h_j \{ t, \overline{\boldsymbol{O}}(t) \}^\prime  \Big[ dN(t) - \rho \left\{ t \mid \overline{\boldsymbol{O}}(t),Y(t) \right\}dt \Big] \right) \mid \boldsymbol{L}   \right\} }_{ = \ 0  \mbox{ by Lemma \ref{martingale} }} \bigg] =0 . \]

 \noindent By \eqref{mart-covar}, the first term in \eqref{mean-gw} is equal to $E \left[  \int_{t=t_1}^{t_2} a_j(t) h_j\{ t, \overline{\boldsymbol{O}}(t) \} \big\{ Y(t) - \boldsymbol{B}(t)^\prime \boldsymbol{\beta} \big\}  dt \right]$. 
Iterated expectation conditioning on $\overline{\boldsymbol{O}}(t)$ shows that this is equal to $E \{ h_j^*(\boldsymbol{O})w_j(\boldsymbol{Z}) \}$, for each $j=1,\ldots,p$.  Therefore $E \Big[ \{ \boldsymbol{g}^*(\boldsymbol{O})-\boldsymbol{h}^*(\boldsymbol{O})  \}^\prime \boldsymbol{w}(\boldsymbol{Z})\Big] =0$.  

Finally, we must show that $E \big[ \big\{\boldsymbol{g}^*(\boldsymbol{O})-\boldsymbol{h}^*(\boldsymbol{O}) \big\} \boldsymbol{S}_{\boldsymbol{\beta}}(\boldsymbol{O})^\prime \big]= \bf{1}_{p \times p}$, where $\boldsymbol{S}_{\boldsymbol{\beta}}(\boldsymbol{O})$ is the observed-data score for $\boldsymbol{\beta}$ and $\bf{1}_{p \times p}$ is the $p \times p$ identity matrix.  Following \citet{tsiatis2006book} Theorem 8.3, we rewrite the left side using iterated expectations and the fact that $\boldsymbol{S}_{\boldsymbol{\beta}}(\boldsymbol{O})=E[ \boldsymbol{S}_{\boldsymbol{\beta}}(\boldsymbol{L}) | \boldsymbol{O}]$, where $\boldsymbol{S}_{\boldsymbol{\beta}}(\boldsymbol{L})$ is the full-data score for $\boldsymbol{\beta}$:
 \begin{align*}
 E  \big[ \big\{ \boldsymbol{g}^*(\boldsymbol{O})  -\boldsymbol{h}^*(\boldsymbol{O}) \big\} \boldsymbol{S}_{\boldsymbol{\beta}}(\boldsymbol{O})^\prime \big] 
 = & \  E \big[ \big\{ \boldsymbol{g}^*(\boldsymbol{O})-\boldsymbol{h}^*(\boldsymbol{O}) \big\} E \{ \boldsymbol{S}_{\boldsymbol{\beta}}(\boldsymbol{L})^\prime \mid \boldsymbol{O} \big\} \big] \\
= & \  E \big[  \big\{ \boldsymbol{g}^*(\boldsymbol{O})-\boldsymbol{h}^*(\boldsymbol{O}) \big\} \boldsymbol{S}_{\boldsymbol{\beta}}(\boldsymbol{L})^\prime  \big] \\
 = & \ E \big[ \underbrace{ E \{ \boldsymbol{g}^*(\boldsymbol{O}) \mid \boldsymbol{L} \} }_{\boldsymbol{\varphi}(\boldsymbol{L})} \boldsymbol{S}_{\boldsymbol{\beta}}(\boldsymbol{L})^\prime  \big] - E \big[ \underbrace{ E \{ \boldsymbol{h}^*(\boldsymbol{O}) \mid \boldsymbol{L} \} }_{= \boldsymbol{0}} \boldsymbol{S}_{\boldsymbol{\beta}}(\boldsymbol{L})^\prime  \big] \\
= & \  E \big[ \boldsymbol{\varphi}(\boldsymbol{L}) \boldsymbol{S}_{\boldsymbol{\beta}}(\boldsymbol{L})^\prime  \big]  = \bf{1}_{p \times p}
 \end{align*}
 
 \noindent where the last equality holds since $\boldsymbol{\varphi}(\boldsymbol{L})$ is a full-data influence function.  This completes the proof.

\end{proof}

\subsection*{Appendix B.4: Large sample distribution of $\widehat{\boldsymbol{\beta}}$} \label{append:remainder}

Here we derive the asymptotic distribution of our augmented inverse intensity-weighted estimator $\widehat{\boldsymbol{\beta}}$. Let $\widehat{P}$ be an estimate of the distribution $P$ that uses the empirical distribution of $\overline{\boldsymbol{O}}(t)$, and let $\widehat{\rho} \{ t \mid Y(t), \overline{\boldsymbol{O}}(t) \}$ and $\widehat{E}\big\{ Y(t) \mid \overline{\boldsymbol{O}}(t)\big\}$ be estimates as in the algorithm in Section 4.2.  Following \citep{kennedy2016empiricalprocesses}, 
$\sqrt{n} \left(\widehat{\boldsymbol{\beta}}- \boldsymbol{\beta} \right)$ can be expanded as the sum of three terms:  a central limit term $\frac{1}{\sqrt{n}} \sum_{i=1}^n \boldsymbol{\varphi} \big(\boldsymbol{O}_i; P \big)$, an empirical process term, and a remainder term $\sqrt{n} \ \boldsymbol{Rem} \big( \widehat{P},P \big)$.  We compute the remainder term below:

\begin{prop3} \label{prop:remainder}
The remainder term $\boldsymbol{Rem} \left(\widehat{P},P \right)$ is given by:
\begin{align}
 E \Bigg( \int_{t=t_1}^{t_2} &   \boldsymbol{V}^{-1}\boldsymbol{B}(t) \left[ \frac{ \rho \left\{ t \mid \overline{\boldsymbol{O}}(t),Y(t) \right\} }{ \widehat{\rho} \left\{ t \mid Y(t), \overline{\boldsymbol{O}}(t) \right\} } - 1 \right] \Big[ E \big\{ Y(t) \mid \overline{\boldsymbol{O}}(t)\big\} - \widehat{E}\big\{ Y(t) \mid \overline{\boldsymbol{O}}(t)\big\}  \Big]  dt \Bigg). \label{remainder_eqn}
 \end{align}
\end{prop3}

\begin{proof} Let $\mathcal{M}$ be our model for the observed data $\boldsymbol{O}$.  In what follows, we will use a $\widetilde{P}$ subscript to denote expectations or intensities under a distribution $\widetilde{P} \in \mathcal{M}$.  Expectations and intensities with no subscript are taken under the true distribution $P$.  Let $\boldsymbol{\psi}: \mathcal{M} \to \mathbb{R}^p$ be the parameter mapping, given by $\boldsymbol{\psi}(\widetilde{P}) = \int_{t=t_1}^{t_2} \boldsymbol{V}^{-1}\boldsymbol{B}(t) E_{\widetilde{P}} \{ Y(t) \} dt$, and for each $\widetilde{P} \in \mathcal{M}$, write $\boldsymbol{\psi} \big( \widetilde{P} \big) = \boldsymbol{\beta}_{\widetilde{P}}$.  Now $\boldsymbol{\psi}(\widetilde{P})$ has a von Mises expansion \citep{vandervaart1998asymptotic} as:
\[ \boldsymbol{\psi}(\widetilde{P}) = \boldsymbol{\psi}(P) -  E \big\{ \boldsymbol{\varphi}(\boldsymbol{O} ; \widetilde{P}) \big\}  + \boldsymbol{Rem}(\widetilde{P}, P) , \]

\noindent which we use to compute $\boldsymbol{Rem} \left(\widetilde{P},P \right)$.  We have:

\begin{align*}
E \big\{ & \boldsymbol{\varphi} \big( \boldsymbol{O} ; \tilde{P} \big) \big\} \\
= & \ \underbrace{ E \Bigg(   \int_{t=t_1}^{t_2}  \boldsymbol{V}^{-1} \boldsymbol{B}(t) \frac{ \big[ Y(t)- E_{\widetilde{P}} \big\{ Y(t) \mid \overline{\boldsymbol{O}}(t) \big\} \big] }{ \rho_{\widetilde{P}} \{ t \mid Y(t), \overline{\boldsymbol{O}}(t) \} } \Big[ dN(t) - \rho \left\{ t \mid \overline{\boldsymbol{O}}(t),Y(t) \right\}dt \Big] \Bigg) }_{=0 }   \\
& \hspace{0.5in}  +  E \Bigg( \int_{t=t_1}^{t_2}  \boldsymbol{V}^{-1} \boldsymbol{B}(t) \frac{ \big[ Y(t)- E_{\widetilde{P}} \big\{ Y(t) \mid \overline{\boldsymbol{O}}(t) \big\} \big]}{ \rho_{\widetilde{P}} \{ t \mid Y(t), \overline{\boldsymbol{O}}(t) \} }  \rho \left\{ t \mid \overline{\boldsymbol{O}}(t),Y(t) \right\} dt \Bigg) \\
 &   \hspace{1in}  + E \Bigg[ \int_{t=t_1}^{t_2}  \boldsymbol{V}^{-1} \boldsymbol{B}(t) E_{\widetilde{P}} \big\{ Y(t) \mid \overline{\boldsymbol{O}}(t) \big\}  dt \Bigg] - \ \boldsymbol{\beta}_{\widetilde{P}}   \\
= & \    E \Bigg\{ \int_{t=t_1}^{t_2} E \left(  \boldsymbol{V}^{-1} \boldsymbol{B}(t) \frac{ \rho \left\{ t \mid \overline{\boldsymbol{O}}(t),Y(t) \right\}}{ \rho_{\widetilde{P}} \{ t \mid Y(t), \overline{\boldsymbol{O}}(t) \}}  \Big[ Y(t)- E_{\widetilde{P}} \big\{ Y(t) \mid \overline{\boldsymbol{O}}(t) \big\} \Big] \mid \overline{\boldsymbol{O}}(t) \right) dt \Bigg\} \\
 &   \hspace{1in}  + E \Bigg[ \int_{t=t_1}^{t_2}  \boldsymbol{V}^{-1} \boldsymbol{B}(t) E_{\widetilde{P}} \big\{ Y(t) \mid \overline{\boldsymbol{O}}(t) \big\}  dt \Bigg] - \ \boldsymbol{\beta}_{\widetilde{P}} \\
 = & \ E \Bigg( \int_{t=t_1}^{t_2}  \boldsymbol{V}^{-1} \boldsymbol{B}(t) \frac{ \rho \left\{ t \mid \overline{\boldsymbol{O}}(t),Y(t) \right\}}{\rho_{\widetilde{P}} \{ t \mid Y(t), \overline{\boldsymbol{O}}(t) \}}  \Big[ E \big\{ Y(t) \mid \overline{\boldsymbol{O}}(t) \big\}- E_{\widetilde{P}} \big\{ Y(t) \mid \overline{\boldsymbol{O}}(t) \big\} \Big] dt \Bigg) \\
 &   \hspace{1in}  + E \Bigg[ \int_{t=t_1}^{t_2}   \boldsymbol{V}^{-1} \boldsymbol{B}(t)  E_{\widetilde{P}} \big\{ Y(t) \mid \overline{\boldsymbol{O}}(t) \big\}   dt \Bigg] - \ \boldsymbol{\beta}_{\widetilde{P}}
\end{align*}

\noindent where in the last equality we used the fact that, by Proposition 2:
\[ \frac{ \rho \left\{ t \mid \overline{\boldsymbol{O}}(t),Y(t) \right\} }{ \rho_{\widetilde{P}} \{ t \mid Y(t), \overline{\boldsymbol{O}}(t) \}} = \frac{ \lambda \left\{ t \mid \overline{\boldsymbol{O}}(t) \right\} E[ \exp\{ \alpha Y(t) \mid \Delta N(t)=1, \overline{\boldsymbol{O}}(t) \} ] }{\lambda_{\widetilde{P}} \left\{ t \mid \overline{\boldsymbol{O}}(t) \right\} E_{\tilde{P}}[ \exp\{ \alpha Y(t) \} | \Delta N(t)=1, \bar{\boldsymbol{O}}(t)  ] } \]

\noindent is a function of $t$ and $\bar{\boldsymbol{O}}(t)$ only.   Therefore:
\begin{align*}
 & \ \boldsymbol{Rem}  \left( \tilde{P}, P \right) \\
 =  & \ E \Bigg( \int_{t=t_1}^{t_2} \boldsymbol{V}^{-1} \boldsymbol{B}(t) \frac{ \rho \left\{ t \mid \overline{\boldsymbol{O}}(t),Y(t) \right\} }{ \rho_{\widetilde{P}} \{ t \mid Y(t), \overline{\boldsymbol{O}}(t) \} }  \Big[ E \big\{ Y(t) \mid \overline{\boldsymbol{O}}(t) \big\}- E_{\widetilde{P}} \big\{ Y(t) \mid \overline{\boldsymbol{O}}(t) \big\} \Big]  dt \Bigg) \\
 &   \hspace{1in}  + E \Bigg[ \int_{t=t_1}^{t_2}   \boldsymbol{V}^{-1} \boldsymbol{B}(t)  E_{\widetilde{P}} \big\{ Y(t) \mid \overline{\boldsymbol{O}}(t) \big\}   dt \Bigg]  \underbrace{ - \boldsymbol{\beta}_{\tilde{P}} + \boldsymbol{\psi}(\tilde{P})}_{=0} - \boldsymbol{\psi}(P) \\
= & \   E \Bigg( \int_{t=t_1}^{t_2}  \boldsymbol{V}^{-1} \boldsymbol{B}(t) \frac{ \rho \left\{ t \mid \overline{\boldsymbol{O}}(t),Y(t) \right\} }{ \rho_{\widetilde{P}} \{ t \mid Y(t), \overline{\boldsymbol{O}}(t) \} }  \Big[ E \big\{ Y(t) \mid \overline{\boldsymbol{O}}(t) \big\}- E_{\widetilde{P}} \big\{ Y(t) \mid \overline{\boldsymbol{O}}(t) \big\} \Big]  dt \Bigg) \\
 &   \ \ \  + E \Bigg[ \int_{t=t_1}^{t_2}   \boldsymbol{V}^{-1} \boldsymbol{B}(t) E_{\widetilde{P}} \big\{ Y(t) \mid \overline{\boldsymbol{O}}(t) \big\} dt \Bigg]  -E \Bigg[ \int_{t=t_1}^{t_2}   \boldsymbol{V}^{-1} \boldsymbol{B}(t) E \big\{ Y(t) \mid \overline{\boldsymbol{O}}(t) \big\}   dt \Bigg]  \\
= & \ E \Bigg( \int_{t=t_1}^{t_2}   \boldsymbol{V}^{-1}\boldsymbol{B}(t) \left[ \frac{ \rho \left\{ t \mid \overline{\boldsymbol{O}}(t),Y(t) \right\} }{ \rho_{\widetilde{P}} \{ t \mid Y(t), \overline{\boldsymbol{O}}(t) \} } - 1 \right] \Big[ E\big\{ Y(t) \mid \overline{\boldsymbol{O}}(t) \big\} - E_{\widetilde{P}} \big\{ Y(t) \mid \overline{\boldsymbol{O}}(t) \big\} \Big]  dt \Bigg) .
\end{align*}
\end{proof}


\bigskip

Next we show sufficient conditions under which the remainder term $\sqrt{n} \boldsymbol{Rem} \left( \widehat{P},P \right)$ is asymptotically negligible.  Suppose the intensity function $\lambda \left\{ t \mid \overline{\boldsymbol{O}}(t) \right\}$ is modeled using a (possibly stratified) Andersen-Gill model as described in Section 4.1.  The partial likelihood estimator $\widehat{\boldsymbol{\gamma}}$  \citep{cox1972regression, cox1975partiallikelihood} and the Breslow estimator $\widehat{\Lambda}_{0,k}(t)$ \citep{breslow1972discussion} are used to estimate $\boldsymbol{\gamma}$ and the cumulative baseline intensity functions $\Lambda_{0,k}(t) = \int_{s=0}^t \lambda_{0,k}(s)ds$ for each assessment $k$.     
Then, kernel smoothing of $\widehat{\Lambda}_{0,k}(t)$ is used to estimate the baseline intensity $\lambda_{0,k}(t)$ as $\widehat{\lambda}_{0,k}(t) \ =$ $\displaystyle{ \frac{1}{h} \sum_j K \left( \frac{ t-T_j}{h} \right) d \widehat{\Lambda}_{0,k}(T_j) }$, where $h$ is a bandwidth, $K( \cdot )$ is a choice of kernel, $T_j$ is a time at which one or more participants had their  $k$th assessment, and $d \widehat{\Lambda}_{0,k}(T_j)$ is the size of the jump in $\widehat{\Lambda}_{0,k}(t)$ at $T_j$.

Our result below uses the asymptotic theory established by several authors.  \citet{andersengill1982} derived the asymptotic distribution of $\widehat{\boldsymbol{\gamma}}$ and $\widehat{\Lambda}_{0}(t)$ in the univariate (unstratified) setting, and \citet{andersenborgan1985countingprocesses} generalized their results to multivariate counting processes such as our stratified model (see also \citet{cooklawless2007book} for more discussion of stratified models).  \citet{ramlauhansen1983smoothing} proposed this kernel-smoothing of the Nelson-Aalen estimator for a univariate counting process and showed the large-sample distribution of the resulting estimator.   \citet{andersenborgan1985countingprocesses},  \citet*{andersen1993bookABGK}, and  \citet{wells1994kernelestimation} extended his results to multivariate versions of the Andersen-Gill model.  Their asymptotic results for the kernel-smoothed $\widehat{\lambda}_{0k}(t)$ are based on the bandwidth $h=h_n$ decreasing at a certain rate as $n$ grows (as in assumption (ii) in Theorem 2 below).

For the model of $dF \left\{ y(t) \mid \Delta N(t)=1,\overline{\boldsymbol{O}}(t) \right\}$, we only impose a certain rate condition.  We then note that the single index model of Section 4.1 \citep{chiang2012new} satisfies this rate.   We let $\boldsymbol{\theta}$ denote the parameters of the model for $dF \left\{ y(t) \mid \Delta N(t)=1,\overline{\boldsymbol{O}}(t) \right\}$.

\medskip

\begin{thm2}
\label{thm:rates} 
Suppose the following:

\begin{enumerate}[(i)]

\item Assumptions 1-4 in the main manuscript hold and the models for $\lambda \left\{ t \mid \overline{\boldsymbol{O}}(t) \right\}$ and \\ $dF \left\{ y(t) \mid \Delta N(t)=1,\overline{\boldsymbol{O}}(t) ; \boldsymbol{\theta} \right\}$ are correctly specified

\item $\lambda \left\{ t \mid \overline{\boldsymbol{O}}(t) \right\}$ is modeled using a (possibly stratified) Andersen-Gill model, $\boldsymbol{\gamma}$ and $\lambda_{0,k}(t)$ are estimated as described above, and the bandwidth $h=h_n$ used for kernel smoothing is chosen using a procedure such that $\lim_{n \to \infty} nh_n^5=d < \infty$

 \item The model for $dF \left\{ y(t) \mid \Delta N(t)=1,\overline{\boldsymbol{O}}(t) \right\}$ and estimators of $E[ \exp\{ \alpha Y(t) \mid \Delta N(t)=1, \overline{\boldsymbol{O}}(t) \} ]$ and $E[ Y(t) \exp\{ \alpha Y(t) \mid \Delta N(t)=1, \overline{\boldsymbol{O}}(t) \} ]$ are such that \\
$\displaystyle{ E \left[ \int_{t=t_1}^{t_2} \Big\{ \widehat{E}[ \exp\{\alpha Y(t)\} \mid \Delta N(t)=1, \overline{\boldsymbol{O}}(t)]  - E[ \exp\{ \alpha Y(t) \mid \Delta N(t)=1, \overline{\boldsymbol{O}}(t) \} ] \Big\}^2 dt \right]}$ and
$E \bigg[ \displaystyle{\int_{t=t_1}^{t_2}} \Big\{ \widehat{E}[ Y(t) \exp\{\alpha Y(t)\} \mid \Delta N(t)=1, \overline{\boldsymbol{O}}(t)]  - E[ Y(t)\exp \{ \alpha Y(t) \} \mid \Delta N(t)=1, \overline{\boldsymbol{O}}(t) ] \Big\}^2 dt \bigg]$ are each $o_p(n^{-1/2})$.
\end{enumerate}
Then the remainder term $\boldsymbol{Rem} \left(\widehat{P},P \right)$ in \eqref{remainder_eqn} is $o_P(n^{-1/2})$.
\end{thm2}


\begin{proof} 
Write $\boldsymbol{Rem} \left(\widehat{P},P \right) = (R_1,\ldots,R_p)^{\prime}$.  We show that, for each $j=1,\ldots,p$, $R_j$ is $o_P(n^{-1/2})$;  then the result for $\boldsymbol{Rem} \left(\widehat{P},P \right)$ follows.

Write $\boldsymbol{V}^{-1}\boldsymbol{B}(t)=(D_1(t),\ldots,D_p(t))^{\prime}$, so that for each $j=1,\ldots,p$, 
\[ R_j=E \Bigg( \int_{t=t_1}^{t_2} D_j(t) \left[ \frac{ \rho \left\{ t \mid \overline{\boldsymbol{O}}(t),Y(t) \right\}  }{  \widehat{\rho} \left\{ t \mid Y(t), \overline{\boldsymbol{O}}(t) \right\} } - 1 \right] 
 \Big[ E \big\{ Y(t) \mid \overline{\boldsymbol{O}}(t) \big\} - \widehat{E} \big[ Y(t) \mid \overline{\boldsymbol{O}}(t)  \big\}  \Big]  dt \Bigg) .\]

By the Cauchy-Schwarz inequality,
\begin{align} 
R_j^2 \leq &  \ E \Bigg( \int_{t=t_1}^{t_2} \left[ \frac{ D_j(t) }{  \widehat{\rho} \left\{ t \mid Y(t), \overline{\boldsymbol{O}}(t) \right\} } \right]^2 \Big[  \rho \left\{ t \mid \overline{\boldsymbol{O}}(t),Y(t) \right\} -  \widehat{\rho} \left\{ t \mid Y(t), \overline{\boldsymbol{O}}(t) \right\} \Big]^2  dt \Bigg)  \label{rho-factor} \\
& \ \ \ \ \ \times \ E \Bigg( \int_{t=t_1}^{t_2} \left[  E \{ Y(t) \mid \overline{\boldsymbol{O}}(t) \} - \widehat{E} \{ Y(t) \mid \overline{\boldsymbol{O}}(t)  \}  \right]^2 dt \Bigg). \label{theta_factor}
\end{align}

\noindent Now we expand:
\begin{align*}
 \widehat{\rho} & \left\{ t \mid Y(t), \overline{\boldsymbol{O}}(t) \right\}   - \rho \left\{ t \mid \overline{\boldsymbol{O}}(t),Y(t) \right\} \\
 = & \  \widehat{ \lambda}_{0k}(t) \exp\{ \widehat{\boldsymbol{\gamma}}^\prime \boldsymbol{Z}(t) \} \exp\{ -\alpha Y(t) \} \widehat{E} \big[ \exp\{ \alpha Y(t) \} \mid \Delta N(t) =1 , \overline{\boldsymbol{O}}(t) ; \widehat{\boldsymbol{\theta}}  \ \big] \  -   \\
& \ \ \ \ \ \   \lambda_{0k}(t) \exp\{ \boldsymbol{\gamma}^\prime \boldsymbol{Z}(t) \} \exp\{ -\alpha Y(t) \} E \big[ \exp\{ \alpha Y(t) \} \mid \Delta N(t) =1 , \overline{\boldsymbol{O}}(t)  \big] \\
 = &  \  \Big\{ \widehat{ \lambda}_{0k}(t)  - \lambda_{0k}(t) \Big\} \exp\{ \widehat{\boldsymbol{\gamma}}^\prime \boldsymbol{Z}(t) \} \exp\{ -\alpha Y(t) \} \widehat{E} \big[ \exp\{ \alpha Y(t) \} \mid \Delta N(t) =1 , \overline{\boldsymbol{O}}(t) \big] + \\
 & \ \ \ \  \lambda_{0k}(t) \Big[  \exp\{ \widehat{\boldsymbol{\gamma}}^\prime \boldsymbol{Z}(t) \} - \exp\{ \boldsymbol{\gamma}^\prime \boldsymbol{Z}(t) \} \Big] \exp\{ -\alpha Y(t) \} \widehat{E} \big[ \exp\{ \alpha Y(t) \} \mid \Delta N(t) =1 , \overline{\boldsymbol{O}}(t)  \big] + \\
 & \ \ \  \lambda_{0k}(t) \exp\{ \boldsymbol{\gamma}^\prime \boldsymbol{Z}(t) \} \exp\{ -\alpha Y(t) \}\Big( \widehat{E} \big[ \exp\{ \alpha Y(t) \} \mid \Delta N(t) =1 , \overline{\boldsymbol{O}}(t)  \big] - \\
 & \hspace{0.5in} E \big[ \exp\{ \alpha Y(t) \} \mid \Delta N(t) =1 , \overline{\boldsymbol{O}}(t)  \big] \Big).
 \end{align*}
 Using this expansion and the inequality $rs \leq (r^2+s^2)/2$, the rate for \eqref{rho-factor} will be determined by the rates of three terms, namely the squares of each of the terms in the expansion above.  
 
 For the factor \eqref{theta_factor}, using Corollary 1, we write:
 
\begin{align*}
 E \{ & Y(t)  \mid \overline{\boldsymbol{O}}(t) \} - \widehat{E} \{ Y(t) \mid \overline{\boldsymbol{O}}(t)  \} \\
 = & \ \frac{ E \big[ Y(t) \exp \{ \alpha Y(t) \} \mid \Delta N(t)=1, \overline{\boldsymbol{O}}(t)  \big] }{ E \big[ \exp \{ \alpha Y(t) \} \mid \Delta N(t)=1, \overline{\boldsymbol{O}}(t) \big]  } - \frac{ \widehat{E} \big[ Y(t) \exp \{ \alpha Y(t) \} \mid \Delta N(t)=1, \overline{\boldsymbol{O}}(t)  \big] }{ \widehat{E} \big[ \exp \{ \alpha Y(t) \} \mid \Delta N(t)=1, \overline{\boldsymbol{O}}(t)  \big]  } \\
 = &  \ \frac{E \big[ Y(t)\exp \{ \alpha Y(t) \} \mid \Delta N(t)=1, \overline{\boldsymbol{O}}(t) \big]}{E \big[ \exp \{ \alpha Y(t) \} \mid \Delta N(t)=1, \overline{\boldsymbol{O}}(t)  \big] \widehat{E} \big[ \exp \{ \alpha Y(t) \} \mid \Delta N(t)=1, \overline{\boldsymbol{O}}(t) \big]} \times \\
 & \ \  \Big( \widehat{E} \big[ \exp \{ \alpha Y(t) \} \mid \Delta N(t)=1, \overline{\boldsymbol{O}}(t) \big]- E \big[  \exp \{ \alpha Y(t) \} \mid \Delta N(t)=1, \overline{\boldsymbol{O}}(t) \big]\Big) \\
 & \ - \ \frac{E \big[ \exp \{ \alpha Y(t) \} \mid \Delta N(t)=1, \overline{\boldsymbol{O}}(t)  \big]}{E \big[ \exp \{ \alpha Y(t) \} \mid \Delta N(t)=1, \overline{\boldsymbol{O}}(t)  \big] \widehat{E} \big[ \exp \{ \alpha Y(t) \} \mid \Delta N(t)=1, \bar{\boldsymbol{O}}(t) \big]} \times \\
 & \ \  \Big( \widehat{E} \big[ Y(t) \exp \{ \alpha Y(t) \} \mid \Delta N(t)=1, \overline{\boldsymbol{O}}(t) \big]- E \big[ Y(t) \exp \{ \alpha Y(t) \} \mid \Delta N(t)=1, \overline{\boldsymbol{O}}(t) \big]\Big)
\end{align*}
The rates for \eqref{theta_factor} will be determined by the rates for the squares of the two terms in this expansion. Combining these results, we have:
 \begin{align}
R_j^2 \leq  \Bigg( & C_{1j}(\widehat{\boldsymbol{\eta}}) \ E \left[ \int_{t=t_1}^{t_2} \Big\{\widehat{ \lambda}_{0k}(t)  - \lambda_{0k}(t) \Big\}^2 dt \right] + \label{Term1} \\
  &  \ \ C_{2j}( \widehat{\boldsymbol{\eta}}) \ E \left[ \int_{t=t_1}^{t_2} \Big[  \exp\{ \widehat{\boldsymbol{\gamma}}^\prime \boldsymbol{Z}(t) \} - \exp\{ \boldsymbol{\gamma} \boldsymbol{Z}(t) \} \Big]^2 dt \right] + \label{Term2} \\
  & \ \ \ C_{3j}(\ \widehat{\boldsymbol{\eta}}) \ E \bigg[ \int_{t=t_1}^{t_2} \Big( \widehat{E} \big[ \exp\{ \alpha Y(t) \} \mid \Delta N(t) =1 , \overline{\boldsymbol{O}}(t)  \big] - \label{Term3} \\
  & \hspace{1.6in} E \big[ \exp\{ \alpha Y(t) \} \mid \Delta N(t) =1 , \overline{\boldsymbol{O}}(t) \big] \Big)^2 dt \bigg] \Bigg) \times \notag \\
   \Bigg\{ & C_{4j}( \widehat{\boldsymbol{\eta}}) \ E \Bigg[ \int_{t=t_1}^{t_2} \bigg(  \widehat{E} \big[ Y(t) \exp\{ \alpha Y(t) \} \mid \Delta N(t) =1 , \overline{\boldsymbol{O}}(t) \big] -  \label{Term4} \\
  & \hspace{1.6in} E \big[ Y(t) \exp\{ \alpha Y(t) \} \mid \Delta N(t) =1 , \overline{\boldsymbol{O}}(t)  \big] \bigg)^2 dt \Bigg] + \notag \\
&  \ \  C_{5j}(\widehat{\boldsymbol{\eta}}) \ E \bigg[ \int_{t=t_1}^{t_2} \Big( \widehat{E} \big[ \exp\{ \alpha Y(t) \} \mid \Delta N(t) =1 , \overline{\boldsymbol{O}}(t)  \big] - \label{Term5} \\
  & \hspace{1.6in} E \big[ \exp\{ \alpha Y(t) \} \mid \Delta N(t) =1 , \overline{\boldsymbol{O}}(t) \big] \Big)^2 dt \bigg] \Bigg\} \notag 
\end{align}

\noindent where $\boldsymbol{\eta}$ denotes all parameters of the models for $\lambda \{ t \mid \overline{\boldsymbol{O}}(t) \}$ and $dF \{ y(t) \mid \Delta N(t)=1, \overline{\boldsymbol{O}}(t) \}$, and where $C_{1j}( \widehat{\boldsymbol{\eta}}), \ldots, C_{5j}( \widehat{\boldsymbol{\eta}})$  are terms that are each $O_P(1)$ under Assumption 2 (positivity assumption) and assumption (i) of Theorem 2. 

We first consider term \eqref{Term1}. Let $X_n(t)= (n b_n)^{1/2} \Big\{ \widehat{\lambda}_{0k}(t)  - \lambda_{0k}(t) \Big\}$.  \citet{wells1994kernelestimation} has shown that, under condition (ii) of Theorem 2, the stochastic processes $\{ X_n(t) : t_1 \leq t \leq t_2 \}_{n=1}^\infty$ converges weakly to a process $\{ X(t) : t_1 \leq t \leq t_2 \}$, where $\{ X(t) : t_1 \leq t \leq t_2 \}$ is a Wiener process plus a term depending on $\lambda_{0k}(t)$.  Therefore it follows by the Continuous Mapping Theorem that $\{ X_n^2(t) : t_1 \leq t \leq t_2 \}_{n=1}^\infty$ converges weakly to $\{ X^2(t) : t_1 \leq t \leq t_2 \}$.  Again by the Continuous Mapping Theorem, $\int_{t=t_1}^{t_2} X_n^2(t) dt = n h_n \left[ \int_{t=t_1}^{t_2}  \big\{ \widehat{\lambda}_{0k}(t)  - \lambda_{0k}(t) \big\}^2 dt \right]$ is $O_p(1)$.  Therefore, the term in \eqref{Term1} is $O_P \{ (nh_n)^{-1} \}$. By assumption (ii), for large $n$, $n h_n$ is approximately $n^{4/5}$, so in particular the convergence rate for \eqref{Term1} is faster than $n^{1/2}$. 

Since $\widehat{\boldsymbol{\gamma}}$ converges to $\boldsymbol{\gamma}$ at root-$n$ rates, the term in \eqref{Term2} is $O_p(n^{-1})$.  

Finally, terms \eqref{Term3}, \eqref{Term4}, and \eqref{Term5} are $o_p(n^{-1/2})$ by assumption (iii). Hence the product that bounds $R_j^2$ is $o_P(n^{-1})$, so that $R_j$ is $o_P(n^{-1/2})$ as claimed.

\end{proof}

\medskip

Results of \citet{chiang2012new} can be used to show that if the conditional outcome distribution $dF \left\{ y(t) \mid \Delta N(t)=1,\overline{\boldsymbol{O}}(t) \right\}$ is modeled using a single index model, then (under conditions on the bandwidth), 
\[ E \left[ \Big( \widehat{E}[ Y(t) \exp\{\alpha Y(t)\} \mid \Delta N(t)=1, \overline{\boldsymbol{O}}(t)]  - E[ Y(t)\exp \{ \alpha Y(t) \} \mid \Delta N(t)=1, \overline{\boldsymbol{O}}(t) ] \Big)^2  \right] \]
and
\[ E \left[ \Big( \widehat{E}[  \exp\{\alpha Y(t)\} \mid \Delta N(t)=1, \overline{\boldsymbol{O}}(t)]  - E[ \exp \{ \alpha Y(t) \} \mid \Delta N(t)=1, \overline{\boldsymbol{O}}(t) ] \Big)^2  \right]\] 
are $O_p(n^{-4/5})$ for each fixed $t$.  It can be further shown that, under an assumption that these are differentiable in $t$ with bounded derivatives on $[t_1,t_2]$, the expectations in assumption $(iii)$ are also $O_p(n^{-4/5})$, so assumption (iii) will hold.

\section*{Appendix C:  Simulation study}

We assessed the finite-sample performance of our estimators in a realistic simulation study based on the ARC data. 
Simulated data was generated separately by treatment arm.  To generate simulated data for participant $i$, we drew their baseline outcome $Y_i(0)$ from the empirical distribution of baseline outcomes in the given arm of the ARC data.  We then iterated between generating times of subsequent assessments, and outcomes at those assessment times.  Given their $k$th assessment time $T_{ik}$ and outcome $Y_i(T_{ik})$, we generated participant $i$'s $(k+1)$st assessment time $T_{i,k+1}$ using Ogata's Thinning Algorithm \citep{ogata1981simulation} as follows: 

\begin{enumerate}

\item Set $\lambda^*$ to be a value that is greater than or equal to $\sup \Big\{ \widehat{\lambda}_{0,k}(t) \exp\{ \widehat{\gamma} \  Y_i(T_{ik}) \} : t \in [0,\tau] \Big\}$, where $\widehat{\lambda}_{0,k}(t)$ and $\widehat{\gamma}$ are the estimates from the stratified Andersen-Gill model that we fit on the ARC data. 

\item Take the start time to be $T_{ik}$.  

\item Draw a potential gap time $s^* \sim Exp(\lambda^*)$, and set $t^* := $ start time $ + s^*$.  Accept the candidate assessment time $t^*$ with probability $ \widehat{\lambda}_{0,k}(t^*) \exp\{ \widehat{\gamma} \ Y_i(T_{ik}) \} /\lambda^*$.

\item  If $t^*$ is accepted, set $T_{i,k+1} := t^*$.  If $t^*$ is rejected, set the new start time to be $t^*$ and return to step 3.  Iterate until a time $t^*$ is accepted, or until $t^* > \tau$, in which case $T_{ik}$ was participant $i$'s final assessment time. 
\end{enumerate}

Outcomes were generated based on a model for $dF \left\{ y(t) \mid \Delta N(t)=1,\overline{\boldsymbol{O}}(t) \right\}$ that we fit on the ARC data.  The model used for generating simulated outcome data was a negative binomial model with predictors of time, lag time since the previous assessment, and outcome value at the previous assessment.  After the time $T_{i,k+1}$ was generated, we obtained the predicted distribution $d\widehat{F} \big\{  y(T_{i,k+1}) \mid \Delta N(T_{i,k+1})=1,\overline{\boldsymbol{O}}_i(T_{i,k+1}) \big\}$ based on this model for $dF \left\{ y(t) \mid \Delta N(t)=1,\overline{\boldsymbol{O}}(t) \right\}$, then drew an outcome $Y_i(T_{i,k+1})$ from this predicted distribution. 

We computed the true value of $\boldsymbol{\beta}$ using the identification formula $\boldsymbol{\beta} = \boldsymbol{V}^{-1} \int_{t=t_1}^{t_2} \boldsymbol{B}(t) \mu(t) dt$ (Proposition 3), under values of $\alpha=-0.6,-0.3,0,0.3,0.6$. 
We simulated a large sample of $N=2,000,000$ participants.  For each time $t \in [t_1,t_2]$, we obtained the predicted value of 
\[ E \{ Y(t) \mid \overline{\boldsymbol{O}}_i(t) \} = \frac{ \sum_{y(t)} y(t) \exp\{ \alpha y(t) \}dF \left\{ y(t) \mid \Delta N(t)=1,\overline{\boldsymbol{O}}_i(t) \right\} }{ \sum_{y(t)} \exp\{ \alpha y(t) \}dF \left\{ y(t) \mid \Delta N(t)=1,\overline{\boldsymbol{O}}_i(t) \right\}} \] 
for each participant.  We then used the approximation $E \{ Y(t) \} \approx \frac{1}{N} \sum_{i=1}^N \widehat{E}\{ Y(t) \mid \overline{\boldsymbol{O}}_i(t) \}$.  We approximated the integral  $\int_{t=t_1}^{t_2} \boldsymbol{B}(t) \mu(t) dt$ using intervals of one day.  The value of $\boldsymbol{\beta}$  we obtain corresponds to the projection of the true curve of means onto spline curves of the form $\boldsymbol{\beta}^\prime \boldsymbol{B}(t)$.

Simulation results for the mean outcome in each treatment arm at 3, 6, 9, and 12 months are shown in Tables \ref{table:sims_trt} and \ref{table:sims_ctrl}. Bias in both arms is close to zero, with slightly larger absolute bias under $\alpha=0.6$ in the control arm.  Coverage of Wald confidence intervals using the jackknife variance estimate is close to the nominal level of $0.95$ in most cases.  Wald confidence intervals using the influence function-based variance estimate tended to under-cover, with coverage between $0.89$ and $0.93$ in most cases.

\begin{table}

\begin{center}

\small

\renewcommand{\arraystretch}{1.5}

\begin{tabular}{ c  c  c  c  c  c  c  }
\Xhline{2pt}
$\alpha$  & Parameter & True Value & Emp Mean &  $|$Bias$|$ & Wald (IF) & Wald (Jackknife) \\
\Xhline{2pt}
\multirow{4}{0.3in}{ -0.6 } 
& $E[Y(3)]$  & 1.529 &  1.542 &  0.013 & 0.910 &  0.958\\
 \cline{2-7} 
 \  & $E[Y(6)]$ & 1.479 & 1.489 & 0.010 &  0.922  &  0.960\\
 \cline{2-7} 
 \ & $E[Y(9)]$ & 1.450 & 1.460 &  0.010 &  0.908  &  0.958 \\
 \cline{2-7} 
 \ & $E[Y(12)]$ & 1.426 &  1.436 &  0.010 &  0.926 &  0.968 \\
 \Xhline{2pt}
\multirow{4}{0.3in}{ -0.3 } 
   & E[Y(3)] & 1.781 & 1.792 & 0.011 & 0.920 &  0.964 \\
 \cline{2-7} 
 \  & $E[Y(6)]$ & 1.718 &  1.724 &  0.006 &  0.924 &  0.958 \\
 \cline{2-7} 
 \ & $E[Y(9)]$ & 1.681 & 1.691 & 0.010 & 0.924 &  0.946 \\
 \cline{2-7} 
 \ & $E[Y(12)]$ & 1.651 &  1.659 &  0.008 &  0.924 & 0.956 \\
\Xhline{2pt}
\multirow{4}{0.3in}{ 0 } 
& E[Y(3)] & 2.094 & 2.103 & 0.009 &  0.912 & 0.936 \\
 \cline{2-7} 
 \  & $E[Y(6)]$ & 2.014 & 2.015 & 0.001 & 0.910  & 0.938 \\
 \cline{2-7} 
 \ & $E[Y(9)]$ & 1.968 & 1.976 & 0.008 &  0.924  & 0.948 \\
 \cline{2-7} 
 \ & $E[Y(12)]$ & 1.930 & 1.936 & 0.006 &  0.928  &  0.958 \\
\Xhline{2pt}
\multirow{4}{0.3in}{ 0.3 } 
& E[Y(3)] & 2.472 & 2.478 & 0.006 &  0.906 &  0.934 \\
 \cline{2-7} 
 \  & $E[Y(6)]$ & 2.372 & 2.368 & 0.004 &  0.898 &  0.934 \\
 \cline{2-7} 
 \ & $E[Y(9)]$ & 2.315 & 2.322 & 0.007 & 0.920 &  0.946 \\
 \cline{2-7} 
 \ & $E[Y(12)]$ & 2.269 &  2.271
 &  0.002 &  0.924  &  0.950 \\
\Xhline{2pt}
\multirow{4}{0.3in}{ 0.6 } 
& E[Y(3)] & 2.907 & 2.907 & 0.000 & 0.876 & 0.928 \\
 \cline{2-7} 
 \  & $E[Y(6)]$ & 2.787 & 2.777 & 0.010 &  0.872  & 0.936 \\
 \cline{2-7} 
 \ & $E[Y(9)]$ & 2.721 &  2.723 &  0.002 & 0.918  & 0.948 \\
 \cline{2-7} 
 \ & $E[Y(12)]$ & 2.667 & 2.663 & 0.004 &  0.900  & 0.942\\
\Xhline{2pt}
 \end{tabular}

 \end{center}
 
\smallskip

\begin{singlespace}

\caption{Treatment arm simulation results (N = 200). Shown are the true values of the four target parameters under each of five different data-generating mechanisms that mimic the treatment arm of the ARC data with $\alpha=-0.6,-0.3,0,0.3,0.6$; the empirical mean and absolute value of the empirical bias of the estimators across 500 simulations; and the coverage of Wald confidence intervals using the influence function-based variance estimate (Wald (IF)) and Wald confidence intervals using the jackknife variance estimate (Wald (Jackknife)). }
 
\label{table:sims_trt}

\end{singlespace}

\end{table}

\begin{table}
\begin{center}

\small

\renewcommand{\arraystretch}{1.5}

\begin{tabular}{ c  c  c  c  c  c  c  }
\Xhline{2pt}
$\alpha$  & Parameter & True Value & Emp Mean & $|$Bias$|$  &  Wald (IF)  &  Wald (Jackknife) \\
\Xhline{2pt}
\multirow{4}{0.3in}{ -0.6 } & $E[Y(3)]$ &
 1.441 & 1.437 & 0.004 &0.918 & 0.958\\
 \cline{2-7} 
 \  & $E[Y(6)]$ &
 1.394 & 1.405 & 0.011  & 0.900 & 0.956\\
 \cline{2-7} 
 \ & $E[Y(9)]$ &
 1.360 & 1.363 & 0.003 & 0.926 & 0.954\\
 \cline{2-7} 
 \ & $E[Y(12)]$&
 1.332 & 1.332 &0.000 & 0.918 & 0.958\\
\Xhline{2pt}
\multirow{4}{0.3in}{ -0.3 }
 & E[Y(3)] 
 & 1.754 & 1.744 & 0.010 & 0.922 & 0.948 \\
 \cline{2-7} 
 \  & $E[Y(6)]$ 
  & 1.691 &  1.697 &  0.006 & 0.928 &  0.956 \\
 \cline{2-7} 
 \ & $E[Y(9)]$ 
  & 1.646 & 1.644 &0.002 &0.924 &  0.948 \\
 \cline{2-7} 
 \ & $E[Y(12)]$ 
  & 1.608 & 1.602 &0.006 &  0.906 &  0.950 \\
\Xhline{2pt}
\multirow{4}{0.3in}{ 0 } 
& E[Y(3)] 
 & 2.165 &2.149 & 0.016& 0.918 & 0.948 \\
 \cline{2-7} 
 \  & $E[Y(6)]$ 
   &2.081 & 2.080&0.001 &  0.940 &  0.956 \\
 \cline{2-7} 
 \ & $E[Y(9)]$ 
  &2.021 & 2.014 &0.007 &  0.908 &  0.920 \\
 \cline{2-7} 
 \ & $E[Y(12)]$  
  &1.971 & 1.957 & 0.014 & 0.922 & 0.950 \\
\Xhline{2pt}
\multirow{4}{0.3in}{ 0.3 } 
& E[Y(3)] 
 & 2.675 & 2.652 &0.023 & 0.882 & 0.940 \\
 \cline{2-7} 
 \  & $E[Y(6)]$ 
   &2.567 & 2.562 &0.005 &  0.914 &  0.962 \\
 \cline{2-7} 
 \ & $E[Y(9)]$ 
  &2.493 & 2.480 & 0.013 & 0.910 & 0.936 \\
 \cline{2-7} 
 \ & $E[Y(12)]$  
  &2.431 & 2.405 & 0.026 & 0.918 & 0.950 \\
\Xhline{2pt}
\multirow{4}{0.3in}{ 0.6 } 
& E[Y(3)] 
 & 3.249 & 3.218  & 0.031 & 0.830 &  0.904 \\
 \cline{2-7} 
 \  & $E[Y(6)]$ 
   & 3.125 & 3.114 &  0.011 &  0.894 & 0.944 \\
 \cline{2-7} 
 \ & $E[Y(9)]$ 
  &3.041 &  3.019 &  0.022 &  0.896 &  0.928 \\
 \cline{2-7} 
 \ & $E[Y(12)]$  
  & 2.971 &  2.931 & 0.040 &  0.888 &  0.948 \\
\Xhline{2pt}
 \end{tabular}

  \end{center}

 \begin{singlespace}

\caption{Control arm simulation results (N = 200). Shown are the true values of the four target parameters under each of five different data-generating mechanisms that mimic the control arm of the ARC data with $\alpha=-0.6,-0.3,0,0.3,0.6$; the empirical mean and absolute value of the empirical bias of the estimators across 500 simulations; and the coverage of Wald confidence intervals using the influence function-based variance estimate (Wald (IF)) and Wald confidence intervals using the jackknife variance estimate (Wald (Jackknife)). }

\label{table:sims_ctrl}

\end{singlespace}

\end{table}

\end{document}